\documentclass[apj]{emulateapj}
\usepackage[colorlinks,linkcolor={blue},citecolor={blue},urlcolor={red}]{hyperref}
\usepackage{mathtools}
\usepackage{epsfig,graphicx,natbib,amsmath,amsfonts,amssymb}
\usepackage{color}
\usepackage{multirow}
\usepackage{booktabs}
\usepackage{amssymb}
\usepackage{threeparttablex} 
\usepackage{booktabs} 
\usepackage{mathrsfs}  
\usepackage{lineno}

\usepackage{longtable}
\usepackage{array} 
\usepackage[dvipsnames]{xcolor}  

\newcommand{\msolar}{${\rm M}_\odot$}

\newcommand{\myemail}{\email{enci.wang@phys.ethz.ch; simon.lilly@phys.ethz.ch}}

\shorttitle{The origin of exponential star-forming disks}
\shortauthors{Wang \& Lilly}

\graphicspath{{fig/}}

\received{October 05, 2021}
\accepted{January 09, 2022}

\begin{document}

\title {The origin of exponential star-forming disks 
}

\author {Enci Wang\altaffilmark{1},
Simon J. Lilly\altaffilmark{1}
} \myemail

\altaffiltext{1}{Department of Physics, ETH Zurich, Wolfgang-Pauli-Strasse 27, CH-8093 Zurich, Switzerland}

\begin{abstract}  

The disk components of galaxies generally show an exponential profile extending over several scale lengths, both in mass and star-formation rate, but the physical origin 
is not well understood. We explore a physical model in which the galactic gas disk is viewed as a ``modified accretion disk" in which coplanar gas inflow, driven by viscous stresses in the disk, provides the fuel for star formation, which progressively removes gas as it flows inwards. 
We show that magnetic stresses from magneto-rotational instability are the most plausible source of the required viscosity, and construct a simple physical model to explore this. A key feature is to link the magnetic field strength to the local star-formation surface density, $B_{\rm tot} \propto \Sigma_{\rm SFR}^\alpha$. This provides a feed-back loop between star-formation and the flow of gas. We find that the model naturally produces stable steady-state exponential disks, as long as $\alpha \sim$ 0.15, the value indicated from spatially-resolved observations of nearby galaxies.  The disk scale-length $h_{\rm R}$ is set by the rate at which the disk is fed, by the normalization of the $B_{\rm tot}-\Sigma_{\rm SFR}$ relation and by the circular velocity of the halo. The angular momentum distribution of the gas and stars within the disk is a consequence of the transfer of angular momentum that is inherent to the operation of an accretion disk, rather than the initial angular momentum of the inflowing material. 
We suggest that magnetic stresses 
likely play a major role in establishing the stable exponential form of galactic disks.

\end{abstract}
\keywords{galaxies: general -- galaxies: structure -- galaxies: formation -- galaxies: magnetic fields --  ISM: magnetic fields}

\section{Introduction}
\label{sec:1}

The surface brightness profiles, $I(r)$, of disk galaxies are generally found to have two components \citep[e.g.][]{de-Vaucouleurs-59, Freeman-70, Kent-84,  Allen-86, Weiner-01, Simard-11, Casasola-17}: an inner spheroidal component (bulge), 
and a highly flattened disk that is observed to have an nearly exponential profile $\log I(r) \propto -r/h_{\rm R}$, where $h_{\rm R}$ is the scale-length of disk. In typical galaxies, the exponential profile of the disk is maintained for four scale-lengths, or more \citep[e.g.][]{Kent-85, Weiner-01, Pohlen-06, Simard-11, Meert-13}, corresponding to a large dynamic range in brightness of 2 dex. 
A roughly exponential stellar disk appears to be quite normal among disk galaxies, although some deviations from the pure exponential function are seen. Based on a complete sample of nearby weakly inclined galaxies,  \cite{Pohlen-06} found that only 10\% of the sample galaxy have a pure exponential disk all the way down to the noise limit, while 30\% show clear downbending features, and and 60\% show upbending features in the surface brightness profiles \citep{Erwin-05, Hunter-06, Meert-15}.  

The characteristic exponential profile is found not only in the radial distribution of the stars, but also in the radial distribution of molecular gas \citep{Bigiel-08, Leroy-09} and, of most importance for this paper, in the radial profile of the star formation rate (SFR) surface density ($\Sigma_{\rm SFR}$) \citep[e.g.][]{Bigiel-08, Wyder-09, Gonzalez-Lopezlira-12, Gonzalez-Delgado-16, Casasola-17, Wang-19}. The $\Sigma_{\rm SFR}(r)$ profile may be traced by the ultraviolet continuum radiation of newly formed massive stars, in the thermal infrared emission of dust, and in the H$\alpha$ emission of ionized gas. 

\begin{figure*}
  \begin{center}
    \epsfig{figure=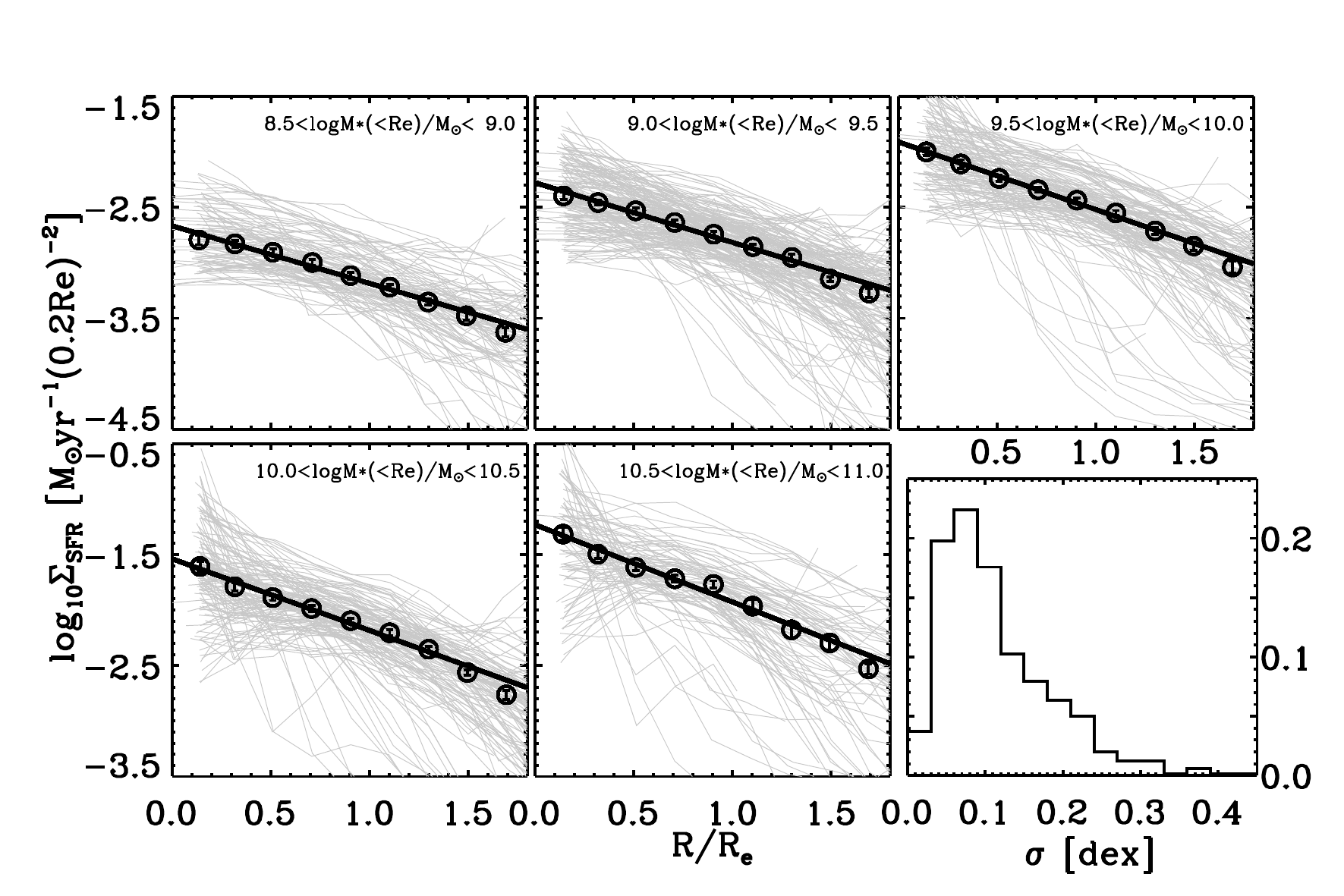,clip=true,width=0.9\textwidth}
    \end{center}
  \caption{The first five panels: the $\Sigma_{\rm SFR}$ profiles for individual Main Sequence galaxies taken from \cite{Wang-19}. In presenting the profiles,  
  we use the normalized radius r/$R_{\rm e}$, as well as the normalized surface density (i.e., a surface density that is computed as the SFR in an area that itself scales as $R_{\rm e}^2$). This ensures that the (visual) integration of a profile on a given surface density-radius diagram reflects the actual integrated quantity in physical terms.
  The sample galaxies are all within $\pm$0.33 dex of the SFMS on the mass-SFR diagram, and excludes mergers and heavily disturbed galaxies \citep[see details in][]{Wang-18}.
  In each stellar mass bin, the black circles show the median profiles of the galaxy population, and the black line shows an exponential profile fit to this median profile. For each individual galaxy, we fit an exponential function to the $\Sigma_{\rm SFR}$ and calculate the r.m.s. deviations of the $\Sigma_{\rm SFR}$ profile from the fitted exponential function. 
  The bottom right panel: the distribution of the r.m.s. deviations of the sample galaxies. 
  }
  \label{fig:0}
\end{figure*}

Figure \ref{fig:0} shows the $\Sigma_{\rm SFR}$ profiles for a large and well-defined sample of 621 SF galaxies, taken from \cite{Wang-19}. These galaxies span the mass range 8.5$<\log M_*(<R_{\rm e})/{\rm M_{\odot}}<$11.0 (split in five mass-bins in the Figure), and are selected to have integrated star-formation rates (within the effective radius $R_{\rm e}$) that are within $\pm$ 0.33 dex of the Main Sequence \citep[e.g.][]{Noeske-07}.  The solid points in each panel show the median of the $\Sigma_{\rm SFR}$ profiles in each mass bin.  This clearly has an exponential form over the range of radius probed by these data.  

Individual galaxies do show variations around these median exponential profiles. These could be due to temporal variations, including the effect of local spiral arms, or other effects. 
The histogram in the bottom right panel of Figure \ref{fig:0} shows the distribution of the r.m.s. deviations of the individual $\Sigma_{\rm SFR}$ profiles relative to pure exponential functions.
More than half of the galaxies have 
r.m.s. deviations of $\sigma < 0.1$ dex, and the vast majority (86\%) have $\sigma < 0.2$ dex.  Exponential star-forming disks seem to be a very common feature of galaxies, especially for galaxies in the low and intermediate stellar mass bins. 

For several decades, efforts have been made to understand the origin of this seemingly universal phenomenon of exponential stellar disks. However, a convincing explanation of the exponential form is still, in our view, lacking. 

The radial mass distribution of a disk is closely linked to the angular momentum distribution of the material, since gas and stars move on more-or-less circular orbits within the disk, the whole system being embedded within a (generally dark-matter dominated) gravitational potential well. Conservation of specific angular momentum is often invoked to explain the angular momentum distribution of the material in galaxy disks.
More than fifty years ago, \cite{Freeman-70} pointed out that a self-gravitating exponential disk has an angular momentum distribution that is almost identical to that of a uniformly rotating sphere of uniform density.  This meshed with the idea \citep{Mestel-63} that each element in a collapsing proto-galaxy might preserve its specific angular momentum during the collapse of the galaxy. This basic concept has been developed in many subsequent discussions that have been based on realistic N-body simulations of the formation of cosmic structure \citep[e.g.][]{Fall-80, Mo-98, Dutton-09}. These have ultimately tried to link the specific angular momentum distribution of the baryonic material in galactic disks to the specific angular momentum distribution of the original material, which is taken to match that of the dark matter particles in the host halos.  

However, the underlying assumption in this picture, that there is no significant exchange or re-distribution of the angular momentum within the baryonic matter, seems somewhat implausible.  It is also hard to see why the disk would maintain the exponential structure while it is being continuously fed by accreting cold gas from the halo.  

Another idea for explaining exponential disks, at least in the stellar mass distribution, is that this profile is the result of radial redistribution, via secular evolution processes, of stars that are formed in the disk.  This idea also has a long history.  \cite{Hohl-71} found that exponential stellar disks are naturally obtained after bar formation, even when the initial disk is not exponential \citep[see also][]{Debattista-06, Foyle-08}.  \cite{Elmegreen-13} found that stellar scattering off of massive clumps in a disk could also lead to the formation of an exponential stellar disk, in a two-dimensional non-self gravitating stellar disk rotating within a fixed halo potential \citep[see also][]{Wu-20}. 
More recently, \cite{Herpich-17} have proposed that the exponential stellar disk can be explained as the maximum entropy state for the distribution of specific angular momentum of stars, if the radial migration of stars is efficient enough.
However, while these models involving mass redistribution can clearly produce the exponential {\it stellar} disks, they do not explain the exponential form of the $\Sigma_{\rm SFR}$ star-formation profiles of disk galaxies \citep[e.g.][]{Wyder-09, Gonzalez-Delgado-16, Casasola-17, Guo-19, Wang-19}. 
Indeed, the similarity of these profiles would have to be something of a coincidence in these redistribution scenarios.  There is no clear reason that the gas or newly formed stars should follow a similar radial distribution to the redistributed stars, since there should be little interaction between cold gas and long-lived stars except through gravity.  

If the $\Sigma_{\rm SFR}(r)$ star-formation profiles of SF galaxies are of exponential form, as we argue they are, then an exponential stellar mass disk can naturally be obtained from the time-integration of $\Sigma_{\rm SFR}(r)$. Radial redistribution of the stars is not therefore required.  We stress however that it could well be occurring if it maintains the exponential form of the mass distribution established by the $\Sigma_{\rm SFR}(r)$ profile \citep[see][]{Vera-Ciro-14}. 

The radial transport of gas within a viscous gas disk may also potentially offer a solution to the exponential disk problem. 
\cite{Lin-87} first proposed that the effective viscosity of a gas disk could redistribute the angular momentum, and thereby cause radial gas inflow.  They then argued that the observed exponential form of stellar disks could be the result of the evolution of a star-forming viscous disk, under the (strong) condition that the timescale of the redistribution of angular momentum should be of the same order of magnitude as the gas depletion time scale in the disk \citep[also see][]{Yoshii-89, Firmani-96, Ferguson-01, Wang-09}. However, it was unclear in this picture what physical mechanism actually produced the viscosity, and why the condition that the timescales for angular momentum transfer and for star formation within the disk should be the same, would be satisfied.  In this picture, the key to understanding the formation of the disk is how the gas inflow happens and the form of the star-formation profile.

In this paper, we will explore further the idea discussed above that the profile of the star-forming disk reflects the operation of viscosity within the gaseous disk.  
We explore the possibility that the establishment of a steady-state star-formation profile might be produced by some kind of feed-back loop operating between star-formation and the inward flow of gas. Such a feed-back loop might conceivably operate through the effect of the star-formation on the viscosity of the gas in the disk. 

Our investigation is carried out in a conceptual framework in which we consider the gas disk of star-forming galaxies to be a ``{\it modified 
accretion disk}''.
Quite apart from the very different physical scale, such a galactic-scale ``accretion disk" would clearly be very different from the classical accretion disks found around black holes, which is why we call it ``modified".  In particular, in a classical accretion disk the mass inflow rate is essentially independent of radius. In a galactic disk, star formation within the galactic disk will continually consume the gas as it moves towards the center. Supernova-driven winds may also act to remove gas from the disk.  The rate of mass inflow within the disk will therefore be far from constant and, assuming that there is no substantial mass sink at the center, it must approach zero towards the center of the system. 

Is such a ``modified accretion disk" picture valid for a galactic disk?  Although the detailed inflow and outflow of gas in galaxies is not well understood, both observations and hydrodynamical simulations have provided insights on the orientation of gas flows in recent years.  
Studies of metal absorption systems suggest that inflows of cool gas onto the system appear to be largely coplanar while supernova-driven outflows are mostly explanar, along the axis of the system
\citep[also see][]{Bordoloi-11, Bouche-12, Kacprzak-12, Schroetter-19}. Consistent with this, hydrodynamical simulations have also shown that the inflowing gas is preferentially coplanar, and that the outflowing gas preferentially leaves the galaxy along the path of least resistance, i.e. along the direction perpendicular to the disk. This further prevents infall of material from regions above and below the plane of the disk \citep{Brook-11, Mitchell-20, DeFelippis-20, Peroux-20, Trapp-21}. As a result, the gas inflowing onto the disk is mostly coplanar and co-rotating with the disk \citep[e.g.][]{Stewart-11, Peroux-20}, while the outflowing gas is ejected perpendicular to the disk \citep{Peroux-20, Trapp-21}. 

These observational and theoretical results support the idea that it is not unreasonable to consider the gas disks of galaxies to be ``modified accretion disks", and we therefore investigate this idea further in this work. In particular we will examine the possible origin for the viscosity that is required for the accretion disk to function, and will establish under what conditions an {\it exponential} star-forming $\Sigma_{\rm SFR}$ disk may be established.  

The basic operation of any viscous accretion disk is to transport mass inward and angular momentum outward \citep[e.g.][]{Lynden-Bell-74, Pringle-81, Lin-87, Yoshii-89, Wang-09}.
The classical viscosity of a gas disk can be produced by cloud-cloud collisions, by turbulence of the gas disk from supernova feedback and/or the motions produced by gravitational instabilities of gas clouds \citep[e.g.][]{Lynden-Bell-74, Pringle-81, Ferguson-01, Stevens-16}. 
In addition to these classical sources of viscosity, magnetic fields can also play an important role in accretion disk dynamics \citep[e.g.][]{Shakura-73, Blandford-NG, Balbus-91}.  Magnetic fields are now generally accepted to be the main source of viscosity in the classical accretion disks around compact objects. 

\cite{Shakura-73} proposed that magnetic turbulence could act as a viscous couple, but argued that nonlinear perturbations are required to disrupt laminar flow. Later, \cite{Balbus-91} found that the combination of a negative gradient in the angular velocity with a weak magnetic field of any plausible astrophysical strength would lead to a dynamical instability. This shearing instability is known as the Magneto-rotational instability \footnote{MRI is also sometimes called Velikhov–Chandrasekhar instability or the Balbus–Hawley instability in the literature.}(MRI). Based on a 3-dimensional magnetohydrodynamical (MHD) simulation, \cite{Hawley-95} found that the transportation of angular momentum is dominated for Keplerian disks by the magnetic stress (or Maxwell stress), rather than by the kinetic stress (or Reynolds stress).  
Magnetic stresses may also play a role in galactic gas disks, provided that there is dynamical coupling between the ionized and neutral material, which is likely. 
The relative role of magnetic and kinetic stresses in shaping the gas distribution in galactic disks has not, to our knowledge, been examined to date.  

In this paper, we will investigate the gas disks of galaxies in this simple ``modified accretion disk" framework in order to try to better understand the formation of the exponential $\Sigma_{\rm SFR}(r)$ profiles of the disks. We wish to understand under what conditions such profiles are established, whether they are stable, and what sets their exponential scale lengths.

The layout of the paper is as follows.  We first reverse-engineer the problem and determine, in Section \ref{sec:2}, the viscous stress that would be {\it required} within such a disk in order to sustain a {\it perfect} exponential $\Sigma_{\rm SFR}(r)$ profile. 
In Section \ref{sec:3}, we then compare this ``required" viscous stress with the expected strength of various sources of kinetic and magnetic stress using (independent) observational properties of galactic disks. We find the magnetic stress is a promising candidate to provide the required viscous stress.  
Motivated by this, we then construct in Section \ref{sec:4} a simple model to explore the operation of a galactic-scale ``modified accretion disk" in which MRI-induced viscosity dominates the viscosity of the gas disk.  
In Section \ref{sec:5}, we perform several runs with this physical model, 
and find that stable exponential star-forming disks with reasonable scale-lengths can easily be established. 
We explore the angular momentum transportation within such disks in Section \ref{sec:5.4X}. 
In Section \ref{sec:6}, we then derive a general scaling relation expected for the scale-length $h_{\rm R}$ of the disks relative to other parameters.  
Finally, in Section \ref{sec:7} we discuss a number of different aspects of our analysis and their implications.


\section{Galactic gas disks as ``modified accretion disks"} \label{sec:2}

\begin{figure*}
  \begin{center}
    \epsfig{figure=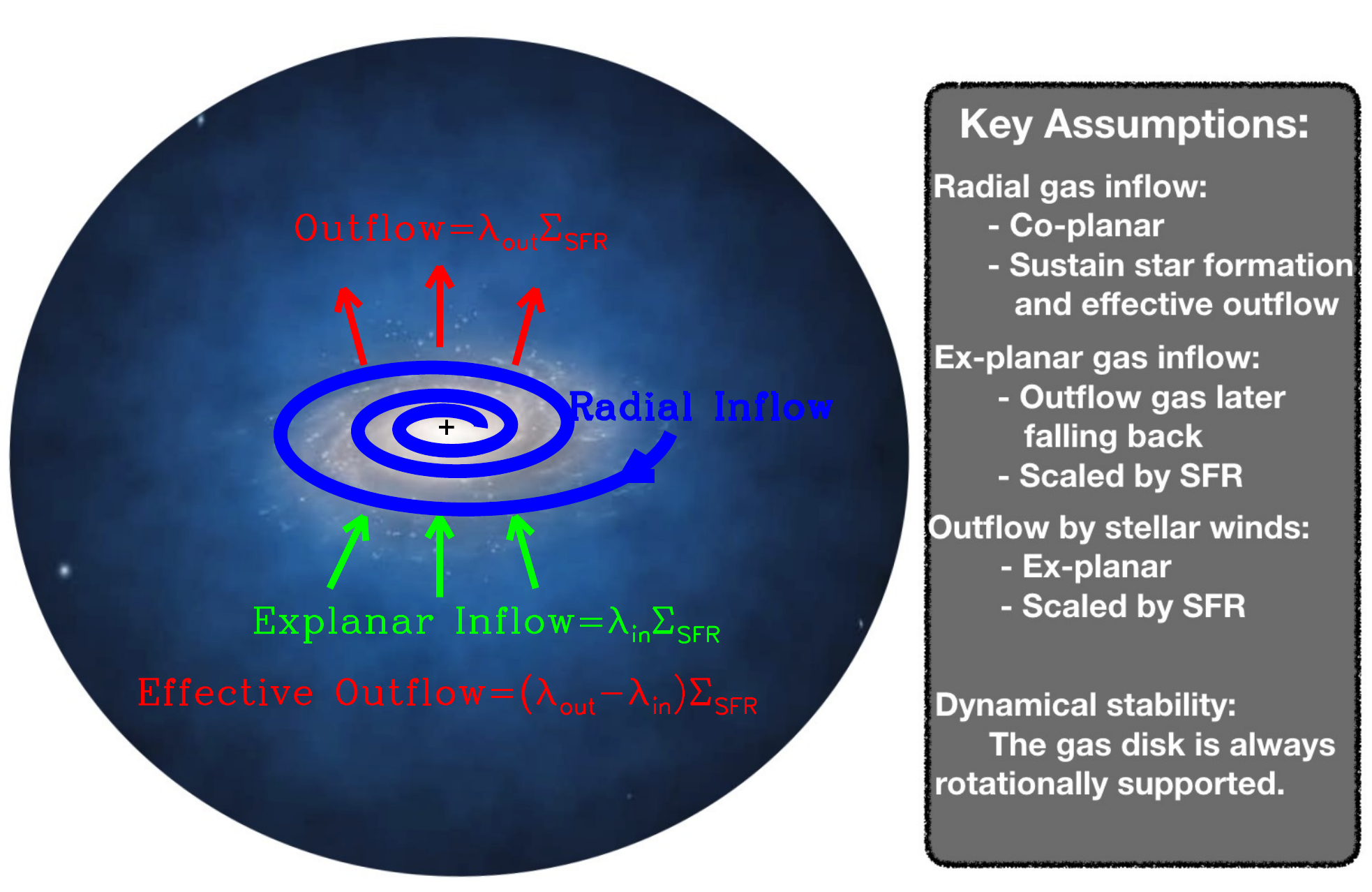, clip=true,width=0.8\textwidth}
    \end{center}
  \caption{Illustration of the modified accretion disk model for spiral galaxies. The SFR surface density is instantaneously determined by the cold gas surface density, which is locally regulated by the gas inflow, wind-driven outflow and star formation.  There are two mode of gas inflow: radial gas inflow (blue curve) and ex-planar gas inflow (green arrows). The inflow is assumed to be dominated by the radial coplanar inflow, which provides the fuel to sustain the star formation and outflow.  The ex-planar inflow is the inflow due to the falling of the ex-planar gas, which is originally from the disk but stripped away by SNe feedback at early times. Therefore, one may expect that the ex-planar inflow rate is less than the outflow rate, i.e. $\lambda_{\rm in}<\lambda_{\rm out}$. We define the effective mass-loading factor, $\lambda = \lambda_{\rm out} - \lambda_{\rm in}$. The ex-planar inflow and outflow should happen in both side of the disk plan, although the arrows here only show one-side of inflow or outflow. 
  We assume that the gas disk is fully rotationally-supported, i.e. the circular velocity of gas always follows the gravitational potential of the system.   
 The original picture\footnote{The original picture is credited by ESO/L Calcada, which is downloaded from https://cerncourier.com/a/the-milky-ways-dark-matter-halo-reappears.} is a schematic picture to depict the  Milky Way embedded in a spherical halo of dark matter (shown in blue).
  }
  \label{fig:1}
\end{figure*}


\subsection{Assumptions of the modified accretion disk} \label{sec:2.1}

In the evolution of the galactic disk, the continuous accretion of cold gas is needed to sustain star formation \citep{Binney-00, Keres-05, Dekel-06, Sancisi-08, Silk-12, Conselice-13, Stern-20} and to maintain the size growth of the disk \citep[e.g.][]{Mosleh-12, Lilly-16}. This accretion of new gas involves smooth accretion from the CGM and mergers with dwarf companions. By using multi-zoom cosmological simulations, \cite{LHuillier-12} quantified these two modes of accretion, and found that most galaxies assemble the majority of their mass through smooth accretion, while mergers are important only for the most massive galaxies at high redshift.  
In this paper, we will therefore only consider the smooth accretion of gas onto our ``modified accretion disk". Mergers and bulk inflow of large gas clouds can lead to the dynamical instability of the gas disk, and are not included here. 

The most important assumption of our ``modified accretion disk" model is that the {\it dominant} flow of gas that fuels star-formation in the disk is the {\it radial} inflow of gas within the plane of the disk. As indicated by recent observations and simulations \citep{Brook-11,Kacprzak-12, Schroetter-19, Peroux-20, Trapp-21}, the coplanar radial inflow of cold gas is expected to dominate the smooth accretion. Coplanar accretion from the surrounding environment can then be represented by a mass-inflow rate of co-rotating gas at some radius. We may regard this radius for convenience to be the nominal outer boundary of our ``modified accretion disk".  

We will assume that any wind-driven outflow from the disk is explanar and that it scales with the local $\Sigma_{\rm SFR}$ via a conventional mass-loading factor $\lambda_{\rm out}$. We will assume that the surface mass loss rate perpendicular to the disk is given by $\lambda_{\rm out}\Sigma_{\rm SFR}$. 

Clearly the effect of a non-zero mass-loading is simply to reduce the mass of stars that are produced when a given mass of gas is removed from the disk. Seen in this way, this has a trivial effect on the model, simply scaling the final stellar mass density.   

The effects of modest amounts of ex-planar {\it inflow} can also be included in an analogous way, if we assume that the surface inflow rate of this ex-planar gas also scales with $\Sigma_{\rm SFR}$.  
This gas could originally have been removed from the gas disk by SNe feedback at early times, perhaps amplified by the addition of further material from the hot halo \citep[e.g.][]{Fraternali-15}. If we can indeed represent this inflow as $\lambda_{\rm in}\Sigma_{\rm SFR}$, then the two effects of outflow and inflow may be combined into a single effective surface mass outflow rate given by $(\lambda_{\rm out}-\lambda_{\rm in})\Sigma_{\rm SFR}$. We therefore define the ``effective" mass-loading factor $\lambda$ as $\lambda = \lambda_{\rm out}-\lambda_{\rm in}$. Throughout the rest of the paper, the mass-loading $\lambda$ in the rest of this paper will refer to this ``effective" outflow.  

In principle, both $\lambda_{\rm in}$ and $\lambda_{\rm out}$ could be a function of galactic radius. However, for simplicity, the effective $\lambda$ will generally be assumed to be the same at all radii.  We may expect that it could vary from galaxy to galaxy and over cosmological timescales. We will however discuss the effect of a radially-dependent $\lambda$ in Section \ref{sec:5.2}. 

A slightly negative value of $\lambda$ is also easily accommodated: effectively, the removal of a given mass of gas from the disk is associated with the production of a {\it larger} mass of stars, leading to a similar scaling of the stellar mass density.  Only as $\lambda$ becomes {\it significantly} negative and approaches $-1$ will the basic assumptions of the modified accretion disk be violated, because in this case the gas for star-formation at a given radius will be dominated by the explanar accretion and not by the coplanar inflow.   Figure \ref{fig:1} illustrates the assumptions about gas-flow in the modified accretion disk model for galactic disks. 

Another important assumption is that the gas in the disk, and also in any explanar outflows or inflows, is always fully rotationally-supported, i.e. that these components all have the tangential velocity equal to the circular velocity of the gravitational potential at that radius.  All three components (resident gas, outflow, inflow) at a given radius in the disk will therefore have the same specific angular momentum, which will be defined by the nominal circular velocity of the potential well and the radius. The removal or addition of gas to the disk therefore does not change the specific angular momentum of the gas in the disk \citep[c.f.][]{Pezzulli-16}. As an aside, this assumption means that the model may not be appropriate for the disks of Irregular galaxies and highly-disturbed Spiral galaxies. 

We further assume that the gas within the disk acts as a ``gas-regulator" system \citep[e.g.][]{Lilly-13, Wang-19} in which the surface star-formation rate $\Sigma_{\rm SFR}$ is instantaneously determined by the gas surface density ($\Sigma_{\rm gas}$) at each point in the disk, via \cite{Kennicutt-98} star-formation law \citep[also see][]{Yoshii-89, Firmani-96, Bouche-10, Schaye-10, Dave-11, Wang-20, Wang-21}.  This star-formation law gives a local star-formation efficiency $\Sigma_{\rm SFR}$/$\Sigma_{\rm gas}$.  Again, for simplicity, this star-formation law is assumed to be time-invariant.  


Finally, for simplicity, we assume that the gravitational potential in which the gas disk evolves does not change with time, at least on the timescales of interest.  In particular, we do not consider the change in the potential due to the emergence of the disk itself.

We will primarily be concerned with the radial profile of {\it star-formation} $\Sigma_{\rm SFR}(r)$ rather than of stellar mass.  Detailed questions about whether stars remain at the galactocentric radius at which they formed or whether any pre-formed stars are brought into the galaxy by mergers therefore need not concern us.

\subsection{Basic Equations} \label{sec:2.1X}

Based on the assumptions given in Section \ref{sec:2.1}, we may then develop the basic equations governing the ``modified accretion disk". The continuity equation for the disk gas mass at a given radius can be written as \citep[e.g.][]{Yoshii-89, Firmani-96}: 
\begin{equation} \label{eq:1}
    \frac{\partial \Sigma_{\rm gas}}{\partial t} = \frac{\partial \Phi}{2\pi r \partial r} - (1-R+\lambda)\cdot \Sigma_{\rm SFR}, 
\end{equation}
where the $\Phi$ is the {\it net} coplanar radial inflow rate at that radius,  and $R$ is the fraction of stellar mass that is returned to the interstellar medium through winds and supernova explosions. We adopt the instantaneous recycling approximation, and take $R =$ 0.4 from stellar population models \citep[e.g.][]{Bruzual-03}. 
The first term on the right-hand side of Equation \ref{eq:1} gives the change of $\Sigma_{\rm gas}$ due to the gradient in the coplanar radial inflow $\Phi(r)$, while the second term gives the effects of star formation and any explanar outflows/inflows (the latter combined into the effective $\lambda$ as discussed in the previous section). 

The change in the stellar mass surface density $\Sigma_*$ can also be trivially written as: 
\begin{equation} \label{eq:2}
    \frac{\partial \Sigma_*}{\partial t} = (1-R)\cdot \Sigma_{\rm SFR}. 
\end{equation}

The conservation of angular momentum of the gas in the disk at a given radius can be written as \citep[also see][]{Yoshii-89, Firmani-96}: 
\begin{equation} \label{eq:3}
    \frac{\partial (\Sigma_{\rm gas}r^2\Omega)}{\partial t} = \frac{\partial (\Phi r^2 \Omega)}{2\pi r\partial r} -r^2\Omega(1-R+\lambda)\Sigma_{\rm SFR} - \frac{\partial \mathscr{G}}{2\pi r \partial r},
\end{equation}
where $\Omega$ is the angular velocity and the $\mathscr{G}$ is the torque due to viscosity within the disk.    The first term on the right-hand side of Equation \ref{eq:3} gives the change of angular momentum surface density due to the coplanar radial inflow, the second term gives the loss of angular momentum from the gas because of the removal of gas by star formation and outflow, and the third term represents the effect of a viscous torque.  
We define the viscous stress as the viscous force per unit length around the circular circumference at a given $r$, and denote this as $W$. The torque $\mathscr{G}$ can then be written as:
\begin{equation} \label{eq:4}
    \mathscr{G} = 2\pi r^2 W. 
\end{equation}

At this point, we have not yet specified the physical origin of the viscous stress.  This could be classical kinetic stress and/or magnetic stress \citep[e.g.][]{Shakura-73, Lynden-Bell-74, Balbus-91, Balbus-99}. 

Combining with Equation \ref{eq:1}, Equation \ref{eq:3} can then be simplified as follows: 
\begin{equation} \label{eq:5}
     2\pi r^3\Sigma_{\rm gas}\cdot \frac{\partial \Omega}{\partial t} = 
\Phi \cdot \frac{\partial (\Omega r^2)}{\partial r} -
\frac{\partial }{\partial r}(2\pi r^2 W) 
\end{equation}
This equation connects the radial inflow $\Phi(r)$, the viscous stress $W(r)$ and the rotation curve $\Omega (r)$.  

An important point to note is that, if the gravitational potential does not evolve with time, i.e. if $\partial \Omega/\partial t \sim 0$, so that the left hand side is zero, then the coplanar radial mass inflow rate $\Phi(r)$ is determined {\it only} by the viscous stress $W$ and the form of the gravitational potential $\Omega(r)$.  Further, in a steady-state situation, Equation \ref{eq:1} directly links $\Phi(r)$ to the $\Sigma_{\rm SFR}$.  In other words, once the rotation curve is specified, there should be a direct and fully-defined relationship between the viscous stress $W(r)$ and $\Sigma_{\rm SFR}$ star-formation profile in any steady-state solution.

However, the quantities in the above equations, including $\Sigma_{\rm gas}$, $\Sigma_{\rm SFR}$, $\Phi$, $\mathscr{G}$ and $W$, may all be functions of both time and galactic radius (recall that we assumed $\Omega(r)$ is however time-invariant).  One of the main goals of this work is therefore to examine whether steady-state solutions with interesting radial profiles are established when starting from arbitrary initial conditions. 

In a companion paper (E. Wang \& S.J. Lilly 2022), we investigate the metal-enrichment of gas disks treated as modified accretion disks. We find that the observed gas-phase metallicity profiles of nearby galaxies can be reproduced by our simple model. This can be treated as an independent consistency check of this model.

\subsection{Comparison with the classical accretion disks around compact objects} \label{sec:2.2}

The ``modified accretion disk" model is clearly different in several important aspects from the classical accretion disks around black holes or other compact objects. These differences are briefly listed below.

\begin{itemize}

\item In the classical disks around compact objects, it can usually be assumed that gas is not lost from the disk, i.e. that the mass inflow rate is independent of radius.
In contrast, the radial mass inflow rate in a {\it galactic} disk will steadily decrease towards the center of the system, due to the effects of star formation and any (net) outflows and will approach zero at the center of the system. In a steady-state situation, the net radial mass inflow rate at any given radius should simply and precisely equate to the integrated rate of star formation (plus any associated net outflow) within that radius (see Equation \ref{eq:1}), plus any central sink term.

\item The accretion disk around a compact object is close to Keplerian, i.e. $\Omega \propto r^{-3/2}$, while the circular rotation velocity of galactic disks first increases in the very inner regions and then flattens to a more-or-less constant value beyond the stellar disk, i.e. $\Omega \propto r^{-1}$ \citep[e.g.][]{de-Blok-08, Miller-11}.

\item The gas in the classical accretion disk around a compact object is usually very hot, due to the release of gravitational potential energy as the gas spirals into inner orbits. However, the gas disks of galaxies are usually cold, being composed of molecular and atomic gas.  The inner regions of the gas disk are usually dominated by molecular gas, and the outer region is dominated by atomic gas \citep{Bigiel-08, Bigiel-10}.

\end{itemize}

\subsection{The radial inflow rate and inflow velocity in steady-state exponential disks} \label{sec:2.3}

\begin{figure*}
  \begin{center}
    \epsfig{figure=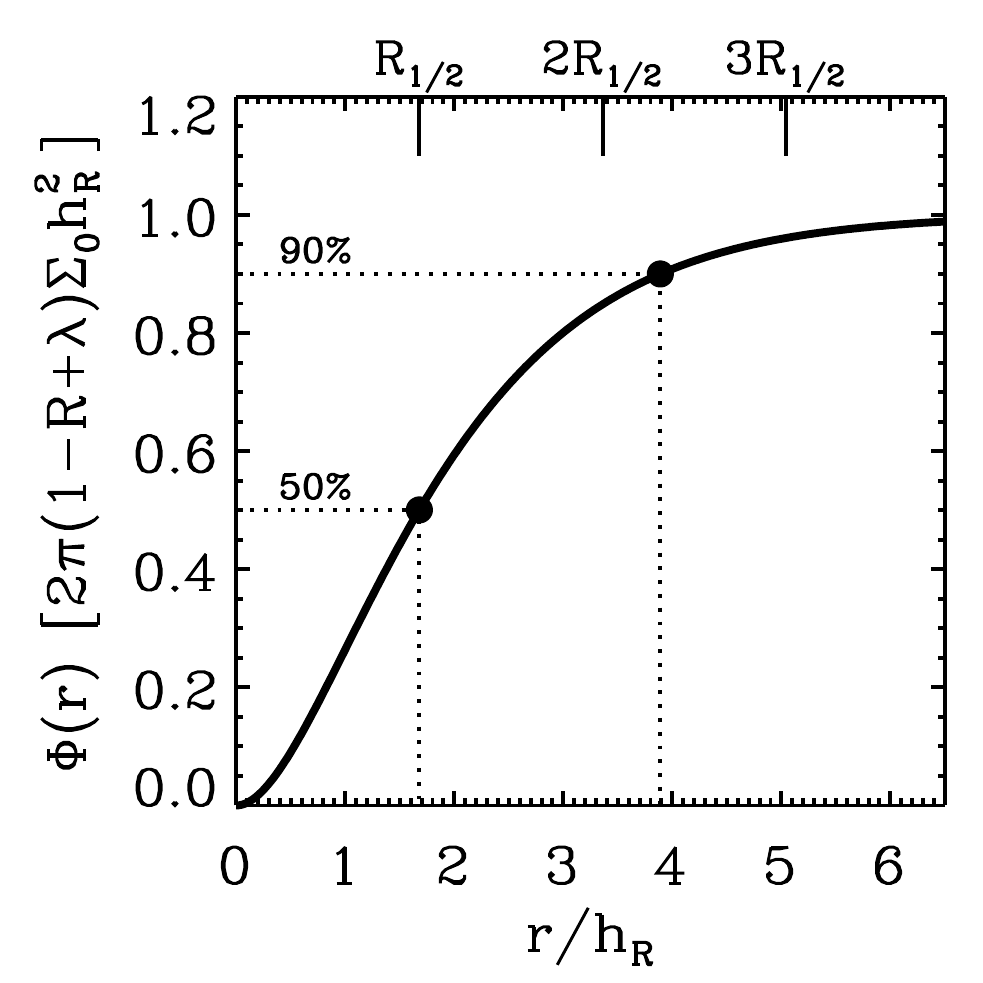,clip=true,width=0.42\textwidth}
    \epsfig{figure=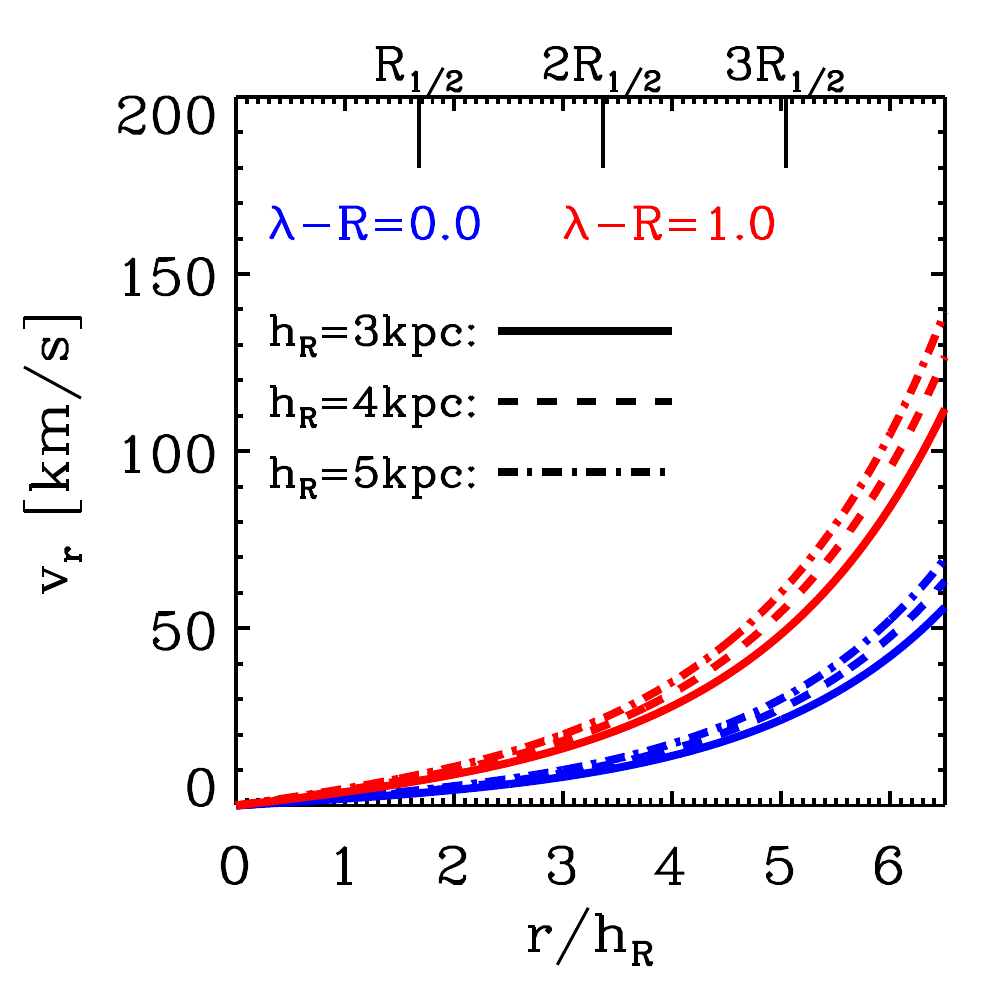,clip=true,width=0.42\textwidth}
    \end{center}
  \caption{
  Left panel: The radial profile of the coplanar inflow rate.
  The shown radial inflow rate is dimensionless,
  which is normalized by $(1-R+\lambda)$SFR. 
  This normalized $\Phi$ is identical to the normalized cumulative SFR with radius. The radii that enclose 50\% and 90\% star formation are indicated by dotted lines. We also show the radius in terms of the half-SFR radius $R_{\rm 1/2}$ ($= 1.68h_{\rm R}$ for an exponential disk) on the top axis.
  Right panel: The radial inflow velocity as a function of radius for a typical disk galaxy of $M_*=3\times10^{10}$\msolar. Different colors indicate different mass-loading factors, and different line-styles indicates different scale-lengths assumed for this galaxy, as denoted in the figure. 
  }
  \label{fig:2}
\end{figure*}

Based on the equations in Section \ref{sec:2.1X}, we now investigate the analytic solutions for the coplanar radial inflow rate in a {\it steady-state} gas disk, i.e. 
\begin{equation} \label{eq:6}
    \frac{\partial \Sigma_{\rm gas}}{\partial t} =0, {\rm and} \ \frac{\partial \Omega}{\partial t} = 0
\end{equation}

Inspired by the existence of exponential star-forming disks \citep[Figure \ref{fig:0}, also][]{Gonzalez-Delgado-16, Casasola-17, Wang-19}, we look at steady-state solutions in which the SFR surface density of the disk $\Sigma_{\rm SFR}(r)$ has an exact exponential decline with radius: 
\begin{equation} \label{eq:7}
    \Sigma_{\rm SFR}(r) = \Sigma_0 \cdot e^{-r/h_{\rm R}}. 
\end{equation}
The purpose of this is to try to ``reverse-engineer" the problem so as to gain insights into the underlying physical processes.

Substituting Equations \ref{eq:6} and \ref{eq:7} into Equation \ref{eq:1}, we can directly obtain the analytic solution of the radial inflow rate as: 
\begin{equation} \label{eq:8}
    \Phi(r) = (1-R+\lambda) {\rm SFR} \cdot [1-(x+1)\cdot \exp(-x)]
    + \Phi_{\rm BH}, 
\end{equation}
where $x$ is a renormalized radius defined as $x=r/h_{\rm R}$.  Here, ${\rm SFR}$ is the {\it total} SFR of the whole disk (${\rm SFR}=2\pi \Sigma_0 h_{\rm R}^2$), and $\Phi_{\rm BH}$ is the inflow rate at the galactic center ($r=0$) onto a central mass sink. As the subscript implies, we consider this to be a black hole, but the nature of the central mass sink is immaterial to the operation of the disk. 

For simplicity, we assume that the inflow rate at the center is a factor $\eta$ of the inflow rate required by the overall star formation and wind-driven outflow of the entire disk, i.e. $\Phi_{\rm BH} = \eta  (1-R+\lambda){\rm SFR}$. The $\eta$ is likely to be of order of, or greater than, $\sim0.001$, the typical ratio of the black halo mass to the stellar mass of galaxies.  Therefore, the Equation \ref{eq:8} can be written as: 
\begin{equation} \label{eq:9}
     \frac{\Phi(r)}{1-R+\lambda} = {\rm SFR} \cdot [1+\eta-(x+1)\cdot \exp(-x)].
\end{equation}

Equation \ref{eq:9} emphasizes an important point made in the previous Section. Once the effective mass-loading factor $\lambda$ (and central sink term $\eta$) are specified, then the required radial inflow rate $\Phi(r)$ is {\it fully determined} once the $\Sigma_{\rm SFR}(r)$ star-formation profile is fixed.   

The left panel of Figure \ref{fig:2} shows the coplanar inflow rate $\Phi(r)$, normalized by the total $(1-R+\lambda)$SFR as a function of $r/h_{\rm R}$, that is obtained with $\eta = 0.0$.   The effect of a non-zero $\eta$ is trivially to displace the vertical origin of the curve.

We note that because we assume constant $\lambda$ and mass return factor $R$, the shape of this normalized $\Phi(r)$ is identical to the normalized {\it cumulative} SFR with radius, integrating out from the center.  
The $\Phi(r)$ therefore monotonically increases with radius, and saturates at large radii ($\sim 4h_{\rm R}$).  Even if the disk extends indefinitely, more than 90\% of the star formation in the exponential disk occurs within the first 4 scale-lengths of the disk. This is equivalent to 2.4 half-SFR radii ($R_{\rm 1/2}$), as shown along the top of the panel. 

The radial inflow velocity $v_{\rm r}$ of the gas in the disk at a given radius can then be written in terms of the mass-inflow rate $\Phi$ at that radius as: 
\begin{equation} \label{eq:10}
     v_{\rm r} = \frac{\Phi}{2\pi r\Sigma_{\rm gas}} = h_{
     \rm R} \cdot \frac{\Phi(x)}{2\pi x \Sigma_{\rm gas}}
\end{equation} 
Based on Equation \ref{eq:9} and \ref{eq:10}, the $v_{\rm r}$ strongly depends on the effective mass-loading factor $\lambda$, the scale-length $h_{\rm R}$ of $\Sigma_{\rm SFR}$ and on the $\Sigma_{\rm gas}$ which follows, via the star-formation law, from the $\Sigma_{\rm SFR}$.

To get a feel for the required form of $v_r(r)$, we show in the right panel of Figure \ref{fig:2} the $v_r$ as a function of radius for a typical Main Sequence galaxy with a stellar mass of $M_*=3\times10^{10}$\msolar.  The SFR of this typical galaxy is set to be 3.5 \msolar${\rm yr}^{-1}$ adopting the star formation main sequence from \cite{Lilly-16} 
at a redshift of zero \citep[also see][]{Noeske-07, Speagle-14, Renzini-15}. In computing the $v_{\rm r}$, the $\Sigma_{\rm gas}$ is obtained using the star formation law from \cite{Kennicutt-98}: 
\begin{equation} \label{eq:11}
    \Sigma_{\rm SFR} = 2.5\times 10^{-4} \cdot (\frac{\Sigma_{\rm gas}}{\rm 1 \ M_{\odot}pc^{-2}})^{1.4} \ \ {\rm M_{\odot}yr^{-1}kpc^{-2}}
\end{equation}
For illustration, we set $h_{\rm R} =$ 3, 4, or 5 kpc, and consider two values of $(\lambda-R) = $0 and 1.  

As shown, the $v_{\rm r}$ is very small within 3$h_{\rm R}$ scale-lengths, being just a few km s$^{-1}$, but increases rapidly at larger radii.  The required $v_{\rm r}$ can be as large as 50-100 km s$^{-1}$ (or more) at 6$h_{\rm R}$, where it becomes very significant compared to the circular velocity ($\sim$200 km s$^{-1}$). In other words, the motion of gas on the disk is strongly deviated from the idealised circular motion. Indeed, based on MHD simulations, \cite{Trapp-21} found that the motion of gas far beyond the stellar disk is no longer rotationally supported. In their simulations, CGM gas particles fall onto the outskirts of the gas disk, conserving their angular momentum. 

A consequence of this is that Equation \ref{eq:3} (and therefore also Equation \ref{eq:5}) may not be applicable in the very outer regions of disks, say beyond 4$h_{\rm R}$ or so.   
We will show, however, in Section \ref{sec:5.4X} that the choice of the nominal outer boundary of the disk, i.e. the radius at which the accreting gas is injected with the rotationally-supported angular velocity, does not in fact have any significant bearing on the $\Sigma_{\rm SFR}$ profile in the inner parts of the disk, where the Equations \ref{eq:3} and \ref{eq:5} should certainly be valid.  We therefore argue that the likely non-applicability of Equation \ref{eq:3} (and therefore also Equation \ref{eq:5}) at these very large radii is not in fact of any practical concern.

It can be seen that Figure \ref{fig:2} predicts that strong radial motions of gas should be present in the outer regions of the gas disks.  These should be potentially detectable in HI velocity maps. In a further parallel paper (E. Wang \& S.J. Lilly 2022, in preparation), we will investigate the kinematic features of radial gas inflow and discuss the degeneracy between the kinematic signatures of radial flows and of warped disks.

\subsection{The required viscous stress in steady-state exponential disks} \label{sec:2.4}

Based on the solution of $\Phi$ in the previous subsection (see Equation \ref{eq:9}), we can now solve Equation \ref{eq:5} to obtain the viscous stress $W$ that would be {\it required} for a steady-state exponential star-forming disk.  For this we need to specify the rotation curve.  

\begin{figure}
  \begin{center}
    \epsfig{figure=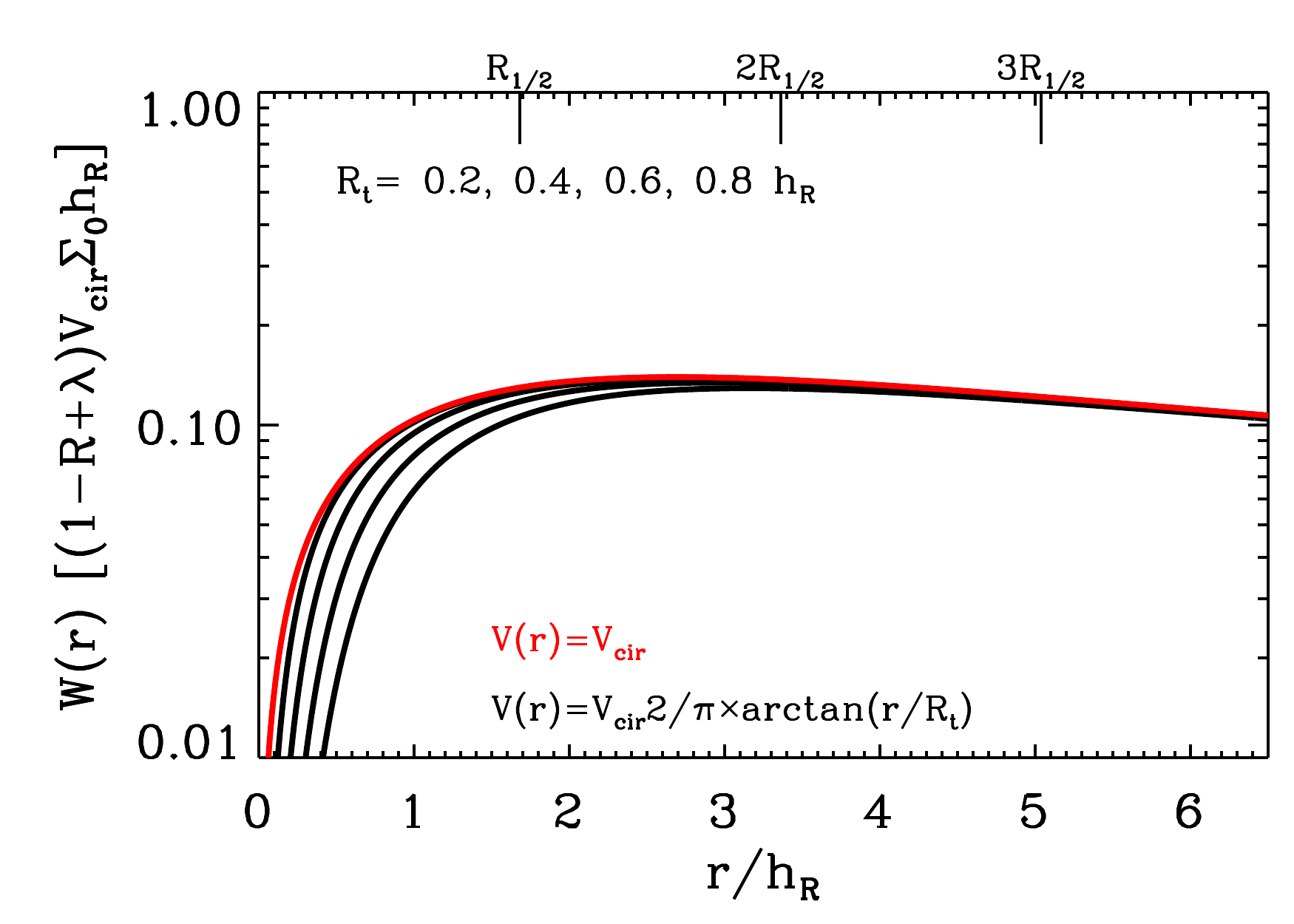,clip=true,width=0.49\textwidth}
    \end{center}
  \caption{The viscous stress that is required to maintain an exponential $\Sigma_{\rm SFR}$. Different lines show the required viscous stress for different assumed rotation curves (i.e. different gravitational potentials). The red line is for a purely flat rotation curve, and the black lines are for rotation curves in {\tt arctan} form as given Equation \ref{eq:13}.  }
  \label{fig:3}
\end{figure}

For analytic simplicity, we may first assume a constant circular velocity, i.e. $\Omega = V_{\rm cir}/r$.  Substituting Equations \ref{eq:6} and \ref{eq:9} into Equation \ref{eq:5}, we can then directly obtain the analytic solution for the viscous stress: 
\begin{equation} \label{eq:12}
\frac{W(r)}{1-R+\lambda} = V_{\rm cir}\Sigma_0 h_{\rm R} \cdot x^{-2}\cdot [(1+\eta)x + (2 +x)\cdot e^{-x} -2]
\end{equation}
The complicated form of Equation \ref{eq:12} already makes it hard to get a direct impression what it looks like.  We therefore show the solution of $W$ (from Equation \ref{eq:12}) with a flat rotation curve (and with sink term $\eta = 0$) as the red curve in Figure \ref{fig:3}.  
The viscous stress sharply increases within the first-scale length of the disk, $h_{\rm R}$, but then becomes nearly constant out to 6 $h_{\rm R}$ or more.

Real galaxies do not of course have flat rotation curves at all radii.   From observations, the rotation curve of disk galaxies can however be quite well represented by a simple {\tt arctan} function: 
\begin{equation} \label{eq:13}
    V_{\phi}(r) = V_{\rm cir} \cdot \frac{2}{\pi} {\rm arctan}(r/R_{\rm t}), 
\end{equation}
where $V_{\rm cir}$ is the maximum rotation velocity, and $R_{\rm t}$ is a turnover radius, characterizing the point between the rising and flat parts of the rotation curve \citep{Courteau-97, de-Blok-08, Miller-11}. The typical observed value of $R_{\rm t}$ is $\sim 0.4 h_{\rm R}$ of the stellar disk \citep{Miller-11}. This is broadly consistent with the $R_{\rm t}$ $\sim 0.5 h_{\rm R}$ that are obtained by assuming the stellar disk dominates the gravitational potential in the inner regions of galaxies ($<h_{\rm R}$). 
The purely flat rotation curve that was considered above corresponds to the extreme case of $R_{\rm t} = 0$. 

Adopting this more realistic rotation curve of Equation \ref{eq:13}, we obtain numerical solutions for the required viscous stress (always for the perfect exponential disk) for different values of $R_{\rm t}$ expressed in terms of $h_{\rm R}$. These are shown as the black lines in Figure \ref{fig:3}.  These numerical solutions resemble the analytic solution in Equation \ref{eq:12}, which as noted is the extreme case of $R_{\rm t} = 0$. The required viscous stresses all sharply increase at small radii and then become nearly flat. The different $R_{\rm t}$ lead to different turnover radii of $W$ in terms of the exponential scale length $h_{\rm R}$.    

An initial conclusion is that, within the framework of our ``modified accretion disk", the viscous stress that is evidently required to maintain a steady-state exponential star-forming disk is remarkably constant over a wide range of radii outside of the central region. We stress that once we have made the basic assumptions of the model (Section \ref{sec:2.1}) this result follows solely from the exponential form of $\Sigma_{\rm SFR}$, and does {\it not} depend on any assumptions about the distribution of $\Sigma_{\rm gas}$ or correspondingly on the star-formation law.   
Motivated by this ``reverse-engineering" indication, we now turn in the next Section to explore possible sources of viscosity that could provide a roughly constant viscous stress at the required level across a wide range of radii.  

\section{The possible physical origins of the viscous stress} \label{sec:3}

\begin{figure*}
  \begin{center}
    \epsfig{figure=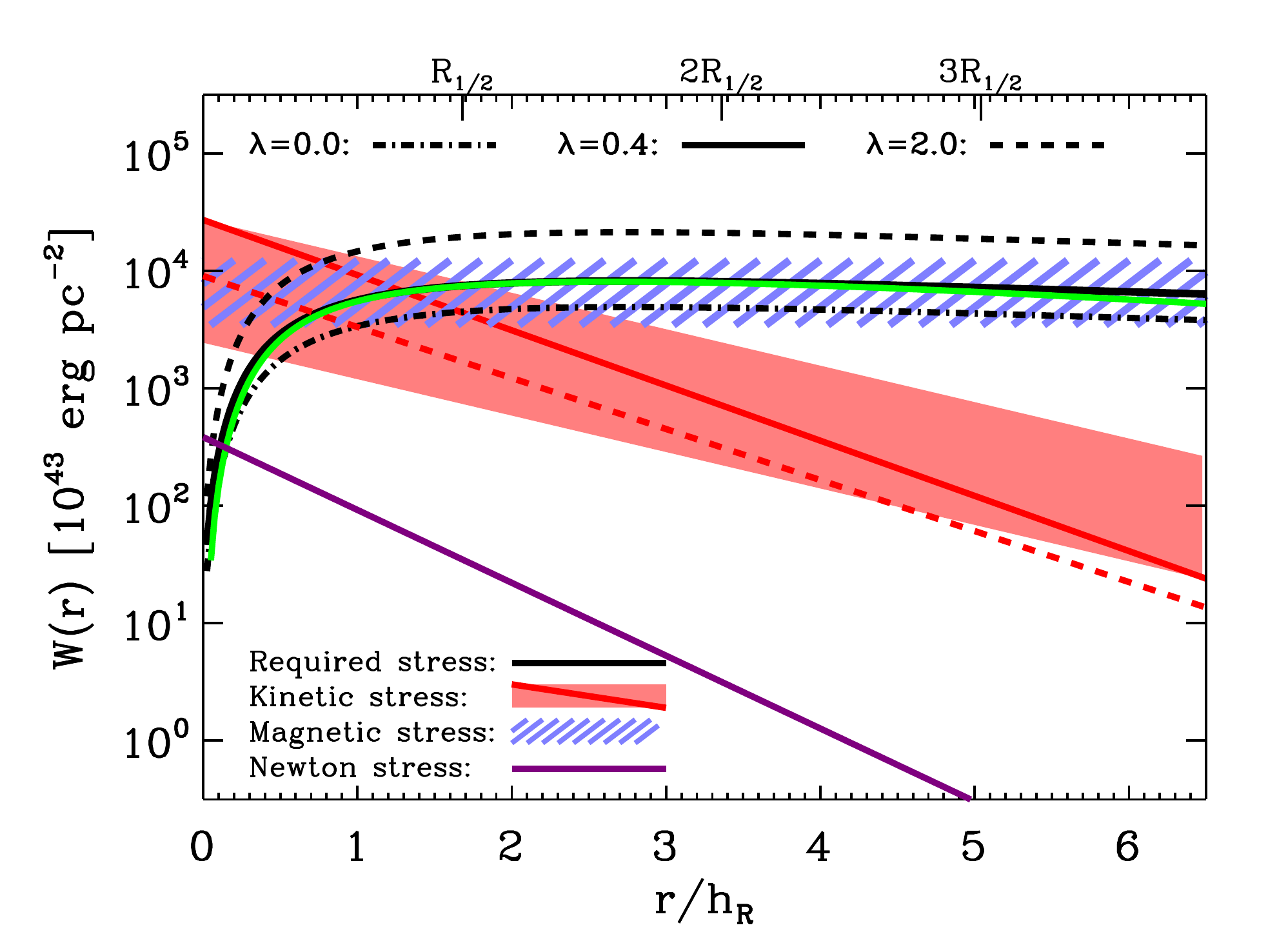,clip=true,width=0.80\textwidth}
    \end{center}
  \caption{Comparison of viscous stresses for a typical disk galaxy with $M_* = 3\times10^{10}$\msolar. The required viscous stress (see Figure \ref{sec:3} or Equation \ref{eq:12}) is shown as the black lines for three different mass-loading factors. The red shaded region and red solid line indicate our estimation of the kinetic stress based on the observed velocity dispersion of cold gas.
  The red dashed line shows a second estimation of kinetic stress based on kinetic turbulence from Type-II SNe feedback.
  The purple solid line shows our estimation of the Newton stress \citep{Wada-02}. 
  The blue hatched region shows a rough estimate of the magnetic stress, where the magnetic field is obtained with an assumption of energy equipartition between the cosmic ray and magnetic field \citep{Beck-19}.  
  The green solid lines show the detailed modeling of magnetic stress from Section \ref{sec:5} with $\lambda=0.4$. }
  \label{fig:4}
\end{figure*}

The viscosity within a differentially rotating accretion disk can come from three main effects: turbulence of the gas in the disk, the magnetic field, and peculiar motions due to the gravitational collapse of structure within the disk \citep{Shakura-73, Lynden-Bell-74, Balbus-91, Balbus-99}. \cite{Balbus-99} related the viscous stress tensor (produced by all of these three processes) to the large-scale mean flow dynamics used in a phenomenological viscous disk model, as done here in Section \ref{sec:2.1}. This enables us to estimate the likely contribution of these three processes.

To keep the three components of the stress tensor in a similar form, \cite{Balbus-99} defined three velocities: the velocity of non-circular motion (${\bf u}$), the Alfven velocity (${\bf u_{\rm A}}$), and the gravitational velocity (${\bf u_{\rm G}}$).  The Alfven velocity is defined as: 
\begin{equation}
    {\bf u}_{\rm A} = \frac{{\bf B}}{\sqrt{4\pi \rho}},
\end{equation}
where ${\bf B}$ is the magnetic field, and $\rho$ is the mass density of the gas disk (note: {\it not} the surface density $\Sigma_{\rm gas}$).  The velocity induced by the collapse of structure in the disk is defined as: 
\begin{equation}
    {\bf u}_{\rm G} = \frac{\nabla \phi_{\rm s}}{\sqrt{4\pi G\rho}}, 
\end{equation}
where $\phi_{\rm s}$ is the self-gravity potential. 

After defining these velocities, the viscous stress tensor in polar coordinates $W_{\rm r\phi}$ can be written as: 
\begin{equation}
  W_{\rm r\phi} = \langle u_{\rm r}u_{\rm \phi} + u_{\rm Gr}u_{
  \rm G\phi} - u_{\rm Ar}u_{\rm A\phi} \rangle.
\end{equation}
The first term is the Reynolds (or kinetic) stress tensor, the second term is the Newton stress tensor, and the third term is the Maxwell (or magnetic) stress tensor \citep[e.g.][]{Hawley-99, Balbus-99}. 

According to \cite{Balbus-99}, the relation between the viscous stress tensor $W_{\rm r\phi}$ and the viscous stress $W$ that was defined in Equation \ref{eq:4} in Section \ref{sec:2.1} is: 
\begin{equation} \label{eq:17}
    W = \Sigma_{\rm gas} \cdot W_{\rm r\phi}. 
\end{equation}

In this section, we will evaluate the viscous stress that is likely to be contributed by these three sources, and compare these with the $W(r)$ that is evidently required to sustain an exponential star-forming disk (see Equation \ref{eq:12} and Figure \ref{fig:3} in Section \ref{sec:2.4}). We can thus determine which, if any, of these three sources is likely to be the dominant source of viscosity in the galactic gas disk. We note that the viscous stress $W$ has the same dimensions as a surface density of energy.

For illustration, we focus on the same typical Main Sequence galaxy that we introduced previously in Section \ref{sec:2.3}, with $M_*=3\times 10^{10}$ \msolar\ and SFR=3.5 \msolar yr$^{-1}$.  We take $R_{\rm t}=2$ kpc, $h_{\rm R}$=4 kpc and $V_{\rm cir}=220$ km s$^{-1}$ for this typical galaxy \citep[e.g.][]{Miller-11, Bovy-12b, Wang-19}. The black solid and dashed lines in Figure \ref{fig:4} show the required viscous stress to maintain the exponential $\Sigma_{\rm SFR}$ (see Figure \ref{fig:3}), for three different values of the mass-loading $\lambda=0.0$, 0.4 and $2.0$.  

\subsection{The Newton stress}

The interaction between rotational shear and the self-gravitation of gas clouds can drive turbulence, even in the absence of spiral arms \citep{Wada-00, Vollmer-02}. Therefore, the Newton stress becomes a potential candidate for viscosity to sustain the turbulence (at least partly) and to transport the angular momentum flux \citep{Wada-02}. However, the  motions induced by gravitational collapse cannot be a local driving mechanism in molecular clouds because of the quick decay of the turbulence \citep[e.g.][]{Mac-Low-04}. 

\cite{Wada-02} estimated the Newton stress tensor for the gas disk of galaxies, which can be written as: 
\begin{equation}
\langle u_{\rm Gr}u_{\rm G\phi} \rangle \simeq 5\cdot \frac{\Sigma_{\rm gas}}{10 \ \rm M_{\odot}pc^{-2}} (\frac{l}{100 \ \rm pc})^2 (\frac{h_z}{100 \ \rm pc})^{-1} ({\rm km\ s^{-1}})^2, 
\end{equation}
where $l$ is the length scale of turbulent perturbations, and $h_{\rm z}$ is the vertical scale-height of the disk.  The purple solid line in Figure \ref{fig:4} shows the resulting  stress $\Sigma_{\rm gas}\langle u_{\rm Gr}u_{\rm G\phi} \rangle$ as a function of radius for the typical Main Sequence galaxy referred to above, assuming $l=100$ pc,  $h_{\rm z}=300$ pc and the Kennicutt star-formation law (see Equation \ref{eq:11}). 

It is clear that the Newton stress is lower than the required viscous stress, by more than an order of magnitude, and, furthermore, it decreases strongly with radius in contrast to the required approximate constancy.  In addition, since the vertical scale-height of the disk typically increases with galactic radius, and can reach $\sim$1 kpc in the outer regions of the disk \citep[e.g.][]{Bacchini-19, Patra-20}, the Newton stress could well be even lower than the one shown in Figure \ref{fig:4}, if a higher $h_{\rm z}$ was to be assumed. 

We can safely conclude that the Newton stress cannot be associated with the production of exponential disks, at least within the general framework that we are considering in this paper.  We will therefore neglect the Newton stress from here on.

\subsection{The Reynolds stress}

The kinetic turbulence of the gas in the disk can be estimated observationally by measuring the velocity dispersion of emission lines. This velocity dispersion in principle includes both thermal and turbulent motions, and therefore gives an upper limit to the turbulent velocity. The velocity dispersion of both atomic and molecular gas in nearby star-forming galaxies decreases with radius \citep[e.g.][]{Boulanger-92, Petric-07, Boomsma-08, Tamburro-09, Bacchini-20}. In the inner regions, the typical velocity dispersion is 15-20 km s$^{-1}$, which well exceeds the expected broadening due to thermal motions alone ($<$8 km s$^{-1}$), while in the outer region of the gas disk, the velocity dispersion is usually 6-10 km s$^{-1}$, comparable to the thermal velocity of warm ($\sim$8000 K) neutral gas.  

We can therefore evaluate an upper limit on the kinetic stress $\Sigma_{\rm gas}\langle u_{\rm r}u_{\rm \phi} \rangle$ by assuming the turbulent velocity ($u_{\rm r}$ and $u_{\phi}$) is in the range of 6-20 km s$^{-1}$ \citep[e.g.][]{Tamburro-09, Bacchini-20}.  The red shaded region of Figure \ref{fig:4} shows the estimated range of the kinetic stress for our typical Main Sequence galaxy. While the kinetic stress has the required strength in the innermost regions, it decreases strongly with radius.  

It is therefore unlikely to be responsible for the nearly flat viscous stress required for exponential star-forming disks, and it quickly falls below the required stress beyond a radius of 2-3$h_{\rm R}$. 
Considering the radial dependence of the velocity dispersion, the decline of the kinetic stress is actually likely to be even steeper, as indicated by the red solid line in Figure \ref{fig:4}.

The supersonic turbulence in the ISM is expected to be rapidly dissipated on timescales of order 10 Myr \citep[e.g.][]{Stone-98,  Padoan-99, Mac-Low-04}. Hence, a continuous source of energy would be needed to sustain the turbulence of the gas.  \cite{Bacchini-20} found that the observed kinetic energy is
remarkably well reproduced across the whole range of the galactic disk by Type-II supernova explosions, with a dimensionless efficiency parameter ($\epsilon_{\rm SN}$) (the efficiency of supernovae in injecting kinetic energy to the ISM) having a value of around $\sim0.02$.  

We can then try to estimate the kinetic stress through an alternative approach, assuming that the kinetic turbulence is dominated by the explosions of SNe.  The energy surface density of kinetic turbulence maintained by the SN explosions, can then be written as \citep[e.g.][]{Tamburro-09, Utomo-19, Bacchini-20}:  
\begin{equation}
    E_{\rm turb,SNe} = \epsilon_{\rm SN}\cdot f_{\rm cc} \cdot \Sigma_{\rm SFR} \cdot E_{\rm SN} \cdot \tau_{d}, 
\end{equation}
where $f_{\rm cc}$ is the number of Type-II SNe that explode per unit of stellar mass formed, $E_{\rm SN}$ is the total energy released by a single SN, and $\tau_{\rm d}$ is the dissipation timescale. We note that the rate of Type-Ia SNe is a few times lower than Type-II SNe \citep{Mannucci-NG}, and therefore it is neglected here. The $f_{\rm cc}$ is $1.3\times 10^{-2}$ \msolar$^{-1}$ for a \cite{Kroupa-01} initial mass function, and $E_{\rm SN}$ is $10^{51}$ erg. 
Adopting $\tau_d=10$ Myr \citep[e.g.][]{Mac-Low-04}, and $\epsilon_{\rm SN}=0.02$ \citep{Bacchini-20}, we can then compute the kinetic stress produced by the turbulent energy surface density as the red dashed line of Figure \ref{fig:4}.    It is self-evident that the kinetic stress estimated in this way, will follow the $\Sigma_{\rm SFR}$ profile and will therefore decrease exponentially with radius in the galaxy, with the same scale-length as $\Sigma_{\rm SFR}$.  

It can be seen that the two estimates of the kinetic stress are broadly consistent. Both approaches fail to match the required viscous stress for the exponential star-forming disk.  We can therefore also reject this as the main origin of viscosity. 

\subsection{The Maxwell stress} \label{sec:3.3}

\cite{Balbus-91} proposed that MRI is likely to be an efficient mechanism to cause magnetic turbulence, and cause the angular momentum transport in classical accretion disks.  Provided that there is coupling between the ionized and non-ionized material this mechanism should also apply in galactic disks.

In this subsection, we briefly review how MRI works. The magnetic forces act like a spring under tension connecting fluid elements. In the presence of a weak axial magnetic field, two radially neighboring fluid elements can be simply treated as two mass points connected by a mass-less spring. The inner fluid element has a larger angular velocity than the outer, causing the spring to stretch.  Hence, the inner fluid element is forced to slow down, and sinks into a lower obit due to the loss of its angular momentum. On the other hand, the outer fluid element speeds up, and moves to a higher orbit due to the increase of its angular momentum.  The spring tension will further increase as the two fluid elements move further apart.  The inner fluid element sinks into lower and lower orbits and the outer fluid element moves to higher and higher orbits.  The behavior of a magnetized differentially rotating fluid is almost exactly analogous to this simple mechanical system \citep[e.g.][]{Balbus-98}. 

Based on a 3-dimensional MHD simulation to investigate the non-linear development of the MRI, \cite{Hawley-95} found that the angular momentum flux is dominated by the magnetic Maxwell stress rather than by the Reynolds stress, and that the total viscous stress is found to be tightly related to the total magnetic field ($B_{\rm tot}$) of the disk: 
\begin{equation} \label{eq:20}
    \Sigma_{\rm gas} \langle W_{\rm r\phi} \rangle \simeq 0.61 \times \langle \frac{B_{\rm tot}^2}{8\pi} \rangle \cdot 2h_{\rm z}
\end{equation}
This empirical expression, obtained from MHD simulations, provides a way to estimate the magnetic stress, if the total strength of the magnetic field and the vertical scale-height of the disk are known. 

Observationally, the magnetic field strength $B_{\rm tot}$ can be estimated from measurements of synchrotron emission by assuming equipartition of energy between the magnetic fields and cosmic rays \citep[e.g.][]{Beck-05, Seta-19}.  Empirically, \cite{Seta-19} found that this method is valid for star-forming spiral galaxies on scales above about 1 kpc, but probably does not hold on smaller scales.  The main uncertainties come from assumptions about the pathlength through the synchrotron-emitting disc and the number ratio of proton/electron (see \cite{Seta-19} for discussion).  An uncertainty of a factor of two in either quantity causes a systematic deviation in the estimated $B_{\rm tot}$ of about 20\%.  We note that the magnetic field strength derived in this approach is the total field strength, including both any small-scale turbulent component and any large-scale ordered component. This $B_{\rm tot}$ is however the one we need to estimate the magnetic stresses from MRI.

With the assumption of energy equipartition,  \cite{Fletcher-NG} found that the $B_{\rm tot}$ shows a large variation in the galaxy population, from a few to a few tens of $\mu G$. The typical $B_{\rm tot}$ given in the recent literature of 21 bright spiral galaxies is 17 $\mu G$, with a standard deviation of 14 $\mu G$. More recently, \cite{Beck-19} measured the $B_{\rm tot}$ averaged in radial rings in the plane of the disk from the radio synchrotron intensity at 4.85 GHz, assuming a proton/electron ratio of 100, and a thickness of the synchrotron-emitting disc of 1 kpc. They found the total field strength (averaged within each galaxy and then between galaxies) to be 13 $\mu G$ with a dispersion of 4 $\mu G$ between galaxies.  For the Milky way, the total equipartition field strength  is  $\sim$10 $\mu G$ at the radius of 5 kpc, decreasing to $\sim$4 $\mu G$ at 15 kpc.

Assuming a $B_{\rm tot} = 13 \pm 4 \ \mu G$, and $h_{\rm z} = 300$ pc \citep{Bacchini-19, Patra-20}, we can then estimate the magnetic stress using Equation \ref{eq:20}.  This is shown by the horizontal blue hatched region in Figure \ref{fig:4}.  We find that the amplitude of the magnetic stress matches rather well with the required stress needed for an exponential star-forming disk. 

We now turn to the expected radial dependence.
The observed total field strength $B_{\rm tot}$ typically decreases with radius by a factor of $\sim$2-3 from the inner regions to the outer edge of galactic disk \citep[e.g.][]{Beck-05, Basu-13, Heesen-14,  Berkhuijsen-16}.  By studying the total magnetic field for five nearby normal disk galaxies at sub-kpc spatial resolution, \cite{Basu-13} found that the field strength is $\sim$20-25 $\mu G$ at the center, and falls to $\sim$10 $\mu G$ in the outer parts.  Interestingly, the vertical scale height $h_{\rm z}$ of gas disks is found to radially increase with radius by a factor of $\sim$3-4 \citep[e.g.][]{Bacchini-19, Patra-20}.  The $h_{\rm z}$ of HI gas is typically 100-300 pc at the center, and 500-1000 pc at the outer region of the disk.  Given these two competing effects, we can then argue that the quantity $B_{\rm tot}^2h_{\rm z}$ is probably only weakly dependent on the radius, as evidently required (from Equations \ref{eq:17} and \ref{eq:20}) to produce a more or less constant $W$.

According to the mechanism of MRI, rotational shear is necessary to produce viscosity \citep{Balbus-91}. For a typical rotation curve of {\tt arctan} form, the angular velocity $\Omega$ increases strongly towards the center of the disk, until a central plateau is reached at ($r\ll R_{\rm t}$). Within this the disk rotates close to a solid body. In this region, the magnetic stress is therefore expected to be low and to therefore be inefficient in momentum transport.  Indeed, using the MHD simulations, \cite{Abramowicz-96} found that the efficiency of angular
momentum transport is reduced by a factor given by the background shear-to-vorticity ratio \citep[also see][]{Hawley-99, Pessah-08}: 
\begin{equation} \label{eq:22}
  f_{\rm s/v} = q/(2-q), 
\end{equation}
where $q$ is defined as: 
\begin{equation} \label{eq:23}
    q = - \frac{\partial \ln \Omega}{\partial \ln r}. 
\end{equation}
In running the model constructed later, we will include this effect by simply multiplying the magnetic stress derived from Equation \ref{eq:20} with the shear-to-vorticity ratio $f_{\rm s/v}$ given by this Equation \ref{eq:22}.

Based on this discussion, we conclude that MRI-induced viscosity is the most promising mechanism to account for the viscous stress that is evidently required to produce a steady-state exponential star-forming disk in which the dominant gas flow is within the plane of the disk.

While magnetic viscosity is promising, we still need to  see whether an exponential disk with reasonable scale length is in fact established from more general initial conditions, and to examine the stability of such a system.
We therefore construct a simple physical model to explore these questions in the next Section of the paper.

To look ahead, the solid green line plotted in Figure \ref{fig:4} shows the computed magnetic stresses obtained in Section \ref{sec:4} and \ref{sec:5} for a typical galaxy with $\lambda=0.4$ (obtained from Equations \ref{eq:20}, \ref{eq:22} and \ref{eq:28}). It is an excellent match to the solid black curve.   We will come back to this point later in Section \ref{sec:4} and \ref{sec:5}.   

\section{A dynamic model of disk formation and evolution using magnetic stresses} \label{sec:4}

Motivated by previous section, in this section we develop a simple physical model of disk formation driven by magnetic stress, which enables us to examine the stability of the system and whether an exponential solution is a natural outcome from arbitrary initial conditions.

\subsection{The logical flow of the model} \label{sec:4.1}

\begin{figure*}
  \begin{center}
    \epsfig{figure=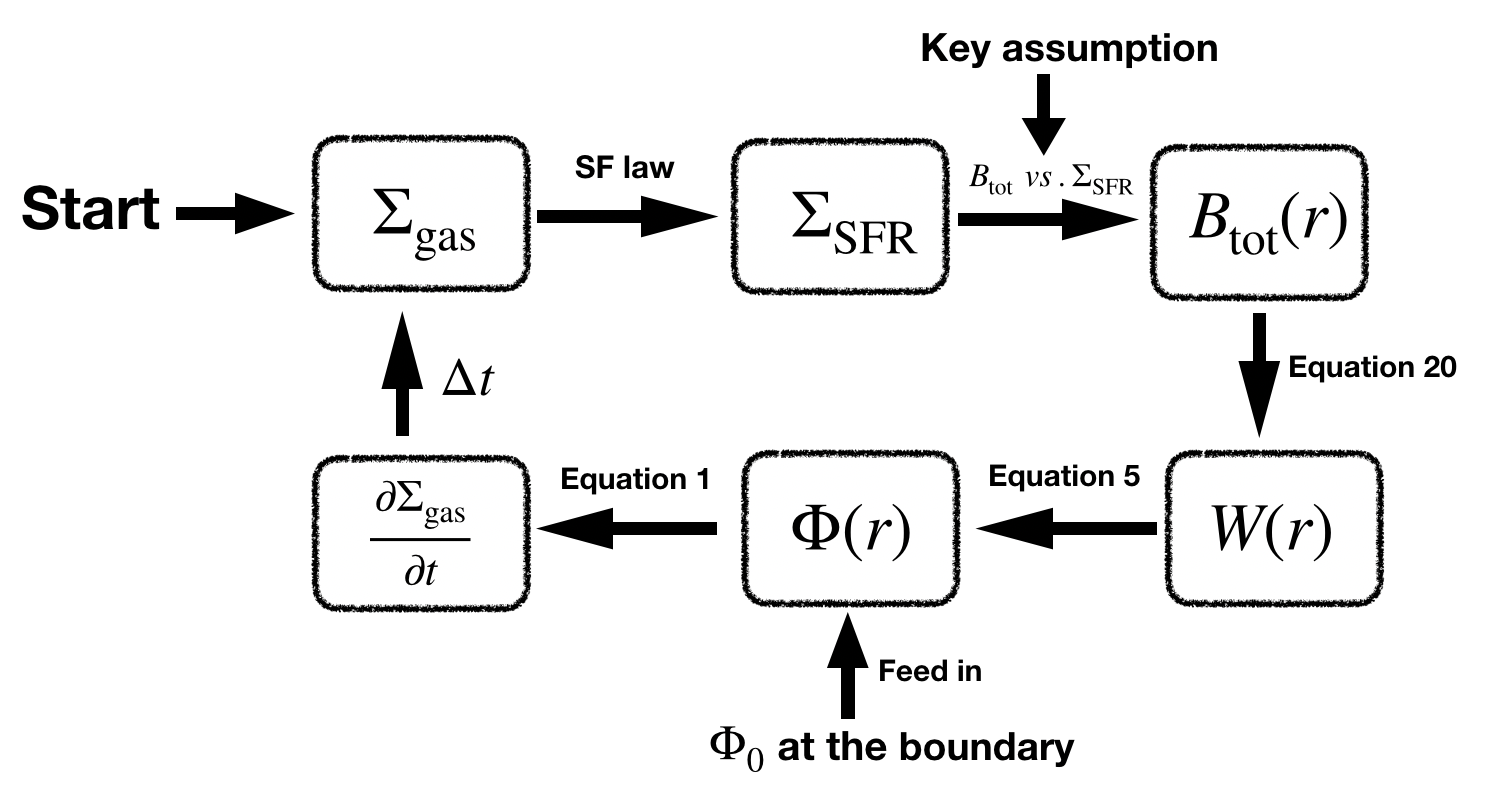,clip=true,width=0.85\textwidth}
    \end{center}
  \caption{ The logical flow of the dynamic model of disk formation.  We start with an initial disk with a given $\Sigma_{\rm gas}$. The $\Sigma_{\rm SFR}$ is known with adopting a star-formation law.  Given the key assumption that the magnetic field is instantaneously determined by $\Sigma_{\rm SFR}$, the field strength of total magnetic field is obtained.  Further, we convert the $B_{\rm tot}$ into viscous stress following Equation \ref{eq:20}. This viscous stress then drives the radial inflow of the accretion disk following the Equation \ref{eq:5}, and determines the instantaneous radial inflow rate.  From the Equation \ref{eq:1},  the change rate of $\Sigma_{\rm gas}$ is then determined. If we set a time-step of $\Delta t$, the gas surface density $\Sigma_{\rm gas}$ can be obtained at the next time-step, which turns on the next loop. } 
  \label{fig:5}
\end{figure*}

The basic idea of the model is to use the magnetic stress to drive the mass and angular momentum flux in the framework of our ``modified accretion disk". Therefore, the key ingredient of the model is how to set the magnetic field of the disk as a function of the radius. 

Observationally, there is a tight correlation between the integrated nonthermal radio emission of a galaxy and its integrated SFR \citep[e.g.][]{Heesen-14, Tabatabaei-17}, indicating a close connection between the magnetic field and star-formation activity.   Motivated by this, we will assume that the local $B_{\rm tot}(r)$ of a galaxy is instantaneously determined by its current star-formation rate at that radius, i.e. $\Sigma_{\rm SFR}(r)$.  The detailed recipe of $B_{\rm tot}$ vs. $\Sigma_{\rm SFR}$  will be developed below in Section \ref{sec:4.3}.  In this subsection, we first establish the basic physical logic of the model. 

Figure \ref{fig:5} shows the logical flow of the model. We will assume azimuthal symmetry for the disk throughout and consider only the radial dependence of the relevant quantities.  We start with an initial disk with some initial $\Sigma_{\rm gas}$ at some initial time $t_0$. The $\Sigma_{\rm SFR}$ is then derived using an assumed star-formation law.  Based on the key assumption that the local magnetic field $B_{\rm tot}(r)$ is instantaneously determined by the $\Sigma_{\rm SFR}(r)$, the field strength of the magnetic field $B_{\rm tot}$ is then obtained across the gas disk.  We then convert the derived $B_{\rm tot}(r)$ into the viscous stress $W(r)$ following Equation \ref{eq:20} as modified by 
Equation \ref{eq:22},
(with an additional assumption about the vertical scale-height of the gas disk $h_{\rm z}$, assumed to be unchanging with time). 

This viscous stress then drives the radial mass inflow rate $\Phi(r)$ at all radii following Equation \ref{eq:5}.   Using this $\Phi(r)$, the change in $\Sigma_{\rm gas}$ after a time-interval of $\Delta t$ can be determined, at all radii, from the continuity equation (Equation \ref{eq:1}). This then gives a new gas surface density profile $\Sigma_{\rm gas}(r)$ at $t_0 + \Delta t$, starting the next loop.   

The $\Phi$ at an outer boundary ($R_{\rm b}$) of the accretion disk, as well as the location of that boundary, are external parameters set by the user.  
The radial inflow rate at $R_{\rm b}$ is the rate of feeding the galactic disk from the CGM, and the location of the outer boundary may be related to the angular momentum of the halo (see Section \ref{sec:5.4X} for discussion). 

We can then let the system evolve for a very long time. We can thereby obtain the evolution of the gas disk in this dynamic model and use it to study both the stability and the convergence towards a steady state equilibrium solution.

A key feature of this model is that the radial gas inflow is driven by the magnetic stress, which is itself linked to the current distribution of gas via the $\Sigma_{\rm SFR}$ of the disk and a $B_{\rm tot}$-$\Sigma_{\rm SFR}$ relation, suggested from observations (see Section \ref{sec:4.3}).  This provides an effective but possibly complex feedback loop between $\Sigma_{\rm gas}$ and $\dot{\Sigma}_{\rm gas}$, i.e. between $\Sigma_{\rm SFR}$ and $\dot{\Sigma}_{\rm SFR}$.  The feedback loop may be quite complex because the magnetic stress, and its effect driving the gas inflow, will both depend on other parameters of the system.   
 
The qualitative action of this feedback loop may be seen as follows.  Rearranging Equation \ref{eq:5} in the case of a fixed potential, we get

\begin{equation} \label{eq:5X}
\Phi = \frac{\partial (2\pi r^2 W)/\partial r} 
{\partial (\Omega r^2)/\partial r}. 
\end{equation}
Suppose that the gradient in $\Sigma_{\rm SFR}$ at some radius $r$ is too steep for some reason, i.e. there is too much star-formation just inside of $r$ compared with just outside of it.  If there is a positive connection between $\Sigma_{\rm SFR}$ and $B_{\rm tot}$, and thus also between $\Sigma_{\rm SFR}$ and $W$, then the partial derivative in the numerator of Equation \ref{eq:5X} will reduce, reducing the inflow rate $\Phi$ at $r$.  This will then have the effect of decreasing the $\Sigma_{\rm SFR}$ interior to $r$, and increasing it outside, thereby reducing the (negative) gradient in $\Sigma_{\rm SFR}$ back towards an equilibrium value. 

Having emphasized the role of this
feedback loop, the tight linkage between $\Phi$ and $W(r)$ discussed earlier (Section \ref{sec:2.1X}) however tells us that steady-state solutions for $\Sigma_{\rm SFR}$ would eventually be established, even without a direct physical coupling between the viscous stress and the $\Sigma_{\rm SFR}$.  Provided that the $W(r)$ has (for some other physical reason) the correct more-or-less flat profile (see Figure \ref{fig:3}), then the same steady-state solution will eventually be established. In essence, at each radius, the disk will be being fed at some rate given uniquely by $W(r)$, and the $\Sigma_{\rm SFR}$ will adapt to this feeding rate, as in any gas regulator system.  However, without a direct connection between $\Sigma_{\rm SFR}$ and $W$, as provided in our model by the magnetic fields, the response time will be much slower, and equal the effective gas depletion timescale \citep{Lilly-13, Wang-19}.

\subsection{The adopted $B_{\rm tot}$ vs. $\Sigma_{\rm SFR}$ relation}   \label{sec:4.3}

Observations can guide us about the form of the $B_{\rm tot}$ vs. $\Sigma_{\rm SFR}$ relation. 
As noted above, there is a tight correlation between the integrated nonthermal radio emission of a galaxy and its integrated SFR \citep[e.g.][]{Heesen-14, Tabatabaei-17}. This observational result is also consistent with the theory of amplification of magnetic fields by a small-scale turbulent dynamo within SF regions \citep[e.g.][]{Gressel-08, Arshakian-09, Schleicher-13}.  

Motivated by this, we simply assume that the local $B_{\rm tot}(r)$ of a modeled galaxy will be instantaneously determined by its current star-formation rate at that radius, i.e. $\Sigma_{\rm SFR}(r)$.  Specifically, the spatially resolved $B_{\rm tot}$-$\Sigma_{\rm SFR}$ {\it within} a single galaxy is assumed to be:  
\begin{equation} \label{eq:25}
   B_{\rm tot}(r) = X\cdot (\Sigma_{\rm SFR}/\Sigma_{\rm SFR,zp})^{\alpha}, 
\end{equation}
where $X$ is the magnetic field strength at an arbitrarily defined zero point SFR surface density ($\Sigma_{\rm SFR,zp}$), and $\alpha$ is the exponent. 

\cite{Heesen-14} examined the relation between the non-thermal radio continuum and the SFR within galaxies for a sample of 17 nearby galaxies.  They found different exponents for the relation between these quantities when using the integrated quantities for galaxies across the galaxy population, and when using spatially resolved quantities {\it within} individual galaxies.  With the assumption of energy equipartition, we would therefore expect the spatially-resolved $B_{\rm tot}$-$\Sigma_{\rm SFR}$ relation to be different from that for integrated measurements of the galaxy as a whole. In this work, the operation of the model requires the spatially-resolved relation, as given in Equation \ref{eq:25}.  

In the work of \cite{Heesen-14}, the $\Sigma_{\rm SFR}$ maps were obtained from a linear combination of GALEX far-UV and Spitzer 24 $\mu$m maps. They measured the exponent between the non-thermal radio emission and $\Sigma_{\rm SFR}$ within individual galaxies, at a spatial resolution of 0.7 kpc, finding an average exponent in this relation of 0.58 with a standard deviation of 0.22 within their sample. With the assumption of energy equipartition \citep[][]{Heesen-14, Beck-19}, we can convert this exponent to the $\alpha$ exponent of the $B_{\rm tot}$-$\Sigma_{\rm SFR}$ relation.  This gives an average value of $\alpha = 0.15$, with a standard deviation of 0.06, in the galaxy population of \cite{Heesen-14}.  This value is also consistent with the spatially-resolved analysis of other works \citep[also see][]{Berkhuijsen-13, Tabatabaei-13a, Tabatabaei-13b}, always assuming energy equipartition. 

For running the model in Section \ref{sec:5} and \ref{sec:5.4X} we will adopt the Equation \ref{eq:25}, choosing suitable values of $X$ and $\alpha$ and exploring also variations around these chosen values. Later, in Section \ref{sec:6}, we will consider the variation of the $B_{\rm tot}$-$\Sigma_{\rm SFR}$ relation across the galaxy population. 


\subsection{Other settings of the model} \label{sec:4.2}

In addition to the crucial $B_{\rm tot}$-$\Sigma_{\rm SFR}$ relation, we need to specify the star formation law, the mass-loading factor, the vertical scale-height of the disk, and the gravitational potential. We will show however that these choices are less critical than the choice of the $B_{\rm tot}$-$\Sigma_{\rm SFR}$ relation.  Furthermore, we have quite independent observational guidance about reasonable choices to make for these other parameters, as discussed here.

Unless explicitly specified, we adopt the Kennicutt star formation law (Equation \ref{eq:11}), and a time- and radially- invariant mass-loading factor $\lambda$ for a given galaxy. We also assume for simplicity that the gravitational potential does not change during the evolution, and adopt the {\tt arctan} form of the rotation curve given by Equation \ref{eq:13}.  Based on \cite{Patra-20}, we assume that $h_{\rm z}$ linearly increases with galactic radius as: 
\begin{equation} \label{eq:hz}
    h_{\rm z} = 150 \ {\rm pc} \times (\frac{r}{R_{\rm z}} + 1.0), 
\end{equation}
where $R_{\rm z}$ characterizes the radial gradient of $h_{\rm z}$. 

When running this model numerically, we need to calculate the spatial derivatives when computing the radial inflow rate $\Phi$ via Equation \ref{eq:5}, and when computing the $\partial \Sigma_{\rm gas}/\partial t$ via Equation \ref{eq:1}.  Truncation errors grow very significantly in computing these derivatives. This is because the errors in the current loop, will be amplified in the next loop in calculating the derivatives. To solve this technical problem, we radially smooth the spatial derivatives at every time-step, by convolving with a Gaussian function with a standard deviation 0.01$R_{\rm b}$. The effect of this smoothing process is only to reduce the growing errors in calculating derivatives. We have examined that a different width of the smoothing kernel does not change any of our results: a narrower smoothing kernel usually needs a higher time resolution (shorter $\Delta t$) and therefore is more computationally time consuming.  

\section{Results of the dynamic numerical model} \label{sec:5}

In this section, we will establish the asymptotic steady-state solutions of $\Sigma_{\rm SFR}(r)$ and $\Sigma_{\rm gas}(r)$ in the disks that are produced by the model and carry out experiments to examine the stability of the solutions and the effects of changing the rate at which the disk is fed.   We will also explore the dependence on the detail settings of the model.
To do this, we first set up a fiducial run of the model with a particular set of all the parameters. Then we try to vary the settings of parameters to explore their effects. 

\subsection{The emergence of exponential SF disks from arbitrary initial conditions}\label{sec:5.1}

\begin{figure*}
  \begin{center}
    \epsfig{figure=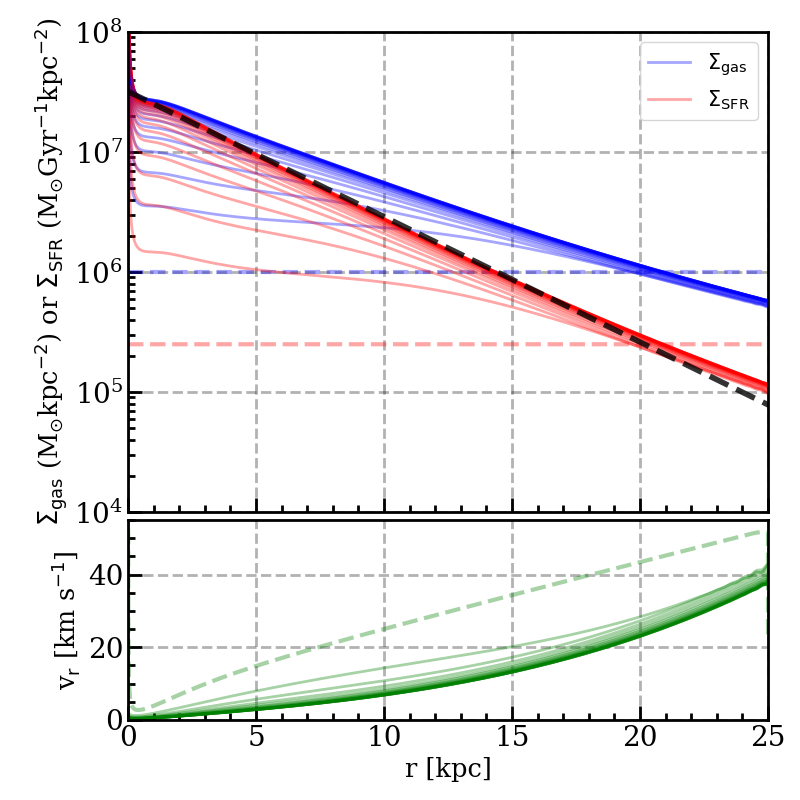,clip=true,width=0.45\textwidth}
    \epsfig{figure=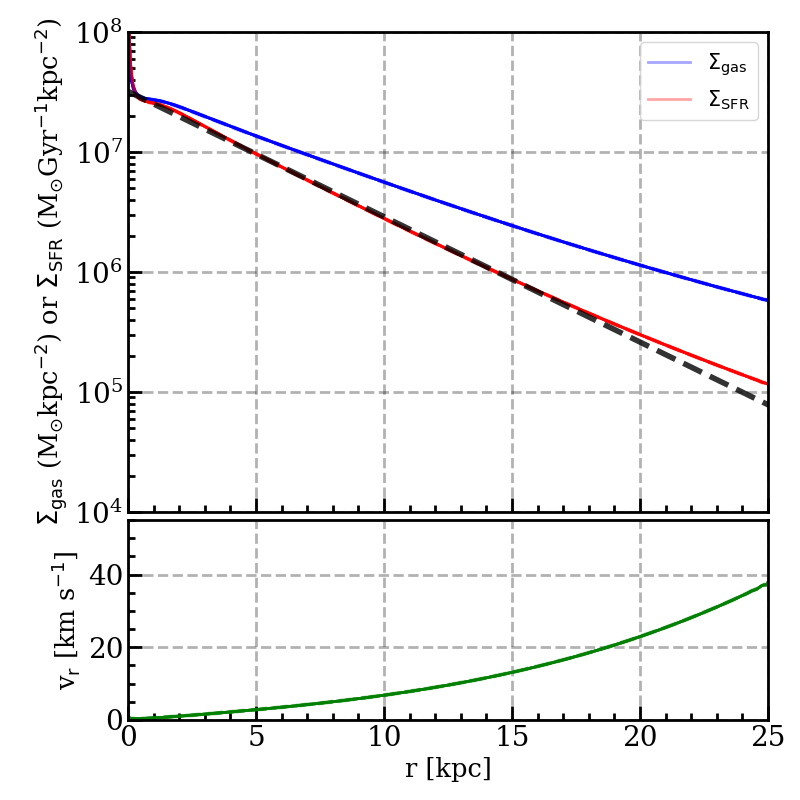,clip=true,width=0.45\textwidth}
    \end{center}
  \caption{ Left: The upper panel shows the evolution of $\Sigma_{\rm SFR}$ (red lines) and $\Sigma_{\rm gas}$ (blue lines) starting from an initial condition of a constant $\Sigma_{\rm gas}$ (dashed lines) and with the disk fed at a rate of 3.5 ${\rm M_{\odot}yr^{-1}}$ at an outer radius of 25 kpc. 
 20 time steps of 300 Myr are shown (6 Gyr in total). The lower panel shows the corresponding radial inflow velocity (see Equation \ref{eq:10}) as a function of radius in the evolution.  All three quantities converge to a steady-state solution, yielding an excellent exponential in $\Sigma_{\rm SFR}$ and thus, given the star-formation law, also in $\Sigma_{\rm gas}$.
  Right panel: Confirmation that the system reaches a steady state.  The model is run for another 3 Gyr.  The further ten red (or blue) lines are shown  overlapped together, indicating that the system has indeed reached a steady-state equilibrium. 
  In both panels, the black dashed line shows the prefect exponential function for comparison, which is determined by two points on the steady-state $\Sigma_{\rm SFR}$ profile at the radius of 5 kpc to 15 kpc. 
  } 
  \label{fig:7}
\end{figure*}

\begin{figure*}
  \begin{center}
    \epsfig{figure=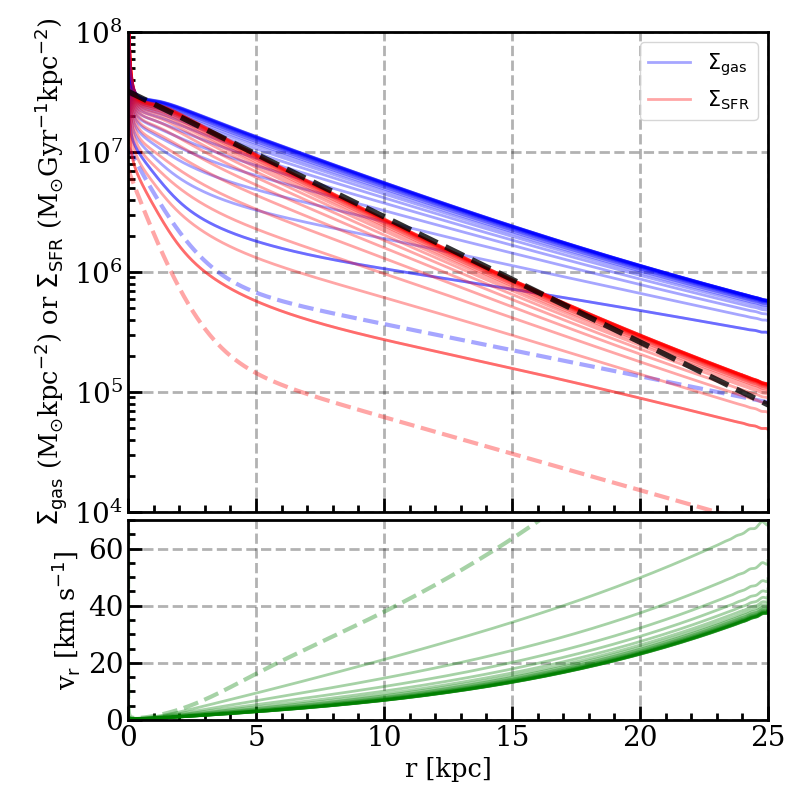,clip=true,width=0.45\textwidth}
    \epsfig{figure=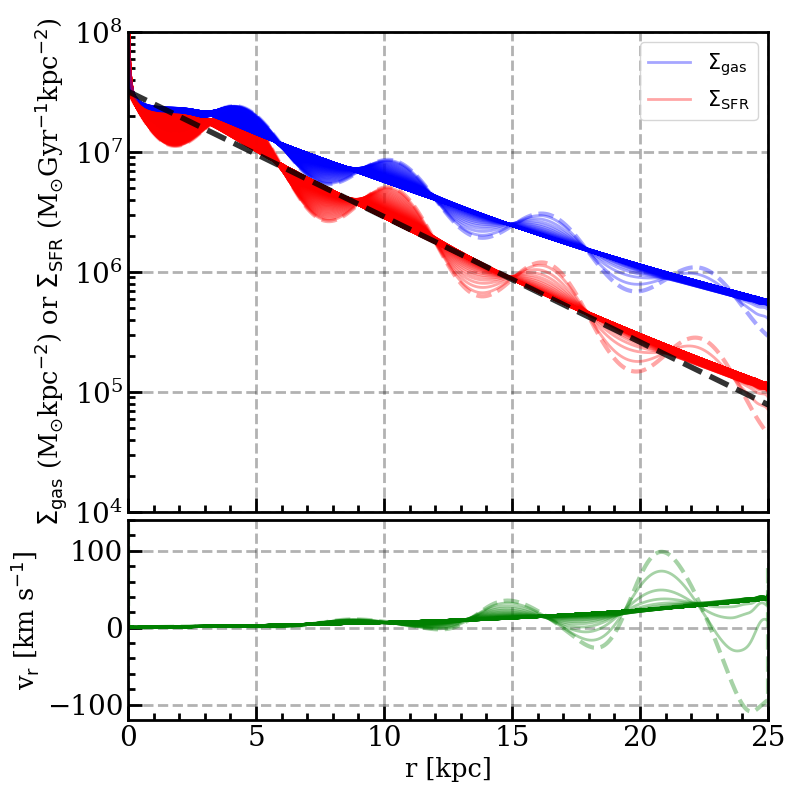,clip=true,width=0.45\textwidth}
    \end{center}
  \caption{  Left panel: The same as the left panel of Figure \ref{fig:7}, but now with a double-exponential $\Sigma_{\rm gas}$ initial gas distribution.  Despite the different starting conditions, the steady-state configuration is the same as in Figure \ref{fig:7}.
  Right panel: The same as the left panel of Figure \ref{fig:7}, but with an initial condition in which the steady-state solution for $\Sigma_{\rm gas}$ from Figure \ref{fig:7} is perturbed with a radial sinusoidal function.  In this case, the model is run for only 1 Gyr, with 50 red (or blue) lines spaced by a time interval of 20 Myr. 
  In both panels, we show for comparison the same black dashed lines as on Figure \ref{fig:7}.  The radial perturbations are rapidly eradicated.} 
  \label{fig:8}
\end{figure*}

We adopt for simplicity a time-invariant relation of $B_{\rm tot}-\Sigma_{\rm SFR}$ in running the model. In other words, we neglect any possible change in $X$ in Equation \ref{eq:25} during the evolution.  

As in Section \ref{sec:3}, we again consider the case of a typical Main Sequence galaxy with $M_* = 3\times10^{10}$ \msolar\ and SFR=3.5 ${\rm M_{\odot}yr^{-1}}$. We therefore set the inflow rate at the outer boundary $R_{\rm b}$ (set to be 25 kpc) as $\Phi_{0}=3.5\ {\rm M_{\odot}yr^{-1}}$, and take the $\lambda-R$ term to be zero (i.e. a modest effective wind-loading of $\lambda = 0.4$). This ensures that the SFR of the system in equilibrium is the desired $\sim$3.5 ${\rm M_{\odot}yr^{-1}}$. In the initial run, there is no central mass sink, i.e. $\eta = 0.0$. For the choice of rotation curve, we set $R_{\rm t}=2$ kpc, $V_{\rm cir}=220$ km s$^{-1}$.  We also set $R_{\rm z}$=10 kpc (see Equation \ref{eq:hz} for settings of $h_{\rm z}$), suggested from observations \citep[see][]{Bacchini-19, Patra-20}.  

In this fiducial run, we assume an (unchanging) $B_{\rm tot}-\Sigma_{\rm SFR}$ relation of Equation \ref{eq:25} with $X$=19.1 $\mu$G, $\Sigma_{\rm SFR,zp}$=0.01 ${\rm M_{\odot}yr^{-1}kpc^{-2}}$ and $\alpha=0.15$, i.e. 
\begin{equation} \label{eq:28}
    B_{\rm tot}(r) = 19.1 \mu G \cdot (\frac{\Sigma_{\rm SFR}}{\rm 0.01 \ M_{\odot}yr^{-1}kpc^{-2}})^{0.15}. 
\end{equation}
The normalization $X$ is taken from Section \ref{sec:6.0} for this typical Main Sequence galaxy, and the value of $\alpha$ (0.15) is, as discussed above, inferred from \cite{Heesen-14} with the assumption of energy equipartition. 

We want to start the model far from any expected steady-state endpoint.  We therefore set the initial condition to be a gas disk with a constant surface density at a level of $\Sigma_{\rm gas}(r)=10^6 \ {\rm M_{\odot}kpc^{-2}}$. 

We then run the fiducial model for 6 Gyr, which is enough for the gas disk to reach a steady-state. The left panel of Figure \ref{fig:7} shows the evolution of $\Sigma_{\rm gas}$ (blue lines) and $\Sigma_{\rm SFR}$ (red lines) in intervals of 300 Myr. 
The corresponding radial inflow velocity $v_{\rm r}$ is also shown in green lines in the bottom small panel. 

As can be seen, after only a few Gyr evolution, the resulting $\Sigma_{\rm SFR}$ profile has a nearly perfect exponential form extending to $\sim$5 scale-lengths, equivalent to a change in $\Sigma_{\rm SFR}$ of more than 2 dex.   
If we keep running the model for another 3 Gyr, there is almost no change in the $\Sigma_{\rm SFR}(r)$, $\Sigma{\rm gas}(r)$ and $v_{\rm r}(r)$ profiles, as shown in the right-hand panel of Figure \ref{fig:7}.  This indicates that the system is indeed in a steady-state.

For comparison, an exact exponential profile is shown on the left panel of Figure \ref{fig:7}, by connecting two points at the radius of 5 kpc and 15 kpc on the final $\Sigma_{\rm SFR}$ profile. This straight line is in the form of: 
\begin{equation} \label{eq:29}
    \Sigma_{\rm SFR}(r) = 0.032 \times \exp(-\frac{r}{\rm 4.16\ kpc})\ {\rm M_{\odot}yr^{-1}kpc^{-2}}.  
\end{equation}

In other words, this very simple model, with observationally-motivated values of the various parameters and with an observationally-motivated $B_{\rm tot}-\Sigma_{\rm SFR}$ relation, yields without any fine-tuning a nearly perfect exponential star-forming disk when starting from very different initial conditions.
The scale length of the resulting exponential disk is $h_{\rm R}=4.16$ kpc - a very sensible value for such a galaxy.  The rather precise exponential form extends over at least 4-5 scale-lengths, i.e. out to a radius of $r \sim 20$ kpc, beyond which there may be a very slight upbend.   We will consider the scale-length of the disk further in Section \ref{sec:5.2}. 

The {\it ab initio} successful production of a rather precise exponential $\Sigma_{\rm SFR}$ disk shown in Figure \ref{fig:7} is the single most important result of this paper. In our view, our modified accretion disk model in which magnetic MRI stresses provide the viscosity gives a natural physical explanation for the striking exponential form of galactic disks.

Our basic model also results in an exponential distribution of the (total) gas surface density. This may appear to be inconsistent with observations \citep{Bigiel-08, Leroy-09, Bigiel-10}.  However, this output gas profile follows entirely from the assumption about the star formation law, and in particular the adoption of the \cite{Kennicutt-98} star formation law, which connects the $\Sigma_{\rm SFR}$ to $\Sigma_{\rm gas}$ with a simple power-law, resulting in an exponential gas disk if the SFR profile is exponential.  We will show, in Section \ref{sec:5.3.4} below, that the steady-state profile of $\Sigma_{\rm SFR}$ does {\it not} depend on the assumed star formation law. Instead, the star formation law determines the profile of the gas disk.  This is a basic feature of any gas regulator system \citep{Lilly-13}: the gas content of the system adjusts itself so as to maintain the SFR, according to whatever star formation law is present. The profile of the gas disk therefore tells us about the star-formation law, and not the other way around, and therefore does not provide a test of the model.

In the very innermost regions of the disk ($r<0.1$ kpc), it can be seen in Figure \ref{fig:7} that there is a pronounced cusp in $\Sigma_{\rm SFR}$.  This is a consequence of the lack of a central mass sink in our initial fiducial run, i.e. we set $\eta = 0.0$. This is not a concern, because this cusp could be easily eliminated by allowing a central mass sink. Such a sink could presumably be associated with the accretion onto a central massive black hole (and/or associated jet-driven outflow), or the formation of a central star-cluster. 

At the radius of 0.1 kpc, the steady-state radial mass inflow rate in the fiducial run is 0.013 \msolar yr$^{-1}$. This corresponds to only $\sim$0.4\% of the total inflow that was injected at the outer boundary of the disk. Such a low central inflow rate implies the accumulation of central masses of order $10^8$ \msolar\ after $10^{10}$ years. This is therefore quite consistent with the expected order of magnitude of $\eta$ (see Section \ref{sec:2.3}).  The small residual inflow rate in the very inner regions of our modified accretion disk model may be directly linked to the feeding of the central black hole. The small and completely reasonable central sink term that is implied by the basic model should be considered another success. 

We explored whether the steady-state exponential form of $\Sigma_{\rm SFR}$ depends on the initial conditions or not. We input an initial $\Sigma_{\rm gas}$ in the form of a double-exponential ($\Sigma_{\rm gas}=10^{7} \cdot \exp(-r/{\rm 1\ kpc}) + 10^{6} \cdot \exp(-r/{\rm 10\ kpc}) \ {\rm M_{\odot}kpc^{-2}}$), and re-ran the model with all the settings unchanged. We show the evolution of $\Sigma_{\rm gas}$ and $\Sigma_{\rm SFR}$ for the running in the left panel of Figure \ref{fig:8}.  For comparison, we also show the same black dashed line (Equation \ref{eq:29}) in Figure \ref{fig:8}. Evidently, the system evolves to the same steady-state (single) exponential solution. 

In a similar way, we also input an initial exponential $\Sigma_{\rm gas}$ gas-disk (similar to the steady-state $\Sigma_{\rm gas}$ distribution) but now with significant sinusoidal radial perturbations. The model was run for only 1 Gyr.  As shown in the right-hand panel of Figure \ref{fig:8}, the bumps and depressions in $\Sigma_{\rm SFR}$ (or equivalently in $\Sigma_{\rm gas}$) are eradicated within this time. The time needed to smooth out the derivations strongly decreases with increasing galactic radius.  This is because the radial motions of gas are more effective at lower $\Sigma_{\rm gas}$. The bottom-right panel of Figure \ref{fig:8} shows the evolution of the inflow velocity $v_{\rm r}$. At the beginning of the evolution, the $v_{\rm r}$ shows large variations at the outer region, due to the action of the magnetic stresses, which leads to this short smoothing timescale. 

\subsection{The effect of changes in the inflow rate at the outer boundary}\label{sec:5.1X}

\begin{figure*}
  \begin{center}
    \epsfig{figure=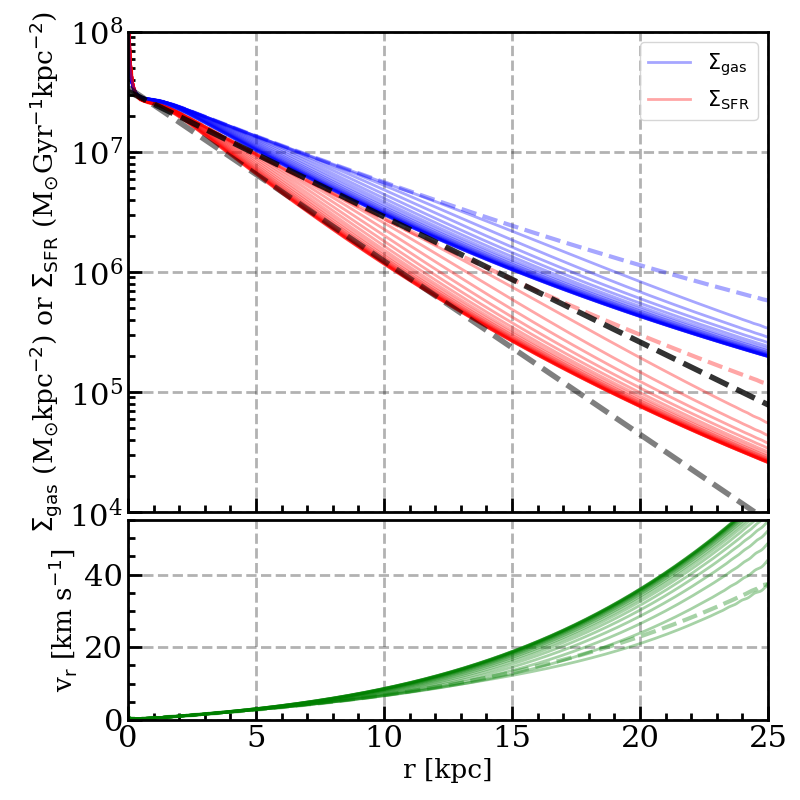,clip=true,width=0.33\textwidth}
    \epsfig{figure=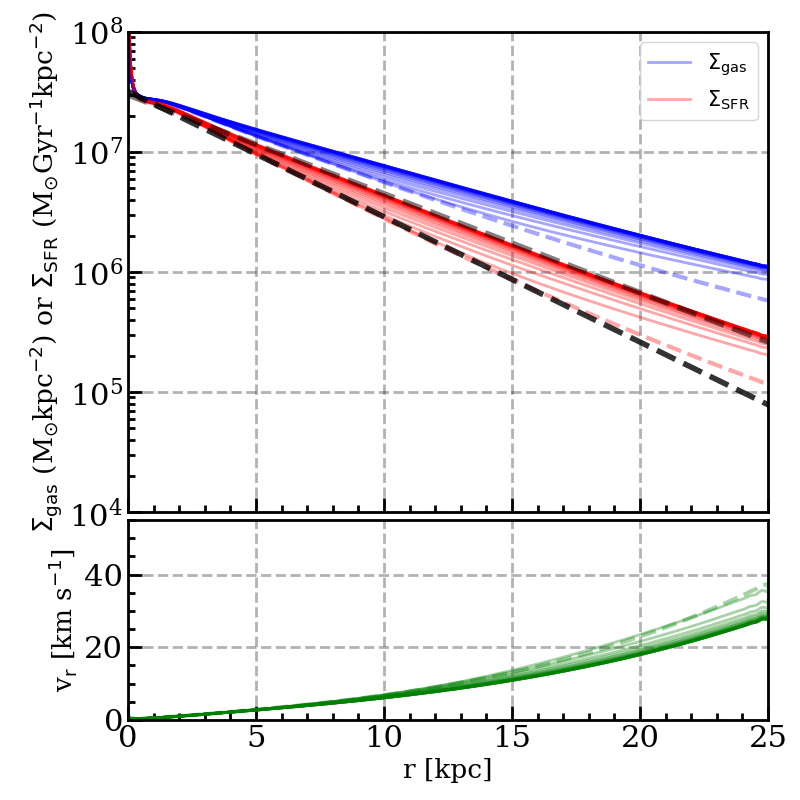,clip=true,width=0.33\textwidth}
    \epsfig{figure=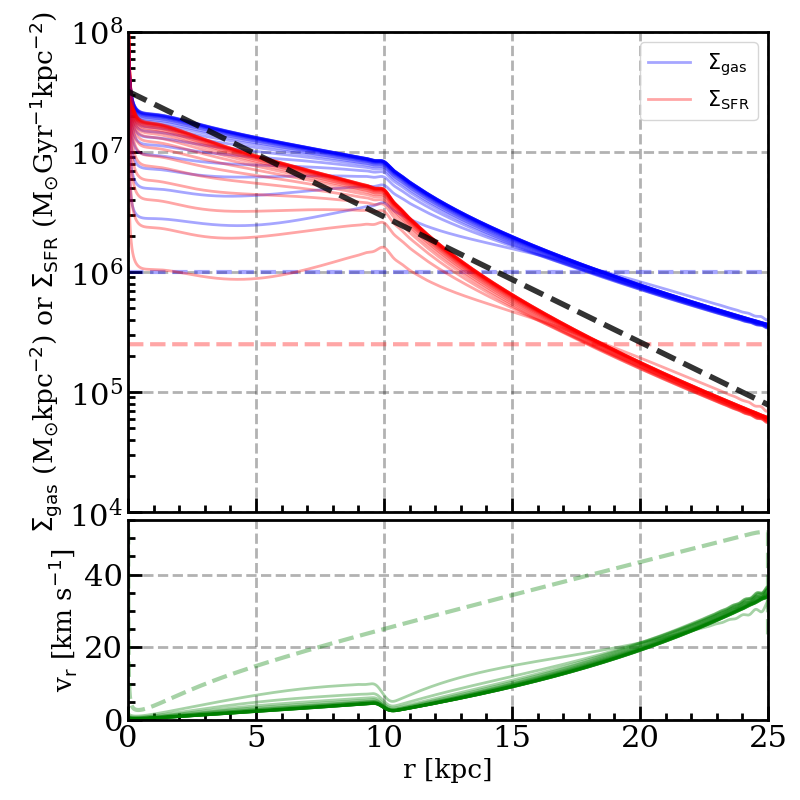,clip=true,width=0.33\textwidth}
    \end{center}
  \caption{ Left panel: The response of $\Sigma_{\rm SFR}$ and $\Sigma_{\rm gas}$ to a sudden decrease of inflow rate at the boundary.  The system starts from the steady-state $\Sigma_{\rm gas}$ obtained with $\Phi_0 = 3.5\ {\rm M_{\odot}yr^{-1}}$, following by a sudden reduction to $\Phi_0 = 2.0 \ {\rm M_{\odot}yr^{-1}}$. The system quickly adjusts to a new exponential equilibrium state extending over 4 scale-lengths (2 dex in $\Sigma_{\rm SFR}(r)$). 
  Middle panel: the response of $\Sigma_{\rm SFR}$ and $\Sigma_{\rm gas}$ to a sudden increase of inflow rate at the boundary from $\Phi_0 = 3.5\ {\rm M_{\odot}yr^{-1}}$, to $\Phi_0 = 5.0 \ {\rm M_{\odot}yr^{-1}}$. 
  Right panel: the evolution of $\Sigma_{\rm SFR}$ and $\Sigma_{\rm gas}$, with inputting a flat $\Sigma_{\rm gas}$ as initial condition and inputting the an inflow rate of 2 ${\rm M_{\odot}yr^{-1}}$ at the boundary and an inflow rate of 1.5 ${\rm M_{\odot}yr^{-1}}$ at 10 kpc.  
  In all three panels, the time interval, the number of  lines and the color-coding are the same as those in the left panel of Figure \ref{fig:7}.  The same black dashed line is shown in all the panels (see Equation \ref{eq:29}). In the left and middle panels, we also show the exponential $\Sigma_{\rm SFR}$ at the new equilibrium as a gray dashed lines for comparison. The $h_{\rm R}$ of the steady-state $\Sigma_{\rm SFR}$ are 3.0 kpc and 5.3 kpc for $\Phi_0=2$ and 5  ${\rm M_{\odot}yr^{-1}}$, respectively.  
   } 
  \label{fig:9}
\end{figure*}

Next we explore the dependence of the resulting $\Sigma_{\rm SFR}$ disk on the rate of accretion at the outer boundary.   We start from the steady-state $\Sigma_{\rm gas}$ solution that is obtained with $\Phi_0 = 3.5\ {\rm M_{\odot}yr^{-1}}$ (see Section \ref{sec:5.1}), but then abruptly change the outer accretion rate, either decreasing it to $\Phi_0 = 2.0$ ${\rm M_{\odot}yr^{-1}}$ or increasing it to $\Phi_0 = 5.0$ ${\rm M_{\odot}yr^{-1}}$, while leaving all other parameters the same.  It should be noted that for the purposes of this experiment we still adopt the same $B_{\rm tot}-\Sigma_{\rm SFR}$ relation as in the fiducial model (Equation \ref{eq:28}).

The left and middle panels of Figure \ref{fig:9} show the evolution of $\Sigma_{\rm gas}$ and $\Sigma_{\rm SFR}$ with these sudden changes in feeding rate. After a few Gyr evolution, the system indeed finds a new steady-state with a different $h_{\rm R}$. The resulting $\Sigma_{\rm SFR}$ still has a closely exponential form over 4-5$h_{\rm R}$, i.e. over a range of 2 dex in $\Sigma_{\rm SFR}$, but now has a changed scalelength $h_{\rm R}$.  

It is clear that the scale-length $h_{\rm R}$, but not the exponential form of the $\Sigma_{\rm SFR}$ profile, depends on the rate of at which the accretion disk is fed at the outer boundary.  We will come back to this point later in Section \ref{sec:6}, where we discuss the connection to the observed mass-size relation for star-forming galaxies.

Finally, we also looked at feeding the accretion disk in a (probably) unrealistic way. We feed the disk at the outer boundary with a rate of 2 ${\rm M_{\odot}yr^{-1}}$, and additionally inject the gas at 10 kpc at a rate of 1.5 ${\rm M_{\odot}yr^{-1}}$. We then run the model for 6 Gyr starting again with a constant initial $\Sigma_{\rm gas}$ as in the fiducial model, again keeping all other settings to be the same. In this case, the system still reaches a steady-state solution, which approximates a double-exponential $\Sigma_{\rm SFR}$ relation, changing scale-length at the intermediate injection point, as shown in the right panel of Figure \ref{fig:9}.  

We conclude that once the physical parameters in the model are established, the feeding rate of the modified accretion disk then becomes a driving factor in determining the scale-length, but not the exponential form, of the disk.

\subsection{The dependence on the detailed values of the model parameters} \label{sec:5.2}

\begin{figure*}
  \begin{center}
      \epsfig{figure=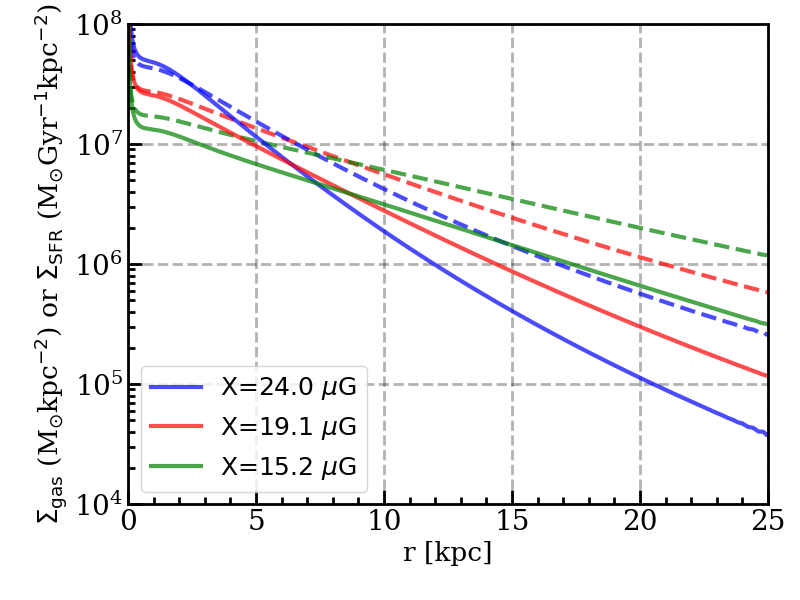,clip=true,width=0.45\textwidth}
      \epsfig{figure=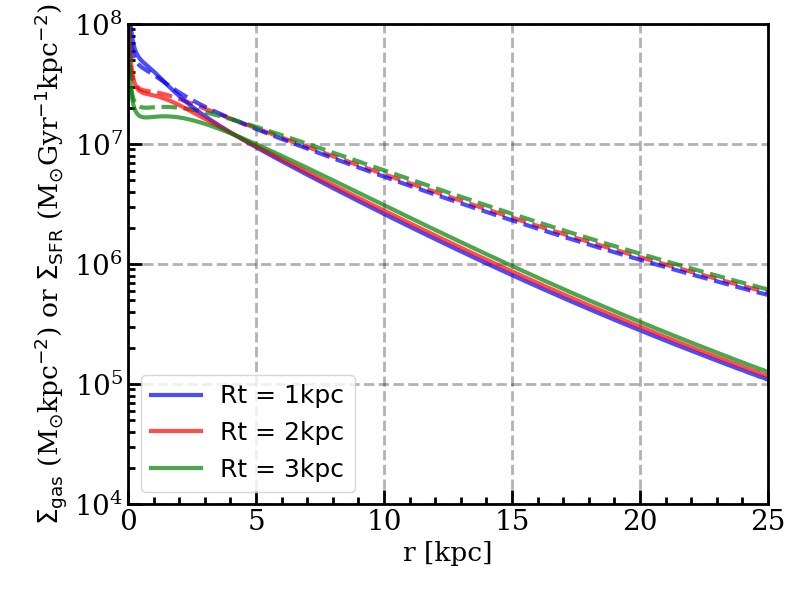,clip=true,width=0.45\textwidth}
     \epsfig{figure=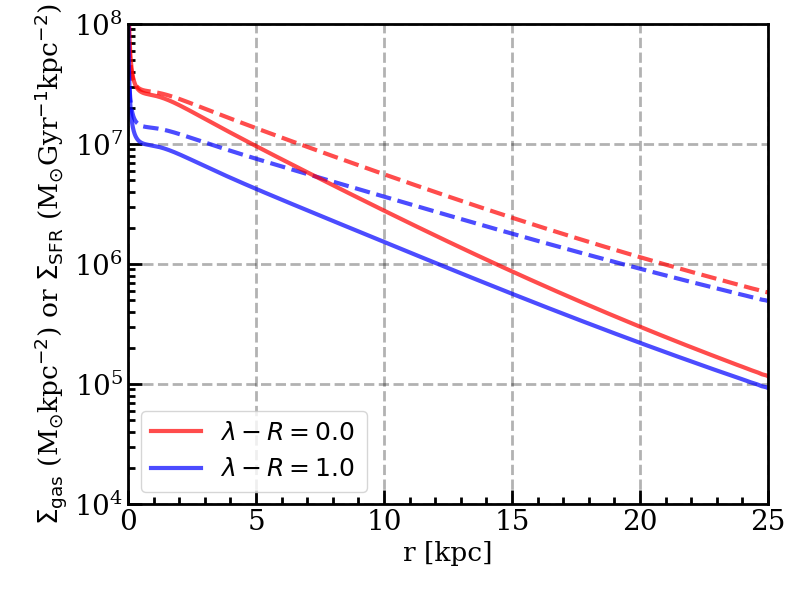,clip=true,width=0.45\textwidth}
     \epsfig{figure=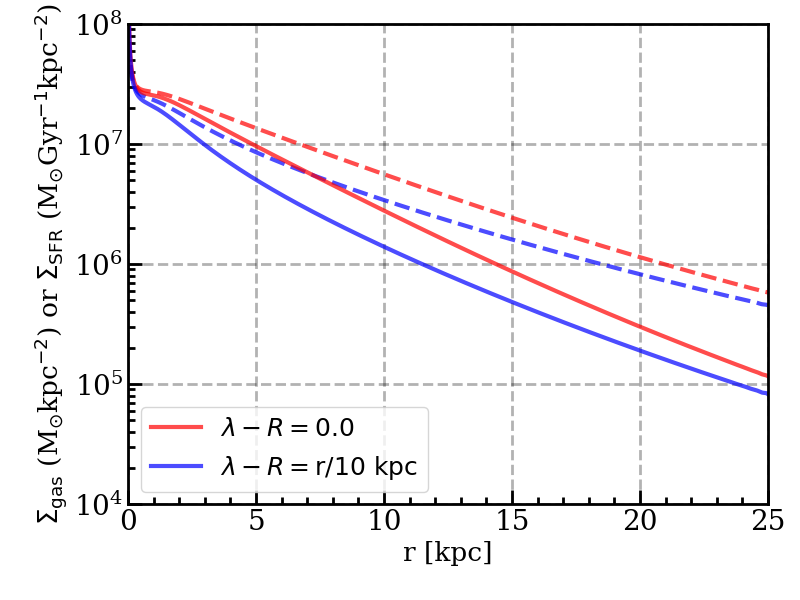,clip=true,width=0.45\textwidth}
     \epsfig{figure=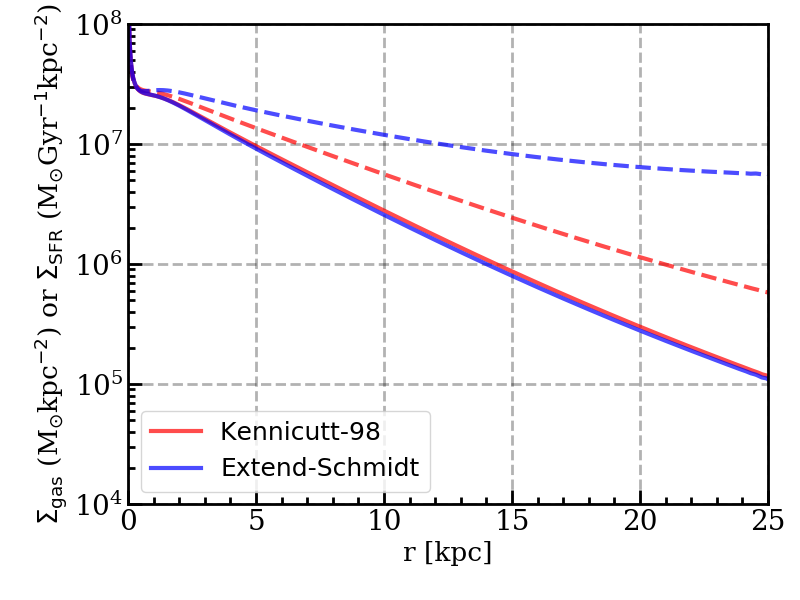,clip=true,width=0.45\textwidth}
    \epsfig{figure=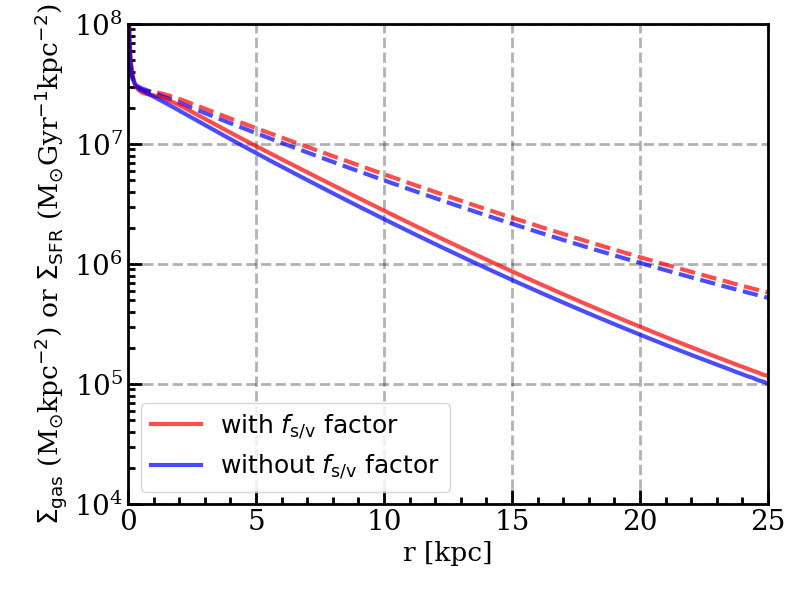,clip=true,width=0.45\textwidth}
    \end{center}
  \caption{These six panels show the dependence of the steady-state $\Sigma_{\rm SFR}$ (and $\Sigma_{\rm gas}$) on, respectively, the overall magnetic field strength $X$ in Equation \ref{eq:28}, the choice of $R_{\rm t}$, the mass-loading factor $\lambda$, on the star-formation law and the inclusion or otherwise of the shear-to-vorticity ratio $f_{\rm s/v}$.  In all six panels of this figure we show the $\Sigma_{\rm SFR}$ as the solid lines, and the $\Sigma_{\rm gas}$ as dashed lines.  The red lines are the same in each panel and denote the steady-state $\Sigma_{\rm SFR}$ (or $\Sigma_{\rm gas}$) of the ``fiducial" run. 
  } 
  \label{fig:10}
\end{figure*}


In this Section, we will explore the effect of altering several of the detailed parameters in the model, including the shape of the galaxy rotation curve via $R_{\rm t}$, the star-formation law linking $\Sigma_{\rm SFR}$ and $\Sigma_{\rm gas}$, the mass-loading factor $\lambda$, the shear-to-vorticity ratio $f_{\rm s/v}$, and the overall normalization and exponent of the $B_{\rm tot}-\Sigma_{\rm SFR}$ relation. 

To do this, we rerun the model several times, changing just one of these parameters around its nominal observationally-motivated value while keeping all the others at the values used in the original fiducial run. We then compare the resulting $\Sigma_{\rm SFR}$ and $\Sigma_{\rm gas}$ profiles with those produced in the original fiducial run. 
\subsubsection{The overall strength of the magnetic fields, and also $v_{\rm cir}$ and $h_{\rm z}$} \label{sec:5.3.1}

The normalization $X$ of the magnetic field strength in Equation \ref{eq:25} was set to be 19.1 $\mu G$ in the fiducial run (Equation \ref{eq:28}).  The effect of a stronger (or weaker) magnetic field was explored by increasing (or decreasing) $X$ by 0.1 dex.  We show these three steady-state $\Sigma_{\rm SFR}$ in the top left panel of Figure \ref{fig:10}.  The scale-length of the resulting $\Sigma_{\rm SFR}$ decreases significantly with increasing 
normalisation $X$, i.e. the overall strength of the magnetic fields at given $\Sigma_{\rm SFR}$. An overall increase of $X$ by 0.1 dex, results in a decrease of $h_{\rm R}$ of $\sim$0.19 dex. 

We note that it is implicit in Equation \ref{eq:5} and Equation \ref{eq:20}
(see the discussion later in Section \ref{sec:6.1}),
that changing the overall field strength $B_{\rm tot}^2$
should have exactly the same effect as varying the $V_{\rm cir}$ or $h_{\rm z}$. Specifically, increasing the $X$ by 0.1 dex is equivalent to increasing $h_{\rm z}$ by 0.2 dex, or decreasing $V_{\rm cir}$ by 0.2 dex. 
We therefore do not present further figures with changing the values of $V_{\rm cir}$ and $h_{\rm z}$ here. 
We will study the basic scaling-relations of our model later in Section \ref{sec:6} in more detail, and return to this point there. 

\subsubsection{The effective mass-loading of the explanar winds $\lambda$}

In the fiducial run, the $\lambda-R$ term was set for convenience to be zero, so that the SFR exactly equals the feeding rate of the disk in the steady-state, i.e. the explanar mass-loss from the disk is set (arbitrarily) to be equal to the mass return from newly formed stars. In the middle left panel of Figure \ref{fig:10}, we show the steady-state $\Sigma_{\rm SFR}$ for a case with stronger outflow, i.e. $\lambda-R = 1$. As shown, the overall $\Sigma_{\rm SFR}$ is now lower than that of the fiducial run, because of the larger mass-loading factor.   However, it is clear that the resulting $\Sigma_{\rm SFR}$ is still in exponential form and that the disk scale-length $h_{\rm R}$ is larger for larger $\lambda$. 

It might seem surprising that the choice of $\lambda$ affects the scale length since it was argued above in Section \ref{sec:2.1} that the effect of $\lambda$ is primarily a scaling between the mass extracted from the gas and the nominal SFR.  However, increasing $\lambda$ reduces the $\Sigma_{\rm SFR}$ that is associated with the extraction of a certain surface mass density of gas from the disk, and thereby reduces the magnetic field.   
In terms of the removal of gas from the disk, changing $\lambda$ has therefore the same effect as changing $X$, as discussed in the previous subsection. 

Although we assumed a constant $\lambda$ in the fiducial run, it might be expected that the $\lambda$ could increase with galactic radius, as the gravitational potential well becomes less deep. We therefore examine the effect of a radially dependent $\lambda$, simply assuming $\lambda-R=r/10$ kpc. The resulting $\Sigma_{\rm SFR}$ is shown in the middle right panel of Figure \ref{fig:10}.  Not surprisingly, this change leads to a not perfect exponential profile in $\Sigma_{\rm SFR}$. Of course, such deviations are seen in real galaxies, with both upbending and downbending in the outer region of the disk \citep[Figure \ref{fig:0}, see also][]{Erwin-05, Pohlen-06}. We will return to this point later in Section \ref{sec:5.3}. 

\subsubsection{The turnover radius of the rotation curve $R_{\rm t}$}

$R_{\rm t}$ is the turnover radius of the rotation curve defined in Equation \ref{eq:13}, set to $R_{\rm t}=2$ kpc in the fiducial run. The results of varying $R_{\rm t}$ (to 1 and 3 kpc respectively) are shown in the top right panel of Figure \ref{fig:10}. Different $R_{\rm t}$ mostly change the steady-state $\Sigma_{\rm SFR}$ (or $\Sigma_{\rm gas}$) in the innermost regions ($<$4 kpc). A smaller $R_{\rm t}$ (1 kpc) results in a more cusp-like profile while a larger $R_{\rm t}$ (3 kpc) results in a core-like profile.  Changing $R_{\rm t}$ means changing the differential angular velocities in the inner region of the gas disk. The $\Sigma_{\rm SFR}$ profiles are essentially unaffected at radii $r>4$ kpc with only very small offsets between the steady-state solutions. These very small systematic offsets are in fact required to offset the differences at the center of the disk center, since the integrated SFR is set by the overall accretion rate, which was kept constant for all three cases. 

\subsubsection{The choice of star-formation law} \label{sec:5.3.4}

We next examine the effect of the adopted star-formation law. We use a different star formation law, the extended-Schmidt law, in which the star formation efficiency (SFE, defined as the SFR divided by gas mass) is proportional to the $\sim \Sigma_*^{1/2}$ \citep{Shi-11}. We compute the SFE as a function of radius using equation 6 of \cite{Shi-11}, assuming the typical Main Sequence galaxy has an exponential stellar disk with scalelength of 4 kpc. We then run the dynamic model with implementing this new (time-invariant) star formation law. The steady-state $\Sigma_{\rm SFR}$ and $\Sigma_{\rm gas}$ are shown in the bottom left panel of Figure \ref{fig:10}. 

As shown, the $\Sigma_{\rm SFR}$ with the new star formation law is essentially identical to the $\Sigma_{\rm SFR}$ of the fiducial run, but the $\Sigma_{\rm gas}$ are therefore quite different. The resulting $\Sigma_{\rm SFR}$ evidently do not depend on the star formation law that we adopted, indicating that the $\Sigma_{\rm SFR}$ is the more fundamental output than $\Sigma_{\rm gas}$. This ultimately reflects the presence of $\Sigma_{\rm SFR}$ rather than $\Sigma_{\rm gas}$ in the angular momentum Equation \ref{eq:3}.  This is also a basic feature of any gas regulator system \citep{Lilly-13, Wang-19, Wang-21}.  The $\Sigma_{\rm gas}$ adjusts to whatever is needed to maintain the required $\Sigma_{\rm SFR}$. 

\subsubsection{The $f_{\rm s/v}$ term in the magnetic stress}

As discussed in Section \ref{sec:3.3}, we adjusted the magnetic stress (Equation \ref{eq:20}) in our fiducial run, by multiplying by the shear-to-vorticity ratio $f_{\rm s/v}$ in Equation \ref{eq:22} \citep{Abramowicz-96, Hawley-99}.  
We show in the bottom right panel of Figure \ref{fig:10} the effect on the steady-state $\Sigma_{\rm SFR}$ profile of turning off this $f_{\rm s/v}$ term. Omitting this reduction factor, the steady-state $\Sigma_{\rm SFR}$ is barely changed over almost all radii (there is a small constant offset of $\sim$0.07 dex) but there is, not surprisingly, a big difference in the innermost parts of the disk where the $f_{\rm s/v}$ term acts in the fiducial model to decrease the viscosity.  

The effect of the $f_{\rm s/v}$ term is to reduce the efficiency of angular momentum (and thus mass) transportation at the disk center.  Without the $f_{\rm s/v}$ reduction factor, the disk is still able to efficiently transport gas into the disk center, resulting in a gas inflow rate at 0.1 kpc that is as high as 0.53 \msolar yr$^{-1}$, corresponding to $\sim$15\% of total inflow rate at the outer boundary. This may be compared with the $\sim$0.4\% discussed above when the $f_{\rm s/v}$ reduction in the magnetic stress is included.

Evidently, {\it without} the $f_{\rm s/v}$ factor, the feeding rate of the central regions of the galaxy would be expected to be $\sim$40 times higher than with it.  In other words, the inclusion of this term evidently allows the disk to have an entirely reasonable mass sink at the center. Omitting it would result in a very high central mass deposition rate.

\subsubsection{The dependence on $\alpha$} \label{sec:5.3}

\begin{figure}
  \begin{center}
    \epsfig{figure=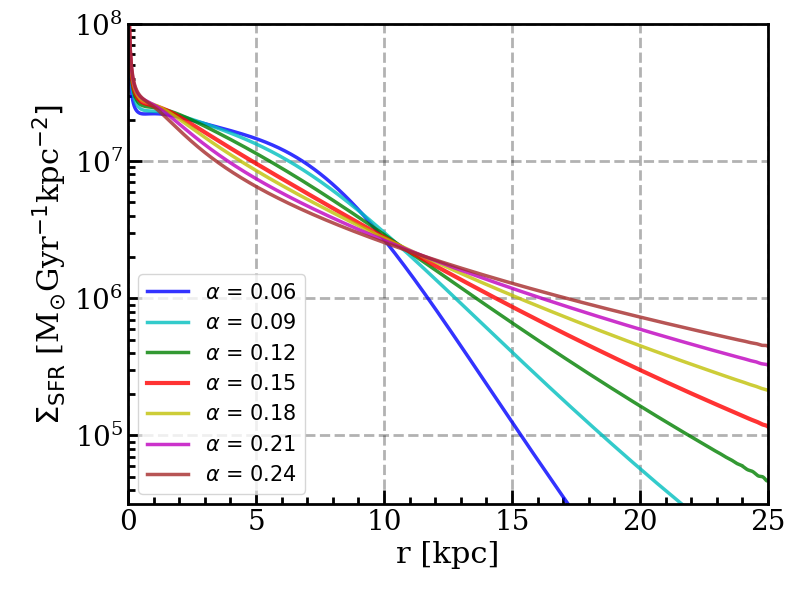,clip=true,width=0.45\textwidth}
    \end{center}
  \caption{The dependence of the steady-state $\Sigma_{\rm SFR}$ profile on the choice of exponent $\alpha$ in the $B_{\rm tot}  - \Sigma_{\rm SFR}$ relation.  All other settings are kept the same as in the fiducial run, which is shown as the red line, with $\alpha=0.15$.}
  \label{fig:11}
\end{figure}

\begin{figure}
  \begin{center}
    \epsfig{figure=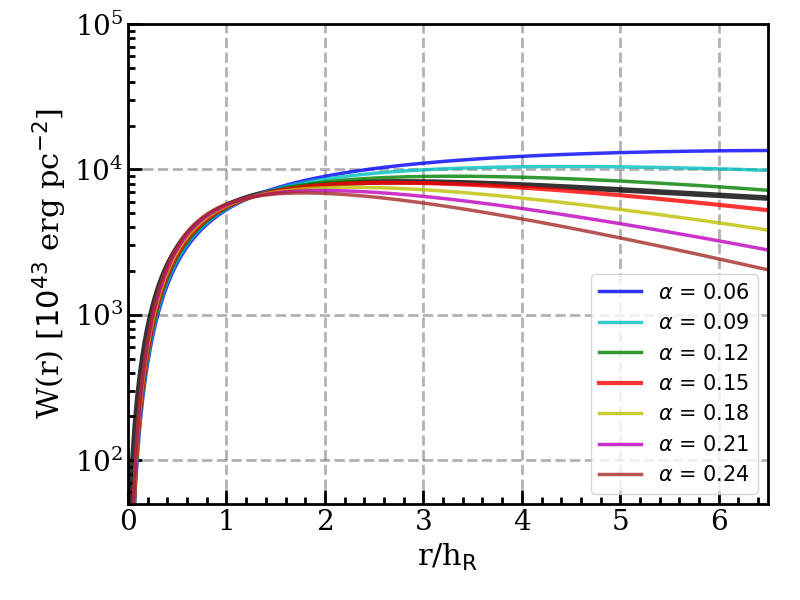,clip=true,width=0.45\textwidth}
    \end{center}
  \caption{ Comparison between the required viscous stress for a perfect exponential $\Sigma_{\rm SFR}$ disk (the black curve) with the viscous stress that is obtained for different values of $\alpha$ for the typical Main Sequence galaxy.  Color-coding of the lines for different $\alpha$ is as on Figure \ref{fig:11}. The mass-loading factor is assumed to be 0.4 in calculating the viscous stresses.  } 
  \label{fig:11X}
\end{figure}



The core of the success of the modified accretion disk model is the circular connection between star-formation and the radial transport of gas (to fuel that star-formation) due to the viscous stresses established by the magnetic fields that are themselves associated with the star-formation.  Not surprisingly therefore, the exponent $\alpha$ in the Equation \ref{eq:25} between the magnetic field strength and the star-formation rate surface density $\Sigma_{\rm SFR}$ {\it within a particular system} plays a key role in the operation of the disk.  In this subsection, we will focus on the effect of $\alpha$ on the steady-state $\Sigma_{\rm SFR}$ profile that is produced by the model.

In previous sections, we set the value of $\alpha$ to be 0.15, the mean value seen in the disk galaxies studied by \cite{Heesen-14}.  However, the inferred $\alpha$ in that study show a significant variation across the galaxy population, with a standard deviation from galaxy to galaxy of 0.06.

We therefore run the model with several different $\alpha$ (in the range of 0.06-0.24), while keeping all other settings to be the same as in the fiducial run. The results are shown in Figure \ref{fig:11}. As can be seen, a single exponential profile is produced {\it only} for a relatively narrow range of $\alpha$ centered on the observationally indicated mid-value of $\sim$0.15.  Specifically, we find $\alpha$ in the range of 0.1-0.2 roughly corresponds to the r.m.s. deviation of the resulting $\Sigma_{\rm SFR}$ from pure exponential function within 0.1 dex. 
The profiles of $\Sigma_{\rm SFR}$ at significantly different $\alpha$ show upbending features (shallower outer regions) for larger $\alpha$, and downbending features (steeper outer region) for smaller $\alpha$. This makes sense: the {\it local} scale-length of $\Sigma_{\rm SFR}(r)$ changes with different $\alpha$ in a way that corresponds to the change of overall scale-length with $X$ that was discussed above in Section \ref{sec:5.3.1}. 

The sensitivity to $\alpha$ is illustrated in Figure \ref{fig:11X}.  This compares our ``required'' $W(r)$ for a perfect exponential disk extracted from Figure \ref{fig:4} with the $W(r)$ that comes from the same $\Sigma_{\rm SFR}(r)$ profile but with different $\alpha$.  Clearly only values close to 0.15 match this curve.
 
This apparent sensitivity to the value of $\alpha$ presents something of a quandary for the model.  On the one hand, the fact that the model only produces good exponential profiles for a small range of $\alpha$ (even possibly smaller than the observed range) might be considered a weakness.  On the other hand, given this evident sensitivity, the fact that this small range is apparently centered on the average value from those (quite independent) observations could be taken as a strong validation of the model. Of course, one might still ask why $\alpha$ takes this particular value in galaxies, but that discussion is beyond the scope of this paper.



It should be noted that not all disks are pure exponentials. Up- and down-bending features are seen in the outer regions for both stars and SFR (see Section \ref{sec:1} and Figure \ref{fig:0}). 
In the context of our model, this variation could perhaps reflect variations in $\alpha$ between different galaxies. 

Finally, as mentioned above, a radial dependence of $\lambda$ or a different radial dependence of the vertical scale height $h_{\rm z}$ would in principle have a similar effect as varying the overall $\alpha$, according to Equation \ref{eq:5}  and Equation \ref{eq:20} (also see the middle right panel of Figure \ref{fig:10}). 

\section{Angular momentum} \label{sec:5.4X}

\begin{figure}
  \begin{center}
    \epsfig{figure=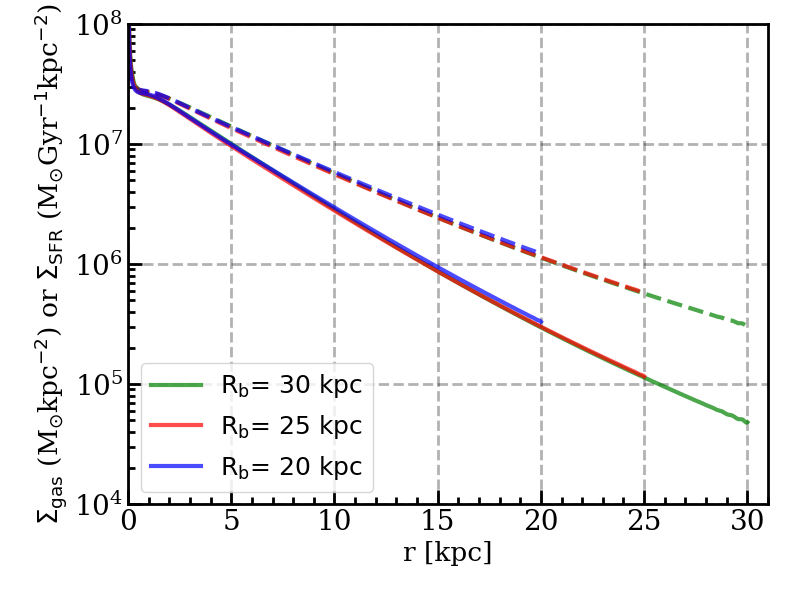,clip=true,width=0.45\textwidth}
    \end{center}
  \caption{The steady-state $\Sigma_{\rm SFR}$ (solid lines) and $\Sigma_{\rm gas}$ (dashed lines) obtained when the disk is fed at different boundaries $R_{\rm b}$. The red (solid and dashed) lines show the result of the fiducial run with $R_{\rm b}=25$ kpc.  The structure of the disk is independent of the choice of outer boundary of the disk, meaning that the results of this paper are insensitive to exactly where the inflowing gas achieves rotational support. 
  } 
  \label{fig:12}
\end{figure}

We now turn to the important question of the angular momentum distribution in the disk. In classical accretion disks around compact objects, the angular momentum is transferred outwards as mass flows inwards, and the same is true also for our modified accretion disk model.  The loss of angular momentum as the gas flows inwards is therefore of considerable interest in relating our modified accretion disk model to attempts to understand the angular momentum distribution of galactic disks on the basis of an assumed {\it conservation} of the specific angular momentum of the disk material. 
In that picture the angular momentum distribution of the disk reflects the angular momentum distribution of the original material, which is generally taken to be the same as the angular momentum distribution of the dark matter in the halo.  In the model presented here, the angular momentum distribution of the disk reflects the effect of the viscous stresses as the gas evolves down towards the center of the disk.

In the fiducial run, we fed the modified accretion disk at a constant mass accretion rate at a particular outer boundary radius $R_{\rm b}$. Since the inflowing material is assumed to be rotationally-supported, i.e. to be rotating at the circular velocity of the gravitational potential at that radius, this new material adds a well-defined angular momentum to the disk.  Because of the approximately flat rotation curve, the rate at which angular momentum is added by this inflow will clearly depend on the choice of the outer boundary $R_{\rm b}$, and will be $\sim$ $R_{\rm b}V_{\rm cir}\Phi_0$. The rate of angular momentum addition will therefore be proportional to $R_{\rm b}$, at least where the rotation curve is flat.

A key question is whether or not the steady-state $\Sigma_{\rm SFR}$ profile of the disk therefore depends on this rate of addition of angular momentum, i.e. on the outer boundary at which (rotationally-supported) mass is injected into the disk. To examine this, we feed the accretion disk at different $R_{\rm b}$, running the model with all other settings the same as in the fiducial run.  The result is shown in Figure \ref{fig:12}.   The $\Sigma_{\rm SFR}$ profiles of different $R_{\rm b}$ are almost completely overlapped together. As encountered before, there is a very minor offset in $\Sigma_{\rm SFR}$, especially for the case of $R_{\rm b}=20$ kpc, but this is due to a trivial truncation effect that arises because the integrated SFR within $R_{\rm b}$ should be exactly equal to the feeding rate in the steady-state (when $\lambda-R$ is set to zero).  Since we feed the system with the same rate but at different radii, the resulting $\Sigma_{\rm SFR}$ for smaller $R_{\rm b}$ should be slightly higher than that for higher $R_{\rm b}$.  

Except for the very minor offset induced by this trivial truncation effect, the final steady-state of the disk (and therefore its angular momentum) evidently does {\it not} depend on the choice of outer boundary at which the disk is fed, i.e. on the rate at which angular momentum was added to the disk.

This is an important point: we discussed in Section \ref{sec:2.3} how the implied $v_{\rm r}$ may be large enough at large radii that the motion of the gas may well not approximate circular motion, possibly invalidating the assumption of rotational support and the applicability of the basic angular momentum Equation \ref{eq:5} for the disk.    However, we know from simulations that the inflowing gas will eventually settle into a rotationally-supported disk at some radius \citep[e.g.][]{Peroux-20, Trapp-21}, where our analysis should be valid, even if it is not fully valid at larger radii.  Knowing that the profile of the star-forming disk does not in fact depend in our model on the precise choice of this outer feeding boundary, we could set the feeding boundary to be this point of rotational-support.   But the fact that the $\Sigma_{\rm SFR}(r)$ profile evidently does {\it not} then depend on the assumed location of this outer boundary of the disk, means that we do not have to worry about exactly where the boundary actually is in practice.

This further emphasizes how the transportation/removal of angular momentum by viscosity is very efficient.  Moving the outer boundary outwards implies that the ``modified accretion disk" removes a larger and larger fraction of the incoming angular momentum through viscous stresses because the angular momentum of the final disk evidently remains the same, independent of the boundary location. 

Clearly any surviving gas that penetrates all the way down to the center of the disk must have lost essentially all of its angular momentum.  An obvious question about the steady-state situation then arises: what fraction of the angular momentum that was deposited in the disk by the incoming gas at the outer boundary is retained in the angular momentum of the newly-formed stars that were formed along the path to the center (or equivalently driven away in a wind)?

\begin{figure}
  \begin{center}
    \epsfig{figure=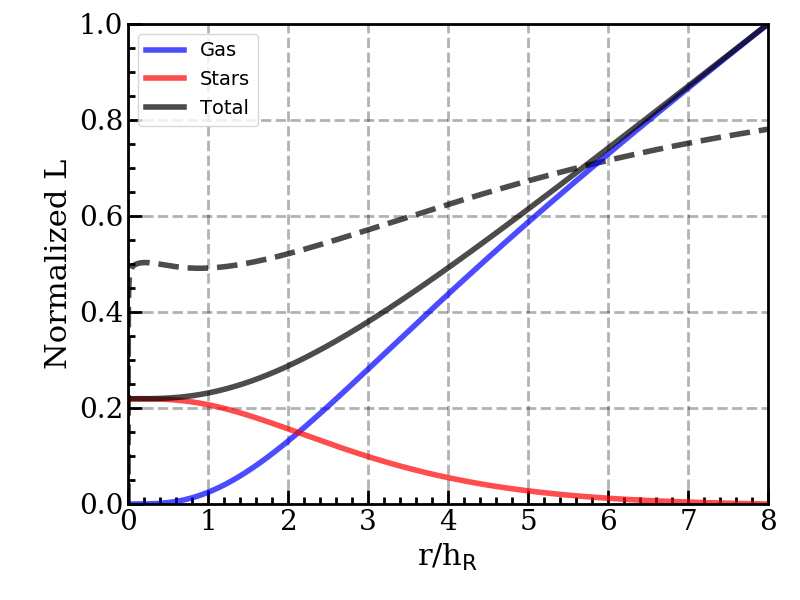,clip=true,width=0.45\textwidth}
    \end{center}
  \caption{
 The loss of angular momentum in the disk, computed for an exponential star-forming disk with an {\tt arctan} rotation curve (assuming $R_{\rm t}=0.5h_{\rm R}$).  The blue curve shows the angular momentum of the surviving gas at radius $r$, normalised to the angular momentum of rotationally-supported gas at 8$h_{\rm R}$. 
 The red curve shows the cumulative angular momentum of the stars (plus any outflowing gas) that have already been formed from this gas element by the time it reaches a given radius, similarly normalized. The sum of these is the total remaining angular momentum (black solid line). 
 The dashed black curve shows the same information in a different way. It shows, as a function of $r$, the fraction of the angular momentum flux of the inflowing gas at radius $r$ that does not appear in the angular momentum flux associated with the formation of stars (or outflow) at all radii within $r$. 
  } 
  \label{fig:12.1}
\end{figure}

We can examine this in very general terms provided that we can assume radial coplanar flow of the gas. For simplicity, we calculate this for an ideal purely exponential star-forming disk with an {\tt arctan} rotation curve.    The blue curve in Figure \ref{fig:12.1} shows the angular momentum of the surviving gas at radius $r$ as it flows down towards the center of the system, normalised to the angular momentum of rotationally-supported gas at 8$h_{\rm R}$. This decreases towards smaller $r$ due to both (a) the loss of mass by star formation and outflow, and (b) the decrease of the specific angular momentum of the gas with $r$ implicit in the disk rotation curve.  The red curve shows the cumulative angular momentum of the stars (plus any outflowing gas) that have {\it already} been formed from this gas element by the time it reaches the radius $r$, similarly normalized.   The sum of these, shown by the black solid line, is the total remaining angular momentum, either in the surviving gas or in the stars (plus outflow) left behind. This also decreases with radius, illustrating the substantial loss of angular momentum from the disk due to the viscous stresses within it.   

The black dashed curve shows the same information in a different way. It shows, as a function of $r$, the fraction of the angular momentum flux into the interior parts of the disk by the inflowing gas at radius $r$ that does not appear in the angular momentum flux associated with the formation of stars (or outflow) at all radii within $r$. This directly gives as a function of $r$ the ``lost" angular momentum that must have been transported outwards by the disk.

For instance, at a radius of 3$h_{\rm R}$ (or 8$h_{\rm R}$),  $\sim$57\% (or $\sim$78\%) of the angular momentum of the inflowing gas must still be lost from the system through outward transport by the viscosity. 
This emphasizes the high levels of angular momentum exchange in an accretion disk. This will be true for any exponential disk in which the dominant gas flow is coplanar flow. The assumption of conservation of (specific) angular momentum of particular gas packets is likely in our view to be a poor one.  We may expect that the specific angular momentum of the stars and gas in galactic disks will be only a fraction (e.g. $20-50 \%$) of the angular momentum of the inflowing gas, because of the extraction of angular momentum by the viscosity.  The precise fraction of the incoming angular momentum that is ``lost" depends on the ratio of the outer boundary (i.e. where the incoming angular momentum is nominally accounted) to the disk scale-length $h_{\rm R}$.

\section{The scaling relations of disk sizes predicted by the model} \label{sec:6}


In Section \ref{sec:5}, we mainly focused on an individual system with parameters appropriate for a typical, reasonably massive, Main Sequence galaxy. In this section, we try to extend our model to the overall star-forming population, and examine whether this simple model can reproduce the observed mass-size relation.  To do this, we need to know how the relevant parameters change with stellar mass.  Some of these may not be well established from observations, such as the scale-height and mass-loading factor. Therefore, before proceeding, we should remind readers of these uncertainties. 

We first show that the $X$ in Equation \ref{eq:25} may well vary from galaxy to galaxy, and that it appears to depend on the integrated SFR of a galaxy.  We then theoretically estimate the expected scaling relation and try to validate this relation of $\Phi_0$-$h_{\rm R}$ (or SFR-$h_{\rm R}$) relation by running the model (see Appendix \ref{sec:A1}). 

\subsection{The dependence of $X$ on SFR} \label{sec:6.0}

\begin{figure}
  \begin{center}
    \epsfig{figure=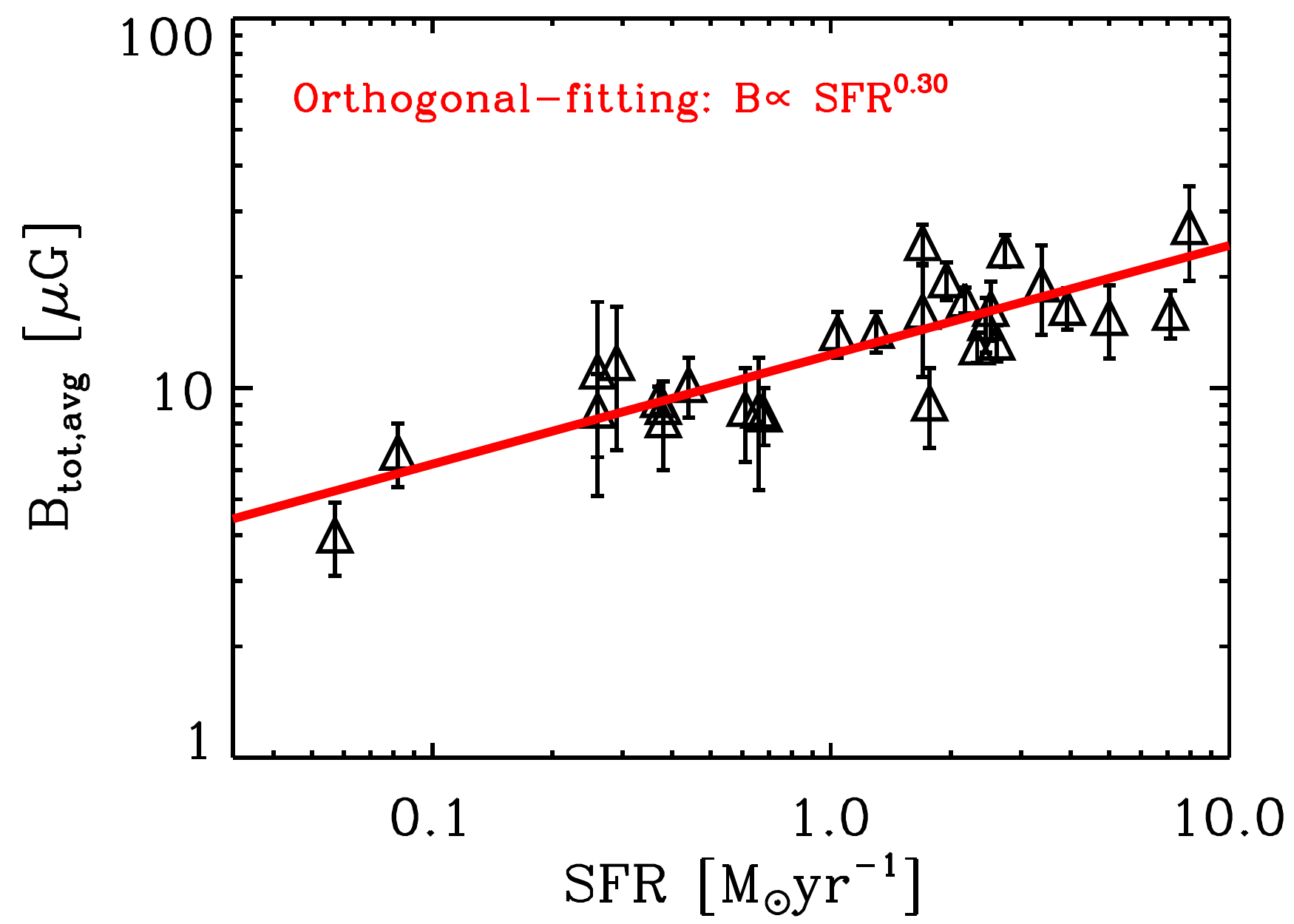,clip=true,width=0.48\textwidth}
    \end{center}
   \caption{The averaged total field strength of {\bf B} vs. the integrated SFR of disk galaxies. The data points are taken from \cite{Tabatabaei-17}. The red line shows the orthogonal regression in logarithmic space represented by Equation \ref{eq:24}. }
   \label{fig:6}
\end{figure}


Figure \ref{fig:6} shows the ``average'' field strength of {\bf B} vs. integrated SFR for a sample of nearby disk galaxies taken from \cite{Tabatabaei-17}. The average field strength can be regarded as the (nonthermal-radio) intensity-weighted field strength.  We denote this averaged total magnetic field as $B_{\rm tot,avg}$, to distinguish it from the radially-varying total field strength $B_{\rm tot}(r)$ that provides the viscosity in our accretion disk. We emphasize again that the {\it total} field includes both ordered and disordered fields. 

An orthogonal regression of the data points in Figure \ref{fig:6} yields:
\begin{equation} \label{eq:24}
B_{\rm tot, avg} = Y \cdot ({\rm \frac{SFR}{ M_{\odot} yr^{-1}}})^{\beta}, 
\end{equation}
where $Y = 12.3\ \mu G$, and the exponent $\beta=0.30$. This exponent of 0.3 also agrees well with the one found in \cite{Heesen-14} among the galaxy population. 

The difference in exponents between the integrated and spatially-resolved $B_{\rm tot}$-SFR relations (i.e. $\beta \sim 0.30$ vs. $\alpha \sim 0.15$) may be due to a smearing out of the radial B-field profile within a galaxy, e.g. by the scattering of the cosmic rays, and/or the increasing significance of a regular magnetic field in the outer regions \citep[e.g.][]{Tabatabaei-13a, Seta-19}. 

The two relations can clearly be formally brought into consistency by making the normalization factor $X$ in Equation \ref{eq:25} to be a function of the overall SFR of the galaxy in question.

To construct a general $B_{\rm tot}-\Sigma_{\rm SFR}$ relation for the disk galaxy population, we therefore need to figure out the dependence of $X$ on the integrated SFR of the galaxy. We could assume for simplicity that $\Sigma_{\rm SFR}$ has an exponential profile (Equation \ref{eq:7}), and that the $B_{\rm tot, avg}$ should therefore be proportional to the $B_{\rm tot}$ at the scale-radius $h_{\rm R}$, i.e $B_{\rm tot,avg}= k \cdot B_{\rm tot}(h_{\rm R})$,  Equation \ref{eq:25} can then be rewritten as: 
\begin{equation} \label{eq:26}
    B_{\rm tot}(r) = Y/k \cdot (\frac{\rm SFR}{\rm  M_{\odot} yr^{-1}})^{\beta} \cdot (\frac{\Sigma_{\rm SFR}(r)}{\Sigma_{\rm SFR}(h_{\rm R})})^{\alpha}, 
\end{equation}
Clearly $Y/k$ and the integrated SFR determine the total field strength at the scale-radius $h_{\rm R}$ of the disk, while $\alpha$ and $h_{\rm R}$ determine the shape of the $B_{\rm tot}(r)$ profile. As we will show later in Section \ref{sec:6.1}, the value of $\beta$ is important to reproduce the observed mass-size relation in our model. 

If $B_{\rm tot,avg}$ is measured within 3$h_{\rm R}$ with $\alpha=0.15$ \citep{Heesen-14} then $k$ has a value of 0.88. The value of $k$ would be slightly lower if $B_{\rm tot,avg}$ is measured within a larger radius.  With this assumption, we obtain the $X=19.1 \ \mu$G value that was used for the typical Main Sequence galaxy referred in Section \ref{sec:5} by setting the $\Sigma_{\rm SFR,zp}$ as $\Sigma_{\rm SFR}(h_{\rm R})\sim 0.01\ {\rm M_{\odot}yr^{-1}kpc^{-2}}$ \citep{Guo-19, Wang-19}.  We note that the value of $X$ is not very sensitive to $\Sigma_{\rm SFR,zp}$ for small $\alpha\sim0.15$. 

\subsection{Theoretical prediction of the $h_{\rm R}$ scaling relation in the model} \label{sec:6.1}

In Section \ref{sec:5.2}, we found that the scale-length of steady-state $\Sigma_{\rm SFR}$ exponential disks depends on the magnetic field strength factor $X$, the mass-loading factor $\lambda$, and the feeding rate $\Phi$ at the boundary. Here we try to understand this dependence analytically. 

In Section \ref{sec:2.4}, we solved the analytic form for the required stress $W$ that was required to maintain a perfect exponential star-forming disk. In Section \ref{sec:5}, we then showed that good exponential disks are obtained by the magnetic stress with our adopted relation between magnetic field strength and $\Sigma_{\rm SFR}$, i.e. with $\alpha \sim 0.15$.

Adopting the Equation \ref{eq:26}, the magnetic stress can be written as: 
\begin{equation} \label{eq:30}
\begin{split}
    W(r) 
    & = f_{\rm s/v} \times 0.61 \times 
    \frac{B_{\rm tot}^2}{8\pi} \cdot 2h_{\rm z}  \\ 
    & = 0.61 f_{\rm s/v} (Y/k)^2  (\frac{\rm SFR}{{\rm 1\ M_{\odot}yr^{-1}}})^{2\beta} (\frac{\Sigma_{\rm SFR}(r)}{\Sigma_{\rm SFR}(h_{\rm R})})^{2\alpha}\cdot \frac{h_{\rm z}}{4\pi}
\end{split}
\end{equation}
Connecting this equation with Equation \ref{eq:12} and {\it ignoring the radially-dependent component}, we have 
\begin{equation} \label{eq:31}
  {\rm SFR}^{2\beta} h_{\rm z} \propto (1-R+\lambda)V_{\rm cir}\cdot \frac{\rm SFR}{h_{\rm R}}. 
\end{equation}
The Equation \ref{eq:31} can be further written as: 
\begin{equation} \label{eq:32}
    h_{\rm R} \propto {\rm SFR}^{1-2\beta} h_{\rm z}^{-1} V_{\rm cir} (1-R+\lambda). 
\end{equation}

The Equation \ref{eq:32} predicts the dependence of the star-formation scale length $h_{\rm R}$ on the values of the various parameters.  As shown in Section \ref{sec:5.2}, an increase of magnetic field strength does indeed lead to a smaller $h_{\rm R}$, as does a decrease of mass-loading factor.  

We now look at how these parameters may vary with the stellar mass of the galaxy, again, being guided by observations as much as possible.  The $V_{\rm cir}$ almost certainly increases with stellar mass, $V_{\rm cir}\propto M_*^{0.25}$ \citep[e.g.][]{McGaugh-00, Stark-09}.  The mass-loading factor $\lambda$ on the other hand is expected to decrease with stellar mass. Based on the hydrodynamical simulation, \cite{Muratov-15} found a nearly time-invariant $\lambda-M_*$ relation of $\lambda \propto M_*^{-0.35}$.  The $1-R+\lambda$ term is more weakly dependent on stellar mass with respect to $\lambda$.
Therefore, the combined effects of $V_{\rm cir}$ and the mass-loading factor $\lambda$ on $h_{\rm R}$ are likely to largely cancel out.  \cite{Wilson-19} found that the $h_{\rm z}$ of galaxies is also nearly constant across a wide range of molecular gas surface density.  

Based on Equation \ref{eq:32}, we therefore expect that, at a given stellar mass and (integrated) SFR, galaxies with stronger magnetic field should have smaller scale-lengths. This could be examined in the future. 

\begin{figure}
  \begin{center}
    \epsfig{figure=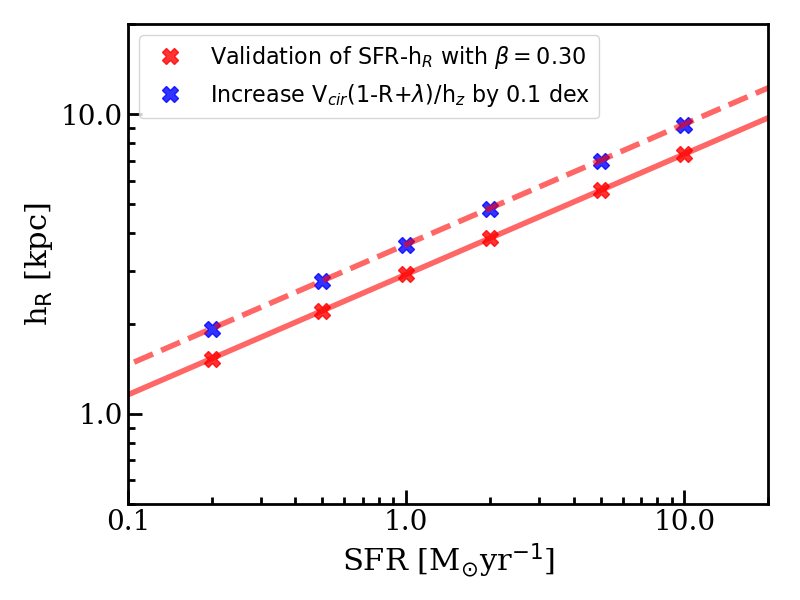,clip=true,width=0.48\textwidth}
    \end{center}
   \caption{ The validation of the theoretical prediction of the $h_{\rm R}$-scaling relation (Equation \ref{eq:32}). The red crosses show the output $h_{\rm R}$ as a function of SFR by running the model with effectively setting $\beta=0.3$ (see detailed settings in Appendix \ref{sec:A1}). The red solid line is the linear fitting of the red crosses.  The blue crosses are similar to the red ones, but increasing the combined parameter $V_{\rm cir}(1-R+\lambda)h_{\rm z}^{-1}$ for 0.1 dex and keeping the amplitude of $B_{\rm tot}(r)$ to be the same for any fixed SFR (see Equation \ref{eq:26}) when running the model. 
   The red dashed line is the vertical shift of red solid line by 0.1 dex, which perfectly agrees with the production of the model (blue crosses). 
   }
   \label{fig:add}
\end{figure}

If we ignore the effects of $h_{\rm z}$, $V_{\rm cir}$ and $\lambda$ as discussed above, the scaling relation between $h_{\rm R}$ and the overall SFR of a galaxy is then expected to be 
\begin{equation} \label{eq:33}
h_{\rm R} \propto {\rm SFR}^{1-2\beta}. 
\end{equation}
The red crosses in Figure \ref{fig:add} show the scaling-relation that is obtained by running the model with (effectively) $\beta=0.3$ (see the detailed presentation in Appendix \ref{sec:A1}), which results in an exponent of 0.4.  This is in good agreement with our theoretical analysis. 

We have also examined the effect of the combined parameter $V_{\rm cir}(1-R+\lambda)h_{\rm z}^{-1}$ on the resulting $h_{\rm R}$, as suggested in Equation \ref{eq:32}. To do this, we run our model having increased this combined parameter by 0.1 dex, but keeping the amplitude of $B_{\rm tot}(r)$ fixed at given SFR for simplicity.  The result is shown in the blue crosses of Figure \ref{fig:add}, which is exactly 0.1 dex higher in than the red crosses.  This is again consistent with the prediction of Equation \ref{eq:32}.  

Observationally, star-forming galaxies are found along a tight sequence in the SFR-$M_*$ diagram, i.e. the so-called star formation Main Sequence (SFMS) \citep[e.g.][]{  Pannella-09, Stark-13, Renzini-15}. The SFMS is sub-linear and can be written as SFR $\propto M_*^{0.8}$.  According to Equation \ref{eq:33}, and taking the observed value of $\beta=0.3$, our model produces $h_{\rm R}\propto M_*^{0.32}$.  This is in good overall agreement with the observed mass-size relation \citep[e.g.][]{Shen-03, vanderWel-14}. 

We note that insertion of the observed value of $\beta \sim 0.3$ in Equation \ref{eq:24} appears to be important in order to reproduce the observed overall mass-size relation.  However, we stress that the scaling relation in this section needs further verification, since (a) we ignore the radially-dependent component in deriving it, and (b) the stellar mass dependence of $h_{\rm z}$ and mass-loading factor needs to be put on firmer ground. 

\section{Discussion} \label{sec:7}

\subsection{The comparison with previous work} \label{sec:7.1}

The early and still pervasive idea for the origin of exponential disks is based on conservation of angular momentum during the formation of the galaxy \citep[e.g.][]{Mestel-63, Freeman-70}, i.e. without any angular momentum exchange via viscosity.  However, this requires a very strong condition on the angular momentum of inflowing gas over a wide range of radii, and it is not known how the disk keeps its exponential form during the evolution.  In our model, the steady-state $\Sigma_{\rm SFR}$ does {\it not} depend on the angular momentum of inflowing gas, nor on the initial condition of the gas surface density. The viscous transportation of angular momentum is efficient in distributing the gas in the disk, and leads to exponential $\Sigma_{\rm SFR}$.  

Other authors have proposed that the exponential stellar disks are formed through secular evolution and the radial redistribution of stars, for instance, by a galactic bar or the stellar scattering by massive clumps \citep{Debattista-06, Foyle-08, Elmegreen-13, Wu-20}. However, it is clear that this cannot (at least directly) explain the exponential form of $\Sigma_{\rm SFR}$ and $\Sigma_{\rm gas}$. In contrast, our simple model naturally explains the exponential form of the star-formation disk $\Sigma_{\rm SFR}(r)$. The exponential form of the stellar (mass) disk presumably then follows from this. Further, the radial form of $\Sigma_{\rm gas}$ will also {\it follow} from this ``required" $\Sigma_{\rm SFR}$, as in any gas-regulator system \citep{Lilly-13}, with the form of $\Sigma_{\rm gas}(r)$ following from the star-formation law.  

The idea of galactic disks as viscous accretion disks was first proposed many years ago by \cite{Lin-87}.  The viscosity of the gas disk redistributes the angular momentum and causes the radial coplanar gas inflow \citep[e.g.][]{Yoshii-89, Firmani-96, Ferguson-01, Wang-09}.  However, these earlier works focused on viscosity derived from classical kinematic turbulence, and required strong and not necessarily justified assumptions about the viscous stress in order to form exponential stellar disks. 

Recently, \cite{Wittenburg-20} proposed an alternative way of forming  exponential disks under the MOND (Milgromian gravity) framework.  They simulated the formation of isolated disk galaxies from the collapse of gas clouds and obtained exponential stellar disks plus a compact bulge.  However, their model does not include the gas accretion in the evolution of galaxies, leading to the likely strong dependence of the output of their model on the initial conditions. 

In this work, we have shown that, based on typical magnetic field strengths in galaxies, it is MRI-induced viscosity that is likely to be the main mechanism of viscosity, rather than the kinematic turbulence, for producing and maintaining the exponential form of $\Sigma_{\rm SFR}$.  Our simple model can produce not only the exponential $\Sigma_{\rm SFR}$, but also produces reasonable scale-lengths.  It even plausibly produces the observed mass-size scaling relation, as shown in Section \ref{sec:6}.

In principle, we could push our model to high redshift and construct the cosmic evolution of the gas disk. However, most of the relevant parameters, such as disk scale-height, circular velocity, mass-loading factor and the $B_{\rm tot}-\Sigma_{\rm SFR}$ relation, are not well determined for high redshift galaxies. We therefore leave this extension of the model to the future.  

In addition, our model produces an exponential star-forming disk, rather than the exponential stellar disk. The question therefore arises whether the time-integration of $\Sigma_{\rm SFR}$ will also be exponential or not if the scalelength $h_{\rm R}$ of SF galaxies evolves significantly with time \citep[e.g.][]{vanderWel-14}.  \cite{Lilly-16} have shown that the exponential stellar disk plus a superposed bulge, and many other properties including the radial gradient of sSFR, are indeed obtained by integrating exponential $\Sigma_{\rm SFR}$ at different epochs, assuming that the evolution of $h_{\rm R}$ and overall SFR follow the relation indicated by observations.  


\subsection{What determines the size of gas disks?} \label{sec:7.2}

One of the key questions in galaxy formation is what determines the scalelengths of galactic disks.  In the framework of the modified accretion disk explored in this paper, the $h_{\rm R}$ depends on the total inflow rate (or SFR), magnetic field, scale-height, circular velocity and mass-loading factor of the gas disk, according to Equation \ref{eq:32}. 
In principle, the angular momentum of inflowing gas would intuitively have an effect on the ``size" of the gas disk. However, our analysis shows that the resulting $\Sigma_{\rm SFR}$ profile does {\it not} depend on the angular momentum of the inflowing gas. 
How to understand this?  As shown in Section \ref{sec:5.4X}, the angular momentum of the inflowing gas probably sets the {\it effective} outer boundary of the viscous accretion disk, i.e. the radius at which the material in the disk can first be considered to be rotationally supported.  At larger radii, the gas in the disk is not rotationally-supported, and is falling towards the center of the galaxy on more or less ballistic trajectories, conserving its angular momentum. But, eventually, rotational support will be achieved \citep[see][]{Trapp-21}.

This transition to rotational support may be seen as effectively setting the outer boundary of the viscous accretion disk.  In this sense, the infall of lower angular momentum results in a smaller boundary radius. 
To this extent, the angular momentum does play a role in determining the ``size'' (i.e. outer boundary) of the gas disk in our model.

However, as we have shown, {\it within} this outer boundary, neither the setting up of a steady-state exponential profile {\it nor} the value of the resulting steady-state scale-length $h_{\rm R}$ will be affected by the location of this outer boundary, i.e. by the angular momentum of the inflowing material.  Instead, we showed that $h_{\rm R}$ is determined primarily by the mass-inflow rate at the boundary $\Phi$, which will also be reflected in the steady-state overall SFR of the disk (as modified by any outflow). Other factors determining $h_{\rm R}$ include the overall strength of the magnetic fields, i.e. the normalization of the $\rm B_{\rm tot} - \Sigma_{\rm SFR}$ relation, the wind-loading factor $\lambda$ and other parameters such as the vertical scale height of the disk.


Hydrodynamical simulations have made great strides in reproducing many properties of galaxies \citep[e.g.][]{Schaye-15, Pillepich-18, Nelson-18}, but there are still some discrepancies between simulated disk galaxies and observed ones.  Based on the EAGLE simulation \citep[e.g.][]{Schaye-15}, \cite{Furlong-15} found that a predicted mass-size relation that is systematically shifted in the sense that the simulated galaxies were 0.2 dex larger (at a given mass) than real ones.
Similarly, by comparing the size-luminosity relations for observed SDSS galaxies and for simulated galaxies from the illustris simulation, \cite{Bottrell-17} also found, after matching the stellar mass, that the simulated galaxies were also roughly twice as large (0.3 dex) on average as real galaxies. Based on the MHD simulation IllustrisTNG, \cite{Genel-18} also found that both SF and quenched galaxies show larger sizes in the simulation, by up to $\sim$0.25 dex when compared with their real counterparts.  

The present work may provide a potential route to resolve these discrepancies. We have shown that a significant fraction of the angular momentum of the incoming angular momentum may be removed by MRI-induced viscous torques.  This reduction of the angular momentum of the disk material should reduce the size of the gas disks in the simulations by a corresponding factor compared with those obtained under the assumption of conservation of angular momentum, i.e. without the inclusion of magnetic MRI torques within the disks.

\section{Summary and Conclusions} \label{sec:8}

How cold gas is accreted onto galactic disks is one of the key things to understand in the formation and evolution of galaxies.  Both observational results and simulations indicate that the direction of gas inflow is preferentially within the plane of the disk, and that the outflow is bi-polar and perpendicular to the disk \citep{Brook-11, Kacprzak-12, Schroetter-19, Peroux-20, Trapp-21}.   Therefore, the galactic gas disk should be treatable as a ``modified accretion disk'' in which cold gas moves inwards towards the inner regions of the disk under the action of viscosity.
Along the way, the gas sustains the star formation in the disk (and any associated outflows of gas perpendicular to the disk).  The basic effect of the viscosity in such a disk is to transport mass inward and angular momentum outward.  

Such a galactic disk differs significantly from a classical accretion disk around a compact object such as a black hole.  Not least, the removal of gas by star-formation (plus any wind outflow from the disk) produces a strong dependence of the mass-inflow rate on radius. The inflow rate tends towards zero  as the center of the disk is approached (depending on the presence of any central sink).  The rotation curve of such a disk will also be quite different from the Keplerian form of a classic accretion disk.
The two basic equations governing such a disk are the continuity equation of gas mass and the transportation of angular momentum by viscosity, given here as Equations \ref{eq:1} and \ref{eq:3}. 

In order to reverse engineer the problem, we first looked at the analytical solution for the radial inflow that is required to produce a steady-state exponential star-forming disk, in Equation \ref{eq:9} (or Figure \ref{fig:2}).  We then obtained the analytical solution for the viscous stress that is required to sustain this radial inflow.  Interestingly, the required viscous stress is set by the star-formation profile and not by the gas profile.  As in any gas regulator system, the latter adjusts to produce the required star-formation profile according to whatever star-formation law is assumed. 

The required viscous stress $W(r)$ to produce an exponential star-forming disk is nearly constant over a wide range of radii, with a sharp drop only near the disk center (see Equation \ref{eq:12} or Figure \ref{fig:3}).  Any $W(r)$ of this form will eventually produce exponential disks, independent of the source of viscosity.

We then examined possible sources of viscous stress that could plausibly account for this nearly-flat behavior of $W(r)$.
Using observational inputs, we evaluated the likely size and radial dependence of the Newton stress, the Reynolds (or kinetic) stress and the Maxwell (or magnetic) stress across the disk of a typical Main Sequence galaxy. We concluded that only the magnetic stress induced by magneto-rotational instability (MRI) plausibly accounts for both the amplitude and the radial dependence of the required viscous stress (see Figure \ref{fig:4}).  This suggests that MRI-induced viscosity may be the dominant mechanism of mass and angular momentum transport in the gas disks of galaxies. 

Motivated by this insight from reverse engineering the origin of exponential disks, we then constructed a forward {\it ab initio} dynamic model of an MRI-driven modified accretion disk. The key part of this model is that the magnetic field strength at a given radius in the galaxy is determined by the instantaneous star-formation rate surface density, $\Sigma_{\rm SFR}$.  Guided by observations, we adopt a relation $B_{\rm tot}(r) \propto \Sigma_{\rm SFR}^{\alpha}$. The exponent $\alpha$ (here defined {\it within} a given system) is indicated on the basis of spatially resolved observations of synchrotron emission and $\Sigma_{\rm SFR}$ in nearby galaxies, to have a value $\alpha \sim 0.15$.

The radial gas inflow is therefore driven by a magnetic stress that is itself linked to the current distribution of gas via the $\Sigma_{\rm SFR}$ profile of the disk and the resulting magnetic field. This provides an effective but possibly complex feedback loop between $\Sigma_{\rm gas}$ and $\dot{\Sigma}_{\rm gas}$, i.e. between $\Sigma_{\rm SFR}$ and $\dot{\Sigma}_{\rm SFR}$. This feedback loop quickly drives the system towards the stable steady-state solutions of $\Sigma_{\rm SFR}(r)$. These solutions are of exponential form for the assumed value of $\alpha$. 

We then explored the properties of this model and found: 

\begin{itemize}

\item When fed at an outer boundary at a constant rate, the dynamic model reaches steady-state solutions within a few Gyr timescale, or less, even when starting from arbitrary and quite different initial conditions. The steady-state equilibrium state does not depend on the initial condition of the gas disk (see Figure \ref{fig:7} and \ref{fig:8}).  

\item The emergent steady-state $\Sigma_{\rm SFR}(r)$ disks have an almost exactly {\it exponential} form over $\sim$4-5 scale-lengths when the model is run with the adopted $B_{\rm tot}(r) - \Sigma_{\rm SFR}$ relation and with other observationally-motivated values for various parameters.   The exponential form of the $\Sigma_{\rm SFR}(r)$ profile evidently does {\it not} depend on the adopted star formation law, the value of the wind mass-loading factor $\lambda$, the overall normalization of $B_{\rm tot}-\Sigma_{\rm SFR}$ relation, or the chosen outer boundary of the gas disk,  or on other details of the model, including the precise form of the disk rotation curve.  Instead, it does depend quite sensitively on the choice of the exponent $\alpha$ in the $\Sigma_{\rm SFR}$ relation. 

\item The value of the exponential scale-length $h_{\rm R}$ depends primarily on the rate of feeding of the disk, on the overall normalization of the $B_{\rm tot}-\Sigma_{\rm SFR}$ relation, and on the overall circular velocity of the rotation curve (see the scaling-relation in Equation \ref{eq:32}).  With observationally-motivated values of the different parameters, the disk scale-lengths are in general comparable to the observed scale-lengths in galaxies, and can broadly reproduce the observed mass-size (${\rm M_*}-h_{\rm R}$) relation of disks.

\item The scale-length of the disk evidently does {\it not} depend on the chosen outer boundary of the disk at which mass is injected at the circular velocity, provided that this lies at several disk scale-lengths from the center of the system.  This means that the model is quite insensitive as to exactly where the inflowing gas transitions from ballistic trajectories to rotational support.   

\item Furthermore, it also means that the size, and thus the angular momentum, of the resultant stellar disk is not closely connected to the angular momentum of the inflowing material. Rather, the angular momentum of the disk is determined by the action of the viscous stresses in the disk in {\it removing} substantial amounts of angular momentum from the inflowing gaseous material.  Depending on where rotational support was achieved (i.e. effectively on the location of the outer boundary of the disk), some 57\% (if at 3$h_{\rm R}$) to 78\% (if at 8$h_{\rm R}$) of the angular momentum is removed from the disk material by the action of these viscous torques.

\item For achieving an exponential profile, the choice of exponent $\alpha$ in the model appears to be critical.  Only a rather small range of $\alpha$ around the mean $\alpha \sim 0.15$ that is indicated (with a dispersion of 0.06) from observations of the galaxy population successfully produces exponential profiles in $\Sigma_{\rm SFR}$ in the model.  The exponent $\alpha$ is not well understood theoretically, and is somewhat indirectly determined observationally on the basis of measurements of synchrotron emission in nearby galaxies. This apparent sensitivity to $\alpha$ is one of the most intriguing aspects of the model.

\end{itemize}

It should be noted that several of the conclusions in this paper do not depend on the applicability of MRI as the source of viscosity.  In particular, the radial flow of gas in the disk (Figure \ref{fig:2}) must be true for any exponential star-forming disk in which the dominant gas flow feeding the star-formation is co-planar.  Further, the requirement of a roughly constant viscous stress $W(r)$ in order to produce exponential disks, as shown in Figure \ref{fig:3}, is also quite independent from the question of the physical origin of the viscosity of the disk.

However, once these more general points are accepted, then MRI-induced viscosity emerges as a natural explanation for the  transportation of mass and angular momentum within galactic disks.  If this MRI-induced viscosity is linked via the magnetic field to the local star-formation surface density $\Sigma_{\rm SFR}$, as suggested by observations, then a feed-back loop can be established that quickly leads to stable steady-state configurations of $\Sigma_{\rm SFR}$.  Impressive steady-state exponential star-forming disks with reasonable scale-lengths are naturally obtained when using observationally-motivated values of the various parameters in the model.  The key parameter in producing exponential disks via MRI in our simple model is the exponent $\alpha$ in the $B_{\rm tot}-\Sigma_{\rm SFR}$ relation.  

While much work remains to be done to pin down the properties of magnetic fields in galaxies, and to test this model in detail using sophisticated magneto-hydrodynamic simulations, this idea may represent a major step towards solving the long-standing puzzle of the origin of the seemingly ubiquitous exponential profiles of star-forming disks. 

\acknowledgments

We thank Steven Balbus, Colin Norman and Scott Tremaine and the anonymous referee for their reading of an earlier version of the manuscript and their perceptive comments. Our interest in this topic was stimulated by illuminating discussions with Alvio Renzini and Gabriele Pezzulli.  

\bibliography{rewritebib.bib}

\begin{thebibliography}{133}
\expandafter\ifx\csname natexlab\endcsname\relax\def\natexlab#1{#1}\fi

\bibitem[{{Abramowicz} {et~al.}(1996){Abramowicz}, {Brandenburg}, \&
  {Lasota}}]{Abramowicz-96}
{Abramowicz}, M., {Brandenburg}, A., \& {Lasota}, J.-P. 1996, \mnras, 281, L21

\bibitem[{{Allen} \& {Martos}(1986)}]{Allen-86}
{Allen}, C., \& {Martos}, M.~A. 1986, RMxAA, 13, 137

\bibitem[{{Arshakian} {et~al.}(2009){Arshakian}, {Beck}, {Krause}, \&
  {Sokoloff}}]{Arshakian-09}
{Arshakian}, T.~G., {Beck}, R., {Krause}, M., \& {Sokoloff}, D. 2009, \aap,
  494, 21

\bibitem[{{Bacchini} {et~al.}(2019){Bacchini}, {Fraternali}, {Iorio}, \&
  {Pezzulli}}]{Bacchini-19}
{Bacchini}, C., {Fraternali}, F., {Iorio}, G., \& {Pezzulli}, G. 2019, \aap,
  622, A64

\bibitem[{{Bacchini} {et~al.}(2020){Bacchini}, {Fraternali}, {Iorio},
  {Pezzulli}, {Marasco}, \& {Nipoti}}]{Bacchini-20}
{Bacchini}, C., {Fraternali}, F., {Iorio}, G., {et~al.} 2020, \aap, 641, A70

\bibitem[{{Balbus} \& {Hawley}(1991)}]{Balbus-91}
{Balbus}, S.~A., \& {Hawley}, J.~F. 1991, \apj, 376, 214

\bibitem[{{Balbus} \& {Hawley}(1998)}]{Balbus-98}
---. 1998, Reviews of Modern Physics, 70, 1

\bibitem[{{Balbus} \& {Papaloizou}(1999)}]{Balbus-99}
{Balbus}, S.~A., \& {Papaloizou}, J. C.~B. 1999, \apj, 521, 650

\bibitem[{{Basu} \& {Roy}(2013)}]{Basu-13}
{Basu}, A., \& {Roy}, S. 2013, \mnras, 433, 1675

\bibitem[{{Beck} {et~al.}(2019){Beck}, {Chamandy}, {Elson}, \&
  {Blackman}}]{Beck-19}
{Beck}, R., {Chamandy}, L., {Elson}, E., \& {Blackman}, E.~G. 2019, Galaxies,
  8, 4

\bibitem[{{Beck} \& {Krause}(2005)}]{Beck-05}
{Beck}, R., \& {Krause}, M. 2005, Astronomische Nachrichten, 326, 414

\bibitem[{{Berkhuijsen} {et~al.}(2013){Berkhuijsen}, {Beck}, \&
  {Tabatabaei}}]{Berkhuijsen-13}
{Berkhuijsen}, E.~M., {Beck}, R., \& {Tabatabaei}, F.~S. 2013, \mnras, 435,
  1598

\bibitem[{{Berkhuijsen} {et~al.}(2016){Berkhuijsen}, {Urbanik}, {Beck}, \&
  {Han}}]{Berkhuijsen-16}
{Berkhuijsen}, E.~M., {Urbanik}, M., {Beck}, R., \& {Han}, J.~L. 2016, \aap,
  588, A114

\bibitem[{{Bigiel} {et~al.}(2010){Bigiel}, {Leroy}, {Walter}, {Blitz},
  {Brinks}, {de Blok}, \& {Madore}}]{Bigiel-10}
{Bigiel}, F., {Leroy}, A., {Walter}, F., {et~al.} 2010, \aj, 140, 1194

\bibitem[{{Bigiel} {et~al.}(2008){Bigiel}, {Leroy}, {Walter}, {Brinks}, {de
  Blok}, {Madore}, \& {Thornley}}]{Bigiel-08}
---. 2008, \aj, 136, 2846

\bibitem[{{Binney} {et~al.}(2000){Binney}, {Dehnen}, \& {Bertelli}}]{Binney-00}
{Binney}, J., {Dehnen}, W., \& {Bertelli}, G. 2000, \mnras, 318, 658

\bibitem[{{Blandford}(1989)}]{Blandford-NG}
{Blandford}, R.~D. 1989, 290, 35

\bibitem[{{Boomsma} {et~al.}(2008){Boomsma}, {Oosterloo}, {Fraternali}, {van
  der Hulst}, \& {Sancisi}}]{Boomsma-08}
{Boomsma}, R., {Oosterloo}, T.~A., {Fraternali}, F., {van der Hulst}, J.~M., \&
  {Sancisi}, R. 2008, \aap, 490, 555

\bibitem[{{Bordoloi} {et~al.}(2011){Bordoloi}, {Lilly}, {Knobel}, {Bolzonella},
  {Kampczyk}, {Carollo}, {Iovino}, {Zucca}, {Contini}, {Kneib}, {Le Fevre},
  {Mainieri}, {Renzini}, {Scodeggio}, {Zamorani}, {Balestra}, {Bardelli},
  {Bongiorno}, {Caputi}, {Cucciati}, {de la Torre}, {de Ravel}, {Garilli},
  {Kova{\v{c}}}, {Lamareille}, {Le Borgne}, {Le Brun}, {Maier}, {Mignoli},
  {Pello}, {Peng}, {Perez Montero}, {Presotto}, {Scarlata}, {Silverman},
  {Tanaka}, {Tasca}, {Tresse}, {Vergani}, {Barnes}, {Cappi}, {Cimatti},
  {Coppa}, {Diener}, {Franzetti}, {Koekemoer}, {L{\'o}pez-Sanjuan},
  {McCracken}, {Moresco}, {Nair}, {Oesch}, {Pozzetti}, \&
  {Welikala}}]{Bordoloi-11}
{Bordoloi}, R., {Lilly}, S.~J., {Knobel}, C., {et~al.} 2011, \apj, 743, 10

\bibitem[{{Bottrell} {et~al.}(2017){Bottrell}, {Torrey}, {Simard}, \&
  {Ellison}}]{Bottrell-17}
{Bottrell}, C., {Torrey}, P., {Simard}, L., \& {Ellison}, S.~L. 2017, \mnras,
  467, 2879

\bibitem[{{Bouch{\'e}} {et~al.}(2012){Bouch{\'e}}, {Hohensee}, {Vargas},
  {Kacprzak}, {Martin}, {Cooke}, \& {Churchill}}]{Bouche-12}
{Bouch{\'e}}, N., {Hohensee}, W., {Vargas}, R., {et~al.} 2012, \mnras, 426, 801

\bibitem[{{Bouch{\'e}} {et~al.}(2010){Bouch{\'e}}, {Dekel}, {Genzel}, {Genel},
  {Cresci}, {F{\"o}rster Schreiber}, {Shapiro}, {Davies}, \&
  {Tacconi}}]{Bouche-10}
{Bouch{\'e}}, N., {Dekel}, A., {Genzel}, R., {et~al.} 2010, \apj, 718, 1001

\bibitem[{{Boulanger} \& {Viallefond}(1992)}]{Boulanger-92}
{Boulanger}, F., \& {Viallefond}, F. 1992, \aap, 266, 37

\bibitem[{{Bovy} {et~al.}(2012){Bovy}, {Allende Prieto}, {Beers}, {Bizyaev},
  {da Costa}, {Cunha}, {Ebelke}, {Eisenstein}, {Frinchaboy}, {Garc{\'\i}a
  P{\'e}rez}, {Girardi}, {Hearty}, {Hogg}, {Holtzman}, {Maia}, {Majewski},
  {Malanushenko}, {Malanushenko}, {M{\'e}sz{\'a}ros}, {Nidever}, {O'Connell},
  {O'Donnell}, {Oravetz}, {Pan}, {Rocha-Pinto}, {Schiavon}, {Schneider},
  {Schultheis}, {Skrutskie}, {Smith}, {Weinberg}, {Wilson}, \&
  {Zasowski}}]{Bovy-12b}
{Bovy}, J., {Allende Prieto}, C., {Beers}, T.~C., {et~al.} 2012, \apj, 759, 131

\bibitem[{{Brook} {et~al.}(2011){Brook}, {Governato}, {Ro{\v{s}}kar},
  {Stinson}, {Brooks}, {Wadsley}, {Quinn}, {Gibson}, {Snaith}, {Pilkington},
  {House}, \& {Pontzen}}]{Brook-11}
{Brook}, C.~B., {Governato}, F., {Ro{\v{s}}kar}, R., {et~al.} 2011, \mnras,
  415, 1051

\bibitem[{{Bruzual} \& {Charlot}(2003)}]{Bruzual-03}
{Bruzual}, G., \& {Charlot}, S. 2003, \mnras, 344, 1000

\bibitem[{{Casasola} {et~al.}(2017){Casasola}, {Cassar{\`a}}, {Bianchi},
  {Verstocken}, {Xilouris}, {Magrini}, {Smith}, {De Looze}, {Galametz},
  {Madden}, {Baes}, {Clark}, {Davies}, {De Vis}, {Evans}, {Fritz}, {Galliano},
  {Jones}, {Mosenkov}, {Viaene}, \& {Ysard}}]{Casasola-17}
{Casasola}, V., {Cassar{\`a}}, L.~P., {Bianchi}, S., {et~al.} 2017, \aap, 605,
  A18

\bibitem[{{Conselice} {et~al.}(2013){Conselice}, {Mortlock}, {Bluck},
  {Gr{\"u}tzbauch}, \& {Duncan}}]{Conselice-13}
{Conselice}, C.~J., {Mortlock}, A., {Bluck}, A. F.~L., {Gr{\"u}tzbauch}, R., \&
  {Duncan}, K. 2013, \mnras, 430, 1051

\bibitem[{{Courteau}(1997)}]{Courteau-97}
{Courteau}, S. 1997, \aj, 114, 2402

\bibitem[{{Dav{\'e}} {et~al.}(2011){Dav{\'e}}, {Finlator}, \&
  {Oppenheimer}}]{Dave-11}
{Dav{\'e}}, R., {Finlator}, K., \& {Oppenheimer}, B.~D. 2011, \mnras, 416, 1354

\bibitem[{{de Blok} {et~al.}(2008){de Blok}, {Walter}, {Brinks},
  {Trachternach}, {Oh}, \& {Kennicutt}}]{de-Blok-08}
{de Blok}, W.~J.~G., {Walter}, F., {Brinks}, E., {et~al.} 2008, \aj, 136, 2648

\bibitem[{{de Vaucouleurs}(1959)}]{de-Vaucouleurs-59}
{de Vaucouleurs}, G. 1959, \apj, 130, 728

\bibitem[{{Debattista} {et~al.}(2006){Debattista}, {Mayer}, {Carollo}, {Moore},
  {Wadsley}, \& {Quinn}}]{Debattista-06}
{Debattista}, V.~P., {Mayer}, L., {Carollo}, C.~M., {et~al.} 2006, \apj, 645,
  209

\bibitem[{{DeFelippis} {et~al.}(2020){DeFelippis}, {Genel}, {Bryan}, {Nelson},
  {Pillepich}, \& {Hernquist}}]{DeFelippis-20}
{DeFelippis}, D., {Genel}, S., {Bryan}, G.~L., {et~al.} 2020, \apj, 895, 17

\bibitem[{{Dekel} \& {Birnboim}(2006)}]{Dekel-06}
{Dekel}, A., \& {Birnboim}, Y. 2006, \mnras, 368, 2

\bibitem[{{Dutton} \& {van den Bosch}(2009)}]{Dutton-09}
{Dutton}, A.~A., \& {van den Bosch}, F.~C. 2009, \mnras, 396, 141

\bibitem[{{Elmegreen} \& {Struck}(2013)}]{Elmegreen-13}
{Elmegreen}, B.~G., \& {Struck}, C. 2013, \apjl, 775, L35

\bibitem[{{Erwin} {et~al.}(2005){Erwin}, {Beckman}, \& {Pohlen}}]{Erwin-05}
{Erwin}, P., {Beckman}, J.~E., \& {Pohlen}, M. 2005, \apjl, 626, L81

\bibitem[{{Fall} \& {Efstathiou}(1980)}]{Fall-80}
{Fall}, S.~M., \& {Efstathiou}, G. 1980, \mnras, 193, 189

\bibitem[{{Ferguson} \& {Clarke}(2001)}]{Ferguson-01}
{Ferguson}, A.~M.~N., \& {Clarke}, C.~J. 2001, \mnras, 325, 781

\bibitem[{{Firmani} {et~al.}(1996){Firmani}, {Hernandez}, \&
  {Gallagher}}]{Firmani-96}
{Firmani}, C., {Hernandez}, X., \& {Gallagher}, J. 1996, \aap, 308, 403

\bibitem[{{Fletcher}(2010)}]{Fletcher-NG}
{Fletcher}, A. 2010, 438, 197

\bibitem[{{Foyle} {et~al.}(2008){Foyle}, {Courteau}, \& {Thacker}}]{Foyle-08}
{Foyle}, K., {Courteau}, S., \& {Thacker}, R.~J. 2008, \mnras, 386, 1821

\bibitem[{{Fraternali} {et~al.}(2015){Fraternali}, {Marasco}, {Armillotta}, \&
  {Marinacci}}]{Fraternali-15}
{Fraternali}, F., {Marasco}, A., {Armillotta}, L., \& {Marinacci}, F. 2015,
  \mnras, 447, L70

\bibitem[{{Freeman}(1970)}]{Freeman-70}
{Freeman}, K.~C. 1970, \apj, 160, 811

\bibitem[{{Furlong} {et~al.}(2015){Furlong}, {Bower}, {Theuns}, {Schaye},
  {Crain}, {Schaller}, {Dalla Vecchia}, {Frenk}, {McCarthy}, {Helly},
  {Jenkins}, \& {Rosas-Guevara}}]{Furlong-15}
{Furlong}, M., {Bower}, R.~G., {Theuns}, T., {et~al.} 2015, \mnras, 450, 4486

\bibitem[{{Genel} {et~al.}(2018){Genel}, {Nelson}, {Pillepich}, {Springel},
  {Pakmor}, {Weinberger}, {Hernquist}, {Naiman}, {Vogelsberger}, {Marinacci},
  \& {Torrey}}]{Genel-18}
{Genel}, S., {Nelson}, D., {Pillepich}, A., {et~al.} 2018, \mnras, 474, 3976

\bibitem[{{Gonz{\'a}lez Delgado} {et~al.}(2016){Gonz{\'a}lez Delgado}, {Cid
  Fernandes}, {P{\'e}rez}, {Garc{\'\i}a-Benito}, {L{\'o}pez Fern{\'a}ndez},
  {Lacerda}, {Cortijo-Ferrero}, {de Amorim}, {Vale Asari}, {S{\'a}nchez},
  {Walcher}, {Wisotzki}, {Mast}, {Alves}, {Ascasibar}, {Bland-Hawthorn},
  {Galbany}, {Kennicutt}, {M{\'a}rquez}, {Masegosa}, {Moll{\'a}},
  {S{\'a}nchez-Bl{\'a}zquez}, \& {V{\'\i}lchez}}]{Gonzalez-Delgado-16}
{Gonz{\'a}lez Delgado}, R.~M., {Cid Fernandes}, R., {P{\'e}rez}, E., {et~al.}
  2016, \aap, 590, A44

\bibitem[{{Gonz{\'a}lez-L{\'o}pezlira}
  {et~al.}(2012){Gonz{\'a}lez-L{\'o}pezlira}, {Pflamm-Altenburg}, \&
  {Kroupa}}]{Gonzalez-Lopezlira-12}
{Gonz{\'a}lez-L{\'o}pezlira}, R.~A., {Pflamm-Altenburg}, J., \& {Kroupa}, P.
  2012, \apj, 761, 124

\bibitem[{{Gressel} {et~al.}(2008){Gressel}, {Elstner}, {Ziegler}, \&
  {R{\"u}diger}}]{Gressel-08}
{Gressel}, O., {Elstner}, D., {Ziegler}, U., \& {R{\"u}diger}, G. 2008, \aap,
  486, L35

\bibitem[{{Guo} {et~al.}(2019){Guo}, {Peng}, {Shao}, {Fu}, {Catinella},
  {Cortese}, {Yuan}, {Yan}, {Zhang}, \& {Dou}}]{Guo-19}
{Guo}, K., {Peng}, Y., {Shao}, L., {et~al.} 2019, \apj, 870, 19

\bibitem[{{Hawley} {et~al.}(1999){Hawley}, {Balbus}, \& {Winters}}]{Hawley-99}
{Hawley}, J.~F., {Balbus}, S.~A., \& {Winters}, W.~F. 1999, \apj, 518, 394

\bibitem[{{Hawley} {et~al.}(1995){Hawley}, {Gammie}, \& {Balbus}}]{Hawley-95}
{Hawley}, J.~F., {Gammie}, C.~F., \& {Balbus}, S.~A. 1995, \apj, 440, 742

\bibitem[{{Heesen} {et~al.}(2014){Heesen}, {Brinks}, {Leroy}, {Heald}, {Braun},
  {Bigiel}, \& {Beck}}]{Heesen-14}
{Heesen}, V., {Brinks}, E., {Leroy}, A.~K., {et~al.} 2014, \aj, 147, 103

\bibitem[{{Herpich} {et~al.}(2017){Herpich}, {Tremaine}, \& {Rix}}]{Herpich-17}
{Herpich}, J., {Tremaine}, S., \& {Rix}, H.-W. 2017, \mnras, 467, 5022

\bibitem[{{Hohl}(1971)}]{Hohl-71}
{Hohl}, F. 1971, \apj, 168, 343

\bibitem[{{Hunter} \& {Elmegreen}(2006)}]{Hunter-06}
{Hunter}, D.~A., \& {Elmegreen}, B.~G. 2006, \apjs, 162, 49

\bibitem[{{Kacprzak} {et~al.}(2012){Kacprzak}, {Churchill}, \&
  {Nielsen}}]{Kacprzak-12}
{Kacprzak}, G.~G., {Churchill}, C.~W., \& {Nielsen}, N.~M. 2012, \apjl, 760, L7

\bibitem[{{Kennicutt}(1998)}]{Kennicutt-98}
{Kennicutt}, Robert~C., J. 1998, \apj, 498, 541

\bibitem[{{Kent}(1984)}]{Kent-84}
{Kent}, S.~M. 1984, \apjs, 56, 105

\bibitem[{{Kent}(1985)}]{Kent-85}
---. 1985, \apjs, 59, 115

\bibitem[{{Kere{\v{s}}} {et~al.}(2005){Kere{\v{s}}}, {Katz}, {Weinberg}, \&
  {Dav{\'e}}}]{Keres-05}
{Kere{\v{s}}}, D., {Katz}, N., {Weinberg}, D.~H., \& {Dav{\'e}}, R. 2005,
  \mnras, 363, 2

\bibitem[{{Kroupa}(2001)}]{Kroupa-01}
{Kroupa}, P. 2001, \mnras, 322, 231

\bibitem[{{Leroy} {et~al.}(2009){Leroy}, {Walter}, {Bigiel}, {Usero}, {Weiss},
  {Brinks}, {de Blok}, {Kennicutt}, {Schuster}, {Kramer}, {Wiesemeyer}, \&
  {Roussel}}]{Leroy-09}
{Leroy}, A.~K., {Walter}, F., {Bigiel}, F., {et~al.} 2009, \aj, 137, 4670

\bibitem[{{L'Huillier} {et~al.}(2012){L'Huillier}, {Combes}, \&
  {Semelin}}]{LHuillier-12}
{L'Huillier}, B., {Combes}, F., \& {Semelin}, B. 2012, \aap, 544, A68

\bibitem[{{Lilly} \& {Carollo}(2016)}]{Lilly-16}
{Lilly}, S.~J., \& {Carollo}, C.~M. 2016, \apj, 833, 1

\bibitem[{{Lilly} {et~al.}(2013){Lilly}, {Carollo}, {Pipino}, {Renzini}, \&
  {Peng}}]{Lilly-13}
{Lilly}, S.~J., {Carollo}, C.~M., {Pipino}, A., {Renzini}, A., \& {Peng}, Y.
  2013, \apj, 772, 119

\bibitem[{{Lin} \& {Pringle}(1987)}]{Lin-87}
{Lin}, D.~N.~C., \& {Pringle}, J.~E. 1987, \apjl, 320, L87

\bibitem[{{Lynden-Bell} \& {Pringle}(1974)}]{Lynden-Bell-74}
{Lynden-Bell}, D., \& {Pringle}, J.~E. 1974, \mnras, 168, 603

\bibitem[{{Mac Low} \& {Klessen}(2004)}]{Mac-Low-04}
{Mac Low}, M.-M., \& {Klessen}, R.~S. 2004, Reviews of Modern Physics, 76, 125

\bibitem[{{Mannucci}(2005)}]{Mannucci-NG}
{Mannucci}, F. 2005, 342, 140

\bibitem[{{McGaugh} {et~al.}(2000){McGaugh}, {Schombert}, {Bothun}, \& {de
  Blok}}]{McGaugh-00}
{McGaugh}, S.~S., {Schombert}, J.~M., {Bothun}, G.~D., \& {de Blok}, W.~J.~G.
  2000, \apjl, 533, L99

\bibitem[{{Meert} {et~al.}(2013){Meert}, {Vikram}, \& {Bernardi}}]{Meert-13}
{Meert}, A., {Vikram}, V., \& {Bernardi}, M. 2013, \mnras, 433, 1344

\bibitem[{{Meert} {et~al.}(2015){Meert}, {Vikram}, \& {Bernardi}}]{Meert-15}
---. 2015, \mnras, 446, 3943

\bibitem[{{Mestel}(1963)}]{Mestel-63}
{Mestel}, L. 1963, \mnras, 126, 553

\bibitem[{{Miller} {et~al.}(2011){Miller}, {Bundy}, {Sullivan}, {Ellis}, \&
  {Treu}}]{Miller-11}
{Miller}, S.~H., {Bundy}, K., {Sullivan}, M., {Ellis}, R.~S., \& {Treu}, T.
  2011, \apj, 741, 115

\bibitem[{{Mitchell} {et~al.}(2020){Mitchell}, {Schaye}, {Bower}, \&
  {Crain}}]{Mitchell-20}
{Mitchell}, P.~D., {Schaye}, J., {Bower}, R.~G., \& {Crain}, R.~A. 2020,
  \mnras, 494, 3971

\bibitem[{{Mo} {et~al.}(1998){Mo}, {Mao}, \& {White}}]{Mo-98}
{Mo}, H.~J., {Mao}, S., \& {White}, S. D.~M. 1998, \mnras, 295, 319

\bibitem[{{Mosleh} {et~al.}(2012){Mosleh}, {Williams}, {Franx}, {Gonzalez},
  {Bouwens}, {Oesch}, {Labbe}, {Illingworth}, \& {Trenti}}]{Mosleh-12}
{Mosleh}, M., {Williams}, R.~J., {Franx}, M., {et~al.} 2012, \apjl, 756, L12

\bibitem[{{Muratov} {et~al.}(2015){Muratov}, {Kere{\v{s}}},
  {Faucher-Gigu{\`e}re}, {Hopkins}, {Quataert}, \& {Murray}}]{Muratov-15}
{Muratov}, A.~L., {Kere{\v{s}}}, D., {Faucher-Gigu{\`e}re}, C.-A., {et~al.}
  2015, \mnras, 454, 2691

\bibitem[{{Nelson} {et~al.}(2018){Nelson}, {Pillepich}, {Springel},
  {Weinberger}, {Hernquist}, {Pakmor}, {Genel}, {Torrey}, {Vogelsberger},
  {Kauffmann}, {Marinacci}, \& {Naiman}}]{Nelson-18}
{Nelson}, D., {Pillepich}, A., {Springel}, V., {et~al.} 2018, \mnras, 475, 624

\bibitem[{{Noeske} {et~al.}(2007){Noeske}, {Weiner}, {Faber}, {Papovich},
  {Koo}, {Somerville}, {Bundy}, {Conselice}, {Newman}, {Schiminovich}, {Le
  Floc'h}, {Coil}, {Rieke}, {Lotz}, {Primack}, {Barmby}, {Cooper}, {Davis},
  {Ellis}, {Fazio}, {Guhathakurta}, {Huang}, {Kassin}, {Martin}, {Phillips},
  {Rich}, {Small}, {Willmer}, \& {Wilson}}]{Noeske-07}
{Noeske}, K.~G., {Weiner}, B.~J., {Faber}, S.~M., {et~al.} 2007, \apjl, 660,
  L43

\bibitem[{{Padoan} \& {Nordlund}(1999)}]{Padoan-99}
{Padoan}, P., \& {Nordlund}, {\r{A}}. 1999, \apj, 526, 279

\bibitem[{{Pannella} {et~al.}(2009){Pannella}, {Carilli}, {Daddi}, {McCracken},
  {Owen}, {Renzini}, {Strazzullo}, {Civano}, {Koekemoer}, {Schinnerer},
  {Scoville}, {Smol{\v{c}}i{\'c}}, {Taniguchi}, {Aussel}, {Kneib}, {Ilbert},
  {Mellier}, {Salvato}, {Thompson}, \& {Willott}}]{Pannella-09}
{Pannella}, M., {Carilli}, C.~L., {Daddi}, E., {et~al.} 2009, \apjl, 698, L116

\bibitem[{{Patra}(2020)}]{Patra-20}
{Patra}, N.~N. 2020, \mnras, 499, 2063

\bibitem[{{P{\'e}roux} {et~al.}(2020){P{\'e}roux}, {Nelson}, {van de Voort},
  {Pillepich}, {Marinacci}, {Vogelsberger}, \& {Hernquist}}]{Peroux-20}
{P{\'e}roux}, C., {Nelson}, D., {van de Voort}, F., {et~al.} 2020, \mnras, 499,
  2462

\bibitem[{{Pessah} {et~al.}(2008){Pessah}, {Chan}, \& {Psaltis}}]{Pessah-08}
{Pessah}, M.~E., {Chan}, C.-K., \& {Psaltis}, D. 2008, \mnras, 383, 683

\bibitem[{{Petric} \& {Rupen}(2007)}]{Petric-07}
{Petric}, A.~O., \& {Rupen}, M.~P. 2007, \aj, 134, 1952

\bibitem[{{Pezzulli} \& {Fraternali}(2016)}]{Pezzulli-16}
{Pezzulli}, G., \& {Fraternali}, F. 2016, \mnras, 455, 2308

\bibitem[{{Pillepich} {et~al.}(2018){Pillepich}, {Springel}, {Nelson}, {Genel},
  {Naiman}, {Pakmor}, {Hernquist}, {Torrey}, {Vogelsberger}, {Weinberger}, \&
  {Marinacci}}]{Pillepich-18}
{Pillepich}, A., {Springel}, V., {Nelson}, D., {et~al.} 2018, \mnras, 473, 4077

\bibitem[{{Pohlen} \& {Trujillo}(2006)}]{Pohlen-06}
{Pohlen}, M., \& {Trujillo}, I. 2006, \aap, 454, 759

\bibitem[{{Pringle}(1981)}]{Pringle-81}
{Pringle}, J.~E. 1981, \araa, 19, 137

\bibitem[{{Renzini} \& {Peng}(2015)}]{Renzini-15}
{Renzini}, A., \& {Peng}, Y.-j. 2015, \apjl, 801, L29

\bibitem[{{Sancisi} {et~al.}(2008){Sancisi}, {Fraternali}, {Oosterloo}, \& {van
  der Hulst}}]{Sancisi-08}
{Sancisi}, R., {Fraternali}, F., {Oosterloo}, T., \& {van der Hulst}, T. 2008,
  \aapr, 15, 189

\bibitem[{{Schaye} {et~al.}(2010){Schaye}, {Dalla Vecchia}, {Booth}, {Wiersma},
  {Theuns}, {Haas}, {Bertone}, {Duffy}, {McCarthy}, \& {van de
  Voort}}]{Schaye-10}
{Schaye}, J., {Dalla Vecchia}, C., {Booth}, C.~M., {et~al.} 2010, \mnras, 402,
  1536

\bibitem[{{Schaye} {et~al.}(2015){Schaye}, {Crain}, {Bower}, {Furlong},
  {Schaller}, {Theuns}, {Dalla Vecchia}, {Frenk}, {McCarthy}, {Helly},
  {Jenkins}, {Rosas-Guevara}, {White}, {Baes}, {Booth}, {Camps}, {Navarro},
  {Qu}, {Rahmati}, {Sawala}, {Thomas}, \& {Trayford}}]{Schaye-15}
{Schaye}, J., {Crain}, R.~A., {Bower}, R.~G., {et~al.} 2015, \mnras, 446, 521

\bibitem[{{Schleicher} \& {Beck}(2013)}]{Schleicher-13}
{Schleicher}, D. R.~G., \& {Beck}, R. 2013, \aap, 556, A142

\bibitem[{{Schroetter} {et~al.}(2019){Schroetter}, {Bouch{\'e}}, {Zabl},
  {Contini}, {Wendt}, {Schaye}, {Mitchell}, {Muzahid}, {Marino}, {Bacon},
  {Lilly}, {Richard}, \& {Wisotzki}}]{Schroetter-19}
{Schroetter}, I., {Bouch{\'e}}, N.~F., {Zabl}, J., {et~al.} 2019, \mnras, 490,
  4368

\bibitem[{{Seta} \& {Beck}(2019)}]{Seta-19}
{Seta}, A., \& {Beck}, R. 2019, Galaxies, 7, 45

\bibitem[{{Shakura} \& {Sunyaev}(1973)}]{Shakura-73}
{Shakura}, N.~I., \& {Sunyaev}, R.~A. 1973, \aap, 500, 33

\bibitem[{{Shen} {et~al.}(2003){Shen}, {Mo}, {White}, {Blanton}, {Kauffmann},
  {Voges}, {Brinkmann}, \& {Csabai}}]{Shen-03}
{Shen}, S., {Mo}, H.~J., {White}, S. D.~M., {et~al.} 2003, \mnras, 343, 978

\bibitem[{{Shi} {et~al.}(2011){Shi}, {Helou}, {Yan}, {Armus}, {Wu}, {Papovich},
  \& {Stierwalt}}]{Shi-11}
{Shi}, Y., {Helou}, G., {Yan}, L., {et~al.} 2011, \apj, 733, 87

\bibitem[{{Silk} \& {Mamon}(2012)}]{Silk-12}
{Silk}, J., \& {Mamon}, G.~A. 2012, Research in Astronomy and Astrophysics, 12,
  917

\bibitem[{{Simard} {et~al.}(2011){Simard}, {Mendel}, {Patton}, {Ellison}, \&
  {McConnachie}}]{Simard-11}
{Simard}, L., {Mendel}, J.~T., {Patton}, D.~R., {Ellison}, S.~L., \&
  {McConnachie}, A.~W. 2011, \apjs, 196, 11

\bibitem[{{Speagle} {et~al.}(2014){Speagle}, {Steinhardt}, {Capak}, \&
  {Silverman}}]{Speagle-14}
{Speagle}, J.~S., {Steinhardt}, C.~L., {Capak}, P.~L., \& {Silverman}, J.~D.
  2014, \apjs, 214, 15

\bibitem[{{Stark} {et~al.}(2013){Stark}, {Schenker}, {Ellis}, {Robertson},
  {McLure}, \& {Dunlop}}]{Stark-13}
{Stark}, D.~P., {Schenker}, M.~A., {Ellis}, R., {et~al.} 2013, \apj, 763, 129

\bibitem[{{Stark} {et~al.}(2009){Stark}, {McGaugh}, \& {Swaters}}]{Stark-09}
{Stark}, D.~V., {McGaugh}, S.~S., \& {Swaters}, R.~A. 2009, \aj, 138, 392

\bibitem[{{Stern} {et~al.}(2020){Stern}, {Fielding}, {Faucher-Gigu{\`e}re}, \&
  {Quataert}}]{Stern-20}
{Stern}, J., {Fielding}, D., {Faucher-Gigu{\`e}re}, C.-A., \& {Quataert}, E.
  2020, \mnras, 492, 6042

\bibitem[{{Stevens} {et~al.}(2016){Stevens}, {Croton}, \& {Mutch}}]{Stevens-16}
{Stevens}, A. R.~H., {Croton}, D.~J., \& {Mutch}, S.~J. 2016, \mnras, 461, 859

\bibitem[{{Stewart} {et~al.}(2011){Stewart}, {Kaufmann}, {Bullock}, {Barton},
  {Maller}, {Diemand}, \& {Wadsley}}]{Stewart-11}
{Stewart}, K.~R., {Kaufmann}, T., {Bullock}, J.~S., {et~al.} 2011, \apj, 738,
  39

\bibitem[{{Stone} {et~al.}(1998){Stone}, {Ostriker}, \& {Gammie}}]{Stone-98}
{Stone}, J.~M., {Ostriker}, E.~C., \& {Gammie}, C.~F. 1998, \apjl, 508, L99

\bibitem[{{Tabatabaei} {et~al.}(2013{\natexlab{a}}){Tabatabaei}, {Berkhuijsen},
  {Frick}, {Beck}, \& {Schinnerer}}]{Tabatabaei-13a}
{Tabatabaei}, F.~S., {Berkhuijsen}, E.~M., {Frick}, P., {Beck}, R., \&
  {Schinnerer}, E. 2013{\natexlab{a}}, \aap, 557, A129

\bibitem[{{Tabatabaei} {et~al.}(2013{\natexlab{b}}){Tabatabaei}, {Schinnerer},
  {Murphy}, {Beck}, {Groves}, {Meidt}, {Krause}, {Rix}, {Sandstrom}, {Crocker},
  {Galametz}, {Helou}, {Wilson}, {Kennicutt}, {Calzetti}, {Draine}, {Aniano},
  {Dale}, {Dumas}, {Engelbracht}, {Gordon}, {Hinz}, {Kreckel}, {Montiel}, \&
  {Roussel}}]{Tabatabaei-13b}
{Tabatabaei}, F.~S., {Schinnerer}, E., {Murphy}, E.~J., {et~al.}
  2013{\natexlab{b}}, \aap, 552, A19

\bibitem[{{Tabatabaei} {et~al.}(2017){Tabatabaei}, {Schinnerer}, {Krause},
  {Dumas}, {Meidt}, {Damas-Segovia}, {Beck}, {Murphy}, {Mulcahy}, {Groves},
  {Bolatto}, {Dale}, {Galametz}, {Sandstrom}, {Boquien}, {Calzetti},
  {Kennicutt}, {Hunt}, {De Looze}, \& {Pellegrini}}]{Tabatabaei-17}
{Tabatabaei}, F.~S., {Schinnerer}, E., {Krause}, M., {et~al.} 2017, \apj, 836,
  185

\bibitem[{{Tamburro} {et~al.}(2009){Tamburro}, {Rix}, {Leroy}, {Mac Low},
  {Walter}, {Kennicutt}, {Brinks}, \& {de Blok}}]{Tamburro-09}
{Tamburro}, D., {Rix}, H.~W., {Leroy}, A.~K., {et~al.} 2009, \aj, 137, 4424

\bibitem[{{Trapp} {et~al.}(2021){Trapp}, {Keres}, {Chan}, {Escala}, {Hummels},
  {Hopkins}, {Faucher-Giguere}, {Murray}, {Quataert}, \& {Wetzel}}]{Trapp-21}
{Trapp}, C., {Keres}, D., {Chan}, T.~K., {et~al.} 2021, arXiv e-prints,
  arXiv:2105.11472

\bibitem[{{Utomo} {et~al.}(2019){Utomo}, {Blitz}, \& {Falgarone}}]{Utomo-19}
{Utomo}, D., {Blitz}, L., \& {Falgarone}, E. 2019, \apj, 871, 17

\bibitem[{{van der Wel} {et~al.}(2014){van der Wel}, {Franx}, {van Dokkum},
  {Skelton}, {Momcheva}, {Whitaker}, {Brammer}, {Bell}, {Rix}, {Wuyts},
  {Ferguson}, {Holden}, {Barro}, {Koekemoer}, {Chang}, {McGrath},
  {H{\"a}ussler}, {Dekel}, {Behroozi}, {Fumagalli}, {Leja}, {Lundgren},
  {Maseda}, {Nelson}, {Wake}, {Patel}, {Labb{\'e}}, {Faber}, {Grogin}, \&
  {Kocevski}}]{vanderWel-14}
{van der Wel}, A., {Franx}, M., {van Dokkum}, P.~G., {et~al.} 2014, \apj, 788,
  28

\bibitem[{{Vera-Ciro} {et~al.}(2014){Vera-Ciro}, {D'Onghia}, {Navarro}, \&
  {Abadi}}]{Vera-Ciro-14}
{Vera-Ciro}, C., {D'Onghia}, E., {Navarro}, J., \& {Abadi}, M. 2014, \apj, 794,
  173

\bibitem[{{Vollmer} \& {Beckert}(2002)}]{Vollmer-02}
{Vollmer}, B., \& {Beckert}, T. 2002, \aap, 382, 872

\bibitem[{{Wada} {et~al.}(2002){Wada}, {Meurer}, \& {Norman}}]{Wada-02}
{Wada}, K., {Meurer}, G., \& {Norman}, C.~A. 2002, \apj, 577, 197

\bibitem[{{Wada} {et~al.}(2000){Wada}, {Spaans}, \& {Kim}}]{Wada-00}
{Wada}, K., {Spaans}, M., \& {Kim}, S. 2000, \apj, 540, 797

\bibitem[{{Wang} \& {Lilly}(2020)}]{Wang-20}
{Wang}, E., \& {Lilly}, S.~J. 2020, \apj, 892, 87

\bibitem[{{Wang} \& {Lilly}(2021)}]{Wang-21}
---. 2021, \apj, 910, 137

\bibitem[{{Wang} {et~al.}(2019){Wang}, {Lilly}, {Pezzulli}, \&
  {Matthee}}]{Wang-19}
{Wang}, E., {Lilly}, S.~J., {Pezzulli}, G., \& {Matthee}, J. 2019, \apj, 877,
  132

\bibitem[{{Wang} {et~al.}(2018){Wang}, {Li}, {Xiao}, {Lin}, {Bershady}, {Law},
  {Merrifield}, {Sanchez}, {Riffel}, {Riffel}, \& {Yan}}]{Wang-18}
{Wang}, E., {Li}, C., {Xiao}, T., {et~al.} 2018, \apj, 856, 137

\bibitem[{{Wang} {et~al.}(2009){Wang}, {Yan}, {Li}, {Chen}, {Xiang}, {Hu},
  {Ge}, \& {Zhang}}]{Wang-09}
{Wang}, J.-M., {Yan}, C.-S., {Li}, Y.-R., {et~al.} 2009, \apjl, 701, L7

\bibitem[{{Weiner} {et~al.}(2001){Weiner}, {Williams}, {van Gorkom}, \&
  {Sellwood}}]{Weiner-01}
{Weiner}, B.~J., {Williams}, T.~B., {van Gorkom}, J.~H., \& {Sellwood}, J.~A.
  2001, \apj, 546, 916

\bibitem[{{Wilson} {et~al.}(2019){Wilson}, {Elmegreen}, {Bemis}, \&
  {Brunetti}}]{Wilson-19}
{Wilson}, C.~D., {Elmegreen}, B.~G., {Bemis}, A., \& {Brunetti}, N. 2019, \apj,
  882, 5

\bibitem[{{Wittenburg} {et~al.}(2020){Wittenburg}, {Kroupa}, \&
  {Famaey}}]{Wittenburg-20}
{Wittenburg}, N., {Kroupa}, P., \& {Famaey}, B. 2020, \apj, 890, 173

\bibitem[{{Wu} {et~al.}(2020){Wu}, {Struck}, {D'Onghia}, \&
  {Elmegreen}}]{Wu-20}
{Wu}, J., {Struck}, C., {D'Onghia}, E., \& {Elmegreen}, B.~G. 2020, \mnras,
  499, 2672

\bibitem[{{Wyder} {et~al.}(2009){Wyder}, {Martin}, {Barlow}, {Foster},
  {Friedman}, {Morrissey}, {Neff}, {Neill}, {Schiminovich}, {Seibert},
  {Bianchi}, {Donas}, {Heckman}, {Lee}, {Madore}, {Milliard}, {Rich}, {Szalay},
  \& {Yi}}]{Wyder-09}
{Wyder}, T.~K., {Martin}, D.~C., {Barlow}, T.~A., {et~al.} 2009, \apj, 696,
  1834

\bibitem[{{Yoshii} \& {Sommer-Larsen}(1989)}]{Yoshii-89}
{Yoshii}, Y., \& {Sommer-Larsen}, J. 1989, \mnras, 236, 779

\end{thebibliography}

\appendix 

\section{The validation of ${\rm \Phi}-h_{\rm R}$ relation through running the model} \label{sec:A1}
\begin{figure*}
  \begin{center}
    \epsfig{figure=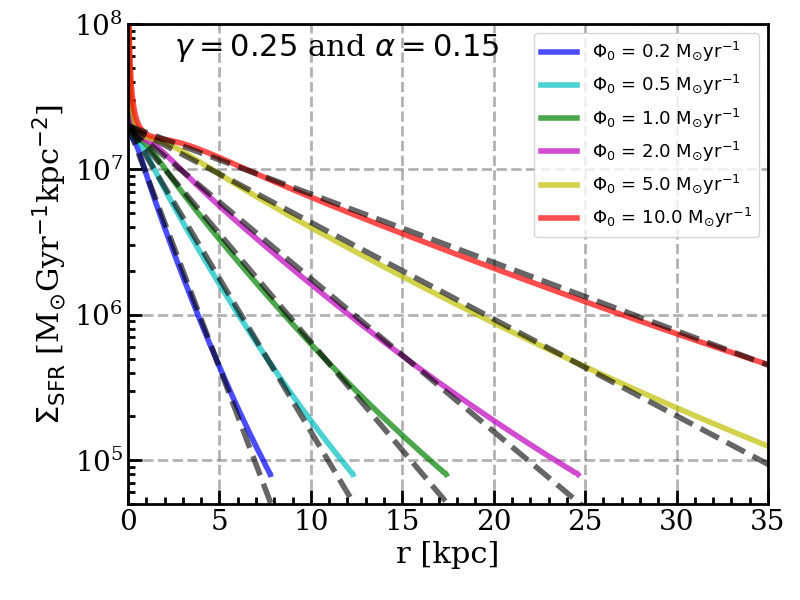,clip=true,width=0.45\textwidth}
    \epsfig{figure=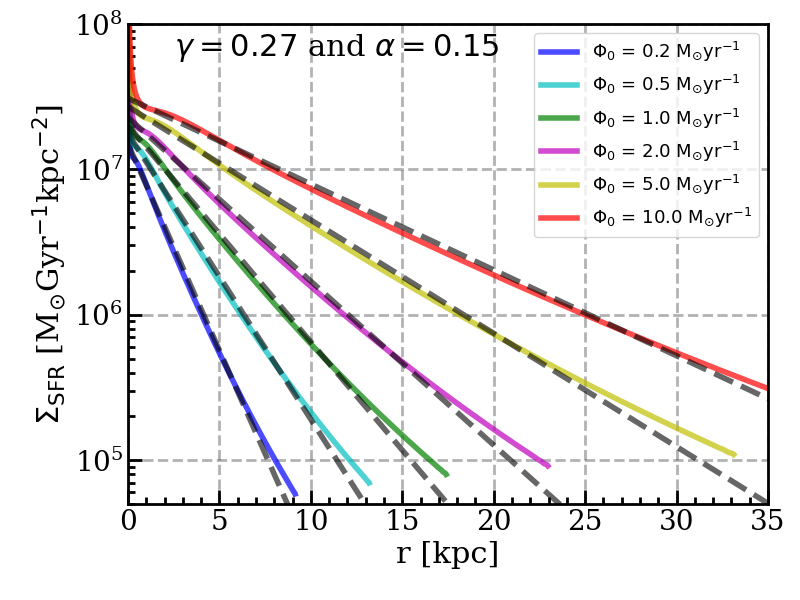,clip=true,width=0.45\textwidth}
    \epsfig{figure=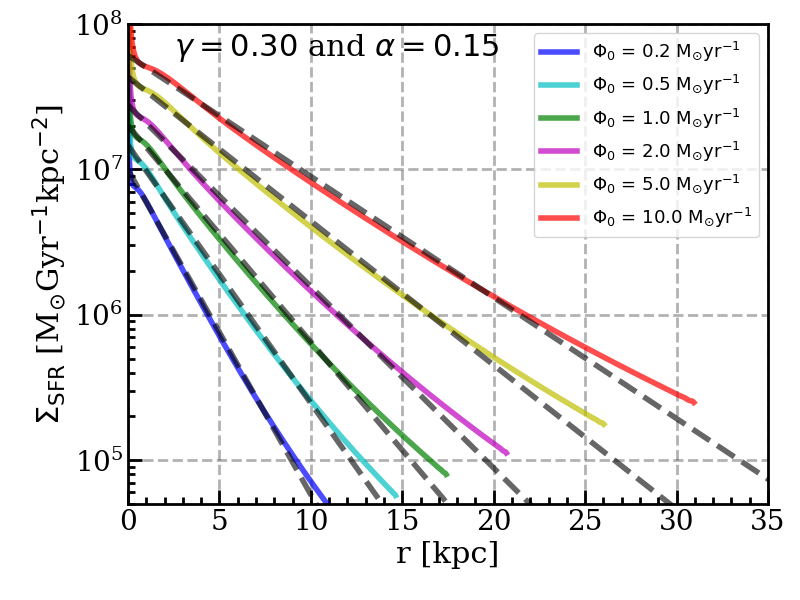,clip=true,width=0.45\textwidth}
    \epsfig{figure=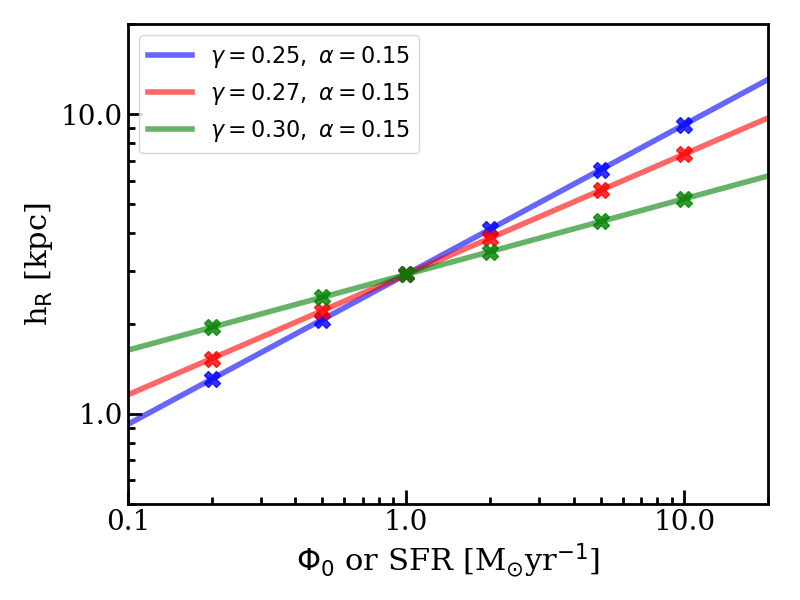,clip=true,width=0.45\textwidth}
    \end{center}
  \caption{Verification of the scaling relation.  The first three panels show the steady-state  $\Sigma_{\rm SFR}(r)$ at a set of different $\Phi_0$ with $\gamma=$0.25, 0.27 and 0.30. In each panel, different colors are for different $\Phi_0$, as denoted in the top right of the panel.  Each colored line of steady-state $\Sigma_{\rm SFR}$ is compared with a gray dashed line, which is  determined by the resulting $\Sigma_{\rm SFR}$ at two radii ($r=0.125R_{\rm b}$ and $0.625R_{\rm b}$). These dashed lines are used to characterize the $h_{\rm R}$ of the steady-state $\Sigma_{\rm SFR}$. 
  Bottom right panel: the $\Phi_0$-$h_{\rm R}$ (or SFR-$h_{\rm R}$) relation for three different $\gamma$ produced by the dynamic model.  Different colors are for different settings of $\gamma$.  The colored solid lines are the linear fittings of these mocked data points.  We note that $\gamma$=0.25, 0.27 and 0.30 correspond to $\beta$= 0.25, 0.30 and 0.375 based on Equation \ref{eq:35}, respectively. 
  } 
  \label{fig:14}
\end{figure*}

In this section, we try to validate the broad scaling-relation obtained in Section \ref{sec:6.1}, i.e. Equation \ref{eq:33}, by numerically running the model. We emphasize that this section does not therefore provide independent new information. 

The Equation \ref{eq:26} was derived assuming an exponential $\Sigma_{\rm SFR}$, and is not therefore strictly applicable for galaxies with other forms of $\Sigma_{\rm SFR}$. It is also more convenient to re-write Equation \ref{eq:26} in terms of the absolute $\Sigma_{\rm SFR}$ rather than the $\Sigma_{\rm SFR}(h_{\rm R})$. We can therefore write our general relation of $B_{\rm tot}-\Sigma_{\rm SFR}$ for the disk galaxies as: 
\begin{equation} \label{eq:27}
B_{\rm tot}(r) = A \times (\frac{\rm SFR}{\rm 1\ M_{\odot}yr^{-1}})^{\gamma } \cdot (\frac{\Sigma_{\rm SFR}(r)}{\rm 0.01\ M_{\odot}yr^{-1}kpc^{-2}})^{\alpha}. 
\end{equation}
The exponent of the integrated SFR is now denoted $\gamma$ in Equation \ref{eq:27}, rather than $\beta$ in Equation \ref{eq:26}. This is because an increase of the integrated SFR is likely along with an increase of $\Sigma_{\rm SFR}$, and therefore $\gamma$ is expected to be slightly less than $\beta$. 
Therefore, Equation \ref{eq:26} is not fully equivalent to Equation \ref{eq:27}. 

In practice, we adopt the Equation \ref{eq:27} in our model, rather than Equation \ref{eq:26} in running the model. In a similar way as in Section \ref{sec:6.1}, we can establish the scaling relation of $h_{\rm R}$ and the integrated SFR with $\gamma$ and $\alpha$, regardless of the other parameters: 
\begin{equation} \label{eq:34}
h_{\rm R} \propto {\rm SFR}^{1-\frac{2(\gamma-\alpha)}{1-4\alpha}}.
\end{equation}
Although the Equation \ref{eq:26} and \ref{eq:27} are not fully equivalent, we can build the relation between $\beta$ and $\gamma$, if we assume the two equations lead to the same scaling relation of ${\rm SFR}-h_{\rm R}$ relation: 
\begin{equation} \label{eq:35}
    \beta = \frac{\gamma-\alpha}{1-4\alpha}.
\end{equation}
We note that the Equation \ref{eq:35} can also be obtained in an alternative way, by assuming that Equations \ref{eq:26} and \ref{eq:27} give the same scaling relation of $B_{\rm tot}-{\rm SFR}$. 

The overall settings are similar as in the fiducial run.  Specifically, the circular velocity and mass-loading factor are set to be the same as in the fiducial run, equivalent to assuming that the effects of these two factors cancel out across galaxies of different stellar mass (or SFR). This also ensures that the feeding rate of the accretion disk is equal to the overall SFR of the system in the steady-state.  We set $R_{\rm t}=0.5h_{\rm R}$, $R_{\rm z}=3h_{\rm R}$ and $R_{\rm b}=6h_{\rm R}$ in the run, which is useful to eliminate the differences introduced by the radial-dependent component.  
We adopt the Equation \ref{eq:27} with $\alpha=0.15$ and $A=14.0 \ \mu G$ (i.e. $Y/k$ in Section \ref{sec:6.0}). We then run the model with three different $\gamma =$ 0.25, 0.27 and 0.30, to see whether our model can reproduce the scaling-relations given in Section \ref{sec:6.1}. 

Since the settings of parameters is connected with the resulting $h_{\rm R}$ in the run, we first run the model iteratively for an individual system. Typically after a few iterations, we find the output $h_{\rm R}$ agrees with the input $h_{\rm R}$ to better than 1\%.  Specifically, we find the model results in an exponential $\Sigma_{\rm SFR}$ with $h_{\rm R}=2.9$ kpc at $\Phi_0=1\ {\rm M_{\odot}yr^{-1}}$.  Then, we can predict the $h_{\rm R}$ for any other $\Phi_0$ based on Equation \ref{eq:34} for a given $\gamma$. Inputting the predicted $h_{\rm R}$ and running the model, we are able to examine whether the output $h_{\rm R}$ follows the Equation \ref{eq:34} or not.

The first three panels of Figure \ref{fig:14} shows the resulting $\Sigma_{\rm SFR}$ for a set of different $\Phi_0$ (different colors) and for different $\gamma$ (different panels).  It is clear that for all the runs, the steady-state $\Sigma_{\rm SFR}$ is very nearly in exponential form.  In each panel, we see a clear dependence of $h_{\rm R}$ on $\Phi_0$ (or SFR).  This dependence becomes weaker with increasing $\gamma$. 
The bottom right panel of Figure \ref{fig:14} shows the $\Phi_0$-$h_{\rm R}$ (or SFR-$h_{\rm R}$) relation for three different $\gamma$ produced by the dynamic model.  The $h_{\rm R}$ is measured at two radii of the resulting $\Sigma_{\rm SFR}$, 0.125$R_{\rm b}$ and 0.625$R_{\rm b}$. 
We perform linear fits to the SFR-$h_{\rm R}$ relations of different $\gamma$.  Strikingly, the relations in the bottom right panel of Figure \ref{fig:14} are $h_{\rm R}\propto {\rm SFR}^{0.50}$, $h_{\rm R}\propto {\rm SFR}^{0.40}$ and $h_{\rm R}\propto {\rm SFR}^{0.25}$ for $\gamma=$0.25, 0.27 and 0.30, respectively. This result is exactly the same as the theoretical analysis in Section \ref{sec:6.1}.  We note that $\gamma$=0.25, 0.27 and 0.30 correspond to $\beta$= 0.25, 0.30 and 0.375 based on Equation \ref{eq:35}, respectively.  The $\beta$ is $\sim$0.3 suggested from the observation (see Figure \ref{fig:6}).  

\end{document}